\documentclass[12pt]{article}
\pdfoutput=1

\usepackage{epsf}
\usepackage{epsfig,graphics}
\usepackage {graphicx}
\usepackage {epsfig}
\usepackage {subfigure}
\usepackage {tabularx} 
\usepackage{rotate}	
\usepackage{slashed}
\usepackage{color}

\usepackage{amsmath}
\usepackage{amsfonts}
\usepackage{amssymb}
\usepackage{graphicx}
\usepackage{cite}

\usepackage{fancyhdr}
\usepackage{hyperref}

\newcommand{\bmat}{\left(\begin{array}}
\newcommand{\emat}{\end{array}\right)}

\def\yzero{\smash{\hbox{$y\kern-4pt\raise1pt\hbox{${}^\circ$}$}}}
\def\p{\partial}
\def\a{\alpha}
\def\b{\beta}
\def\g{\gamma}
\def\d{\delta}
\def\beq{\begin{equation}}
\def\eeq{\end{equation}}
\def\beqa{\begin{eqnarray}}
\def\eeqa{\end{eqnarray}}
\def\Om{\Omega}
\def\om{\omega}
\def\th{\theta}

\def\-{\hphantom{-}}

\def\s2{\frac{1}{\sqrt2}}

\def\oh{\frac{1}{2}}
\def\beq{\begin{equation}}
\def\eeq{\end{equation}}
\def\beqa{\begin{eqnarray}}
\def\eeqa{\end{eqnarray}}

\def\IF{\relax{\rm I\kern-.18em F}}
\def\II{\relax{\rm I\kern-.18em I}}
\def\IP{\relax{\rm I\kern-.18em P}}
\def\IC{\relax\hbox{\kern.25em$\inbar\kern-.3em{\rm C}$}}
\def\IR{\relax{\rm I\kern-.18em R}}

\def\cc{{\cal C}}

\def\cf{{\cal F}}

\def\Dsl{\,\raise.15ex\hbox{/}\mkern-13.5mu D} 
\def\IZ{\mathbb{Z}}

\def\CH {{\cal H}}
\def\CK {{\cal K}}
\def\CM {{\cal M}}
\def\CR {{\cal R}}
\def\CN {{\cal N}}
\def\CF {{\cal F}}

\def\CQ {{\cal Q}}
\def\re{\mbox{Re}}
\def\im{\mbox{Im}}

\def\be{\begin{equation}}
\def\ee{\end{equation}}
\def\bea{\begin{eqnarray}}
\def\eea{\end{eqnarray}}
\def\bes{\begin{subequations}}
\def\ees{\end{subequations}}
\def\raw{\rightarrow}
\def\lraw{\leftrightarrow}
\def\Raw{\Rightarrow}

\def\IC{\mathbb{C}}

\def\IZ{\mathbb{Z}}
\def\IR{\mathbb{R}}
\def\IP{\mathbb{P}}

\def\oh{\frac{1}{2}}
\def\a{{\alpha}}
\def\b{{\beta}}
\def\d{{\delta}}
\def\eps{{\epsilon}}
\def\th{{\theta}}
\def\Lam{{\Lambda}}
\def\lam{{\lambda}}
\def\Om{{\Omega}}
\def\om{{\omega}}
\def\sig{{\sigma}}

\def\g{{\gamma}}

\def\p{{\partial}}

\catcode`\@=11   
\newdimen\@rotdimen
\newbox\@rotbox  

\def\@vspec#1{\special{ps:#1}}
\def\@rotstart#1{\@vspec{gsave currentpoint currentpoint translate
   #1 neg exch neg exch translate}}
\def\@rotfinish{\@vspec{currentpoint grestore moveto}}
\def\@rotr#1{\@rotdimen=\ht#1\advance\@rotdimen by\dp#1%
   \hbox to\@rotdimen{\hskip\ht#1\vbox to\wd#1{\@rotstart{90 rotate}%
   \box#1\vss}\hss}\@rotfinish}
\def\@rotl#1{\@rotdimen=\ht#1\advance\@rotdimen by\dp#1%
   \hbox to\@rotdimen{\vbox to\wd#1{\vskip\wd#1\@rotstart{270 rotate}%
   \box#1\vss}\hss}\@rotfinish}%
\def\@rotu#1{\@rotdimen=\ht#1\advance\@rotdimen by\dp#1%
   \hbox to\wd#1{\hskip\wd#1\vbox to\@rotdimen{\vskip\@rotdimen
   \@rotstart{-1 dup scale}\box#1\vss}\hss}\@rotfinish}%
\def\@rotf#1{\hbox to\wd#1{\hskip\wd#1\@rotstart{-1 1 scale}%
   \box#1\hss}\@rotfinish}%
\def\rotate{\@ifnextchar[{\@rotate}{\@rotate[l]}}
\def\@rotate[#1]#2{\setbox\@rotbox=\hbox{#2}\@nameuse{@rot#1}\@rotbox}

\catcode`\@=12

\topmargin
-1.5cm
\textwidth
15.5cm
\textheight
23.5cm
\oddsidemargin
0.7cm
\evensidemargin
0.7cm

\begin{document}

\makeatletter
\@addtoreset{equation}{section}
\makeatother
\renewcommand{\theequation}{\thesection.\arabic{equation}}
\pagestyle{empty}
\vspace{-0.2cm}
\rightline{ IFT-UAM/CSIC-18-016}
\vspace{1.2cm}
\begin{center}


\LARGE{ The Type IIA  Flux Potential,\\ 
4-forms and Freed-Witten anomalies  \\ [13mm]}

  \large{Alvaro Herr\'aez$,^{a,b}$ Luis E. Ib\'a\~nez$,^{a,b}$ Fernando Marchesano$^{a}$ and Gianluca Zoccarato$^{c,d}$ \\[6mm]}
  \small{
${}^a$ Instituto de F\'{\i}sica Te\'orica UAM-CSIC, Cantoblanco, 28049 Madrid, Spain \\[1mm] 
${}^b$Departamento de F\'{\i}sica Te\'orica, Universidad Aut\'onoma de Madrid, 28049 Madrid, Spain  \\[1mm] 
${}^c$Department of Physics, University of Wisconsin, Madison, WI 53706, USA \\[1mm] 
${}^d$ Institute for Advanced Study, The Hong Kong University
  of Science and Technology, \\ Hong Kong, China
  \\[6mm]} 
\small{\bf Abstract} \\[6mm]
\end{center}
\begin{center}
\begin{minipage}[h]{15.22cm}
We compute the full classical 4d scalar potential of type IIA Calabi-Yau orientifolds in the presence of fluxes and D6-branes. We show that it can be written as a bilinear form $V = Z^{AB} \rho_A\rho_B$, where the $\rho_A$ are in one-to-one correspondence with the 4-form fluxes of the 4d effective theory. The $\rho_A$ only depend on the internal fluxes, the axions and the topological data of the compactification, and are fully determined by the Freed-Witten anomalies of branes that appear as 4d string defects. The quadratic form $Z^{AB}$ only depends on the saxionic partners of these axions. In general, the $\rho_A$ can be seen as the basic invariants under the discrete shift symmetries of the 4d effective theory, and therefore the building blocks of any flux-dependent quantity. All these polynomials may be obtained by derivation from one of them, associated to 
a universal 4-form. The standard $\CN=1$ supergravity flux superpotential is uniquely determined from this {\it master polynomial},  and vice versa.

\end{minipage}
\end{center}
\newpage
\setcounter{page}{1}
\pagestyle{plain}
\renewcommand{\thefootnote}{\arabic{footnote}}
\setcounter{footnote}{0}



\tableofcontents


\section{Introduction}

The existence of plenty of quantised flux degrees of freedom  supports the idea of a large {\it landscape} of vacuum solutions
in string theory.  This fact  led to the proposal of Bousso and Polchinski  \cite{BP} (building on previous ideas of  Brown and Teitelboim \cite{BT}) 
for an understanding of the smallness of the cosmological constant $\Lambda_4$. They argued that in string theory there are plenty of 
non-propagating 3-forms  $C_3^A$ from the RR and NS closed string sector.  Although they do not propagate their (quantised) 
fields strengths $F_4^A$ contribute to the vacuum energy, so that the scalar potential of the observed physics would have a structure\footnote{For pioneering and more recent work on 4-forms see \cite{pioneros,morerecent,KS,Groh:2012tf,KLS,Marchesano:2014mla,Dudas:2014pva,Bielleman:2015ina,Garcia-Valdecasas:2016voz,Carta:2016ynn,Valenzuela:2016yny,Farakos:2017jme} and references therein.}
\beq
V_{BP}\ =\ \sum_{A,B}   Z_{AB}F_4^AF_4^B \  + \  \Lambda_0
\label{BP}
\eeq
where the sums run over all quantised  4-form fluxes $F_4^A$, and $Z_{AB}$ is a positive definite metric depending on all moduli. Here $\Lambda_0$ is some, large (of order $M_p$) and typically negative {\it bare} contribution to the cosmological constant. They showed  that for a sufficiently large number of
fluxes,  there are choices  resulting in a cosmological constant exponentially small.  This approach assumes that the moduli are somehow fixed, so
one actually needs a mechanism for fixing all moduli before addressing the c.c. issue.  

A lot of work, starting with the work in KKLT \cite{KKLT} (see also \cite{LVS,Douglas:2006es,Quevedo:2014xia}), has been dedicated to study full moduli fixing in the context of Type IIB orientifolds. The  complex structure and dilaton fields are fixed by the generic presence of NS and RR closed string fluxes, whereas K\"ahler moduli are fixed by non-perturbative effects. In this way one obtains AdS vacua which must be later on up-lifted by the addition in the background of anti-D-branes or other mechanism providing a positive energy.  In the IIB route map the precise form of the scalar potential in eq.(\ref{BP}) is not obvious. In particular the role of the 4-forms as in the BP mechanism is not apparent although in principle the c.c. may be made small by an analogous mechanism.

The case of  the flux scalar potential for Type IIA orientifolds has been less explored. It has the shortcoming that the mathematical structure of the compactification geometry in the presence of general fluxes is non-trivial. On the other hand, in Calabi-Yau orientifold compactifications the standard RR and NS flux superpotential involves {\it both} K\"ahler moduli and
complex structure fields, offering the possibility of fixing all moduli just by fluxes, without resorting to any non-perturbative effects. Indeed, examples of 
AdS vacua with all moduli fixed have been obtained in the literature, both SUSY and non-SUSY
\cite{DeWolfe:2005uu,Camara:2005dc,Villadoro:2005cu,moredual}. No   dS vacua have been obtained with just standard 
RR and NS fluxes, although generalised non-geometric \cite{Shelton:2005cf} and S-dual \cite{moredual} fluxes could perhaps allow for such vacua, see 
\cite{guarino,Danielsson:2012by}.  Still, the study of Type IIA orientifold 
vacua has been so far much more incomplete. 

In the present paper we revisit and study in detail the structure of the flux potential in Type IIA orientifolds, extending previous analysis in various directions.  A prominent role in our results is played by the four-dimensional 4-forms of the theory, which appear in  a form reminiscent of that  in the BP mechanism. In fact we find that the part of the action density relevant to the scalar potential has the qualitative structure 
\beq
-\ Z_{AB}F_4^AF_4^B\  + \ 2 F_4^A{\rho}_A \ -\  Z^{AB}{\rho}_A{\rho}_B \ ,
\eeq
where the index $A$ runs over all the fluxes of the compactification or equivalently over the four-dimensional four-forms $F_4^A$. On the one hand, $\rho_A$ are integer polynomials on the unit-period axions of the theory, whose coefficients depend only on flux quanta and other topological data. On the other hand, the tensor $Z_{AB}$ depend only on the saxions of the theory. After integrating the equations of motion for the 4-forms one obtains for the scalar potential an expression of the form
\beq
V\ =\ \frac {1}{8\kappa_4^2}Z^{AB}\rho_A\rho_B \ . 
\eeq
In fact one can perform an axion dependent rotation ${\bf R}$ so that $\vec{q}={\bf R^t}\vec{\rho}$ is a vector containing only the different flux quanta. Then the new metric  is given by ${\bf Z'}={\bf R^t}{\bf Z}{\bf R}$  and depends both on the axions and the saxions of the compactification. One then has a bilinear factorised structure for the scalar potential, reminiscent of the BP structure. There are however a number of differences with respect to BP, in particular the metric ${\bf Z'}$ is not positive definite. This bilinear structure in terms of 4-forms was already shown to appear  for the piece of the potential coming from RR and NS in \cite{Bielleman:2015ina}. In this paper we show that the full potential, including the contribution due to the presence of localised sources may still be written in this form. In this generalisation, that also extends the analysis in \cite{Carta:2016ynn}, the inclusion of open string moduli and fluxes is particularly delicate and requires a careful treatment of the redefinition of the open and closed string axions. In fact, we find a expression for the 4d holomorphic variables that differs from previous proposals in the literature \cite{Carta:2016ynn,Grimm:2011dx,Kerstan:2011dy}, but which is essential to obtain a holomorphic flux superpotential. 

This factorised bilinear structure makes more transparent the discrete symmetries of the effective theory. In particular we find that the transformation of axions and fluxes under discrete shift symmetries are encapsulated in the rotation matrix ${\bf R}$. Moreover, the fact that one can write $\vec{\rho} = {\bf R}^{t\, -1} \vec{q}$ is a consequence of gauge invariance at the microscopic level, and can be translated into the anomalies developed by the different branes of the theory. In particular the matrix {\bf R} is specified by the Freed-Witten anomalies \cite{Freed:1999vc,Maldacena:2001xj} of 4d strings coupling to the axions in the presence of RR and NS fluxes. Those anomalies are cured by 4d domain walls ending on such 4d strings \cite{BerasaluceGonzalez:2012zn}, and coupling to the 4d 3-forms of the effective theory. 

The present formulation of the Type IIA orientifold vacua may be considered an alternative way to the standard $\CN=1$ supergravity formulae that provides the scalar potential from a K\"ahler potential and a superpotential $W$. This alternative is appropriate when all the scalar fields come along with axion-like scalars featuring discrete shift symmetries. In the 4-form formulation one can obtain the full scalar potential from the moduli metrics and one of the $\rho$'s, which we dub as the master axion polynomial $\rho_0$, that couples to the universal 4-form $F_4^ 0$ present in any compactification.  Interestingly, we find that the superpotential may be directly obtained from $\rho_0$ as
\beq
{ W = e^{i s^\lambda \p_{\phi^\lambda} } \rho_0\,.
}
\label{superpo1}
\eeq
where $\phi^\lambda$ denotes all the axions in the theory and $s^\lambda$ the corresponding saxion. This applies to both closed- and
open-string axions. Furthermore, all the rest of the polynomials may be obtained from $\rho_0$ by derivation with respect to the axions.
So $\rho_0$ may be considered as the generator of all the 4-form coupling to axions and carries all the information contained in the
superpotential.

The above factorised quadratic expression for the scalar potential may have interesting applications.  It shows  explicitly the discrete shift axion
symmetries of the theory, which is an important ingredient in theories of F-term \cite{Marchesano:2014mla} axion-monodromy inflation models \cite{Silverstein:2008sg,McAllister:2008hb,Baumann:2014nda,Westphal:2014ana}. The structure we obtain here is a generalisation of the Kaloper-Sorbo model \cite{KS} to the multiple axion case of string theory, including at the same time closed and open string axions.
The inclusion in our analysis of open string moduli may be particularly interesting since it has been argued \cite{BIPV,Valenzuela:2016yny,BVW} that such moduli are less 
constrained by {\it swampland} arguments in their use as large field inflatons.  The quadratic expression that we obtain may also be useful to search for 
minima in flux potentials. The fact that all dependence on axions goes  through the $\rho$-polynomials facilitates the analysis of minima 
in the axion directions. Also, since the theory is invariant under axion shift symmetries,  the values of the saxion moduli at the minima are rational functions 
of these $\rho$'s \cite{Valenzuela:2016yny}.  Up to now only AdS minima have been found in this class of Type IIA orientifolds in the absence of open string moduli. Our formulae open the way to a systematic search of minima including open string moduli and fluxes.  In particular they may play a role in the systematic search for dS vacua in compactifications including open string moduli. 

The structure of the rest of this paper is as follows. In Section \ref{s:4forms} we introduce the effective action of Type IIA Calabi-Yau orientifold compactifications 
in the presence of closed- and open-string fluxes, keeping track of the 4-forms appearing upon dimensional reduction. This is done both for the closed string action and for the DBI+CS action associated to the background D6-branes. In Section \ref{s:bilinear} we describe how the full classical scalar flux potential  may be written as a bilinear form on the $\rho$-polynomials, both in the presence of closed- and open-string moduli and fluxes. In Section \ref{s:FW} we analyse the discrete symmetries of these theories and show how the discrete symmetries shifting axions and fluxes may be understood in terms of 4d strings Freed-Witten anomalies. In Section \ref{s:master} we show how the superpotential may be derived starting from the master polynomial $\rho_0$  as in eq.(\ref{superpo1}) and how all the rest of the $\rho$ axion polynomials may be obtained by derivation of $\rho_0$ with respect to all the axions. In Section \ref{s:aux} we describe how the $\CN=1$ supergravity moduli auxiliary fields may also be expressed in terms of the $\rho$ polynomials and the metrics. We leave Section \ref{s:outlook} for our last remarks. 

Several technical details are relegated to the appendices. In Appendix \ref{ap:dimred} we perform the dimensional reduction giving rise to the 4d four-form action. In Appendix \ref{ap:sugra} we recover the type IIA flux potential as an F-term potential from the standard $\CN=1$ supergravity formula and the K\"ahler and superpotentials used in the main text. In Appendix \ref{ap:micro} we discuss the case where D6-brane position moduli are periodic directions in moduli space, and can be treated as 4d axions. In Appendix \ref{ap:metric} we show how the potential bilinear structure  is also present in toroidal orientifolds with metric fluxes, and draw an interesting connection between the Bianchi identities and the invertibility of $Z^{AB}$. In Appendix \ref{ap:sym} we illustrate how the discrete symmetries of the NS axions in the same orientifold  are obtained as a subgroup of its $SL(2, \IZ)^3$ duality group.

\section{Four-forms and Type IIA orientifolds}\label{s:4forms}

Let us consider type IIA string theory compactified in an orientifold of $\mathbb R^{1,3}\times \mathcal M_6$ with $\mathcal M_6$ a compact Calabi-Yau 3-fold. Following the standard construction \cite{Ibanez:2012zz,Blumenhagen:2005mu,Blumenhagen:2006ci,Marchesano:2007de}, we take the orientifold action to be generated by $\Omega_p (-1)^{F_L}\CR$, where $\Omega_p$ is the worldsheet parity reversal operator, ${F_L}$ is the space-time fermion number for the left-movers, and $\CR$ is an internal anti-holomorphic involution of the Calabi-Yau. This involution acts on the K\"ahler 2-form $J$ and the holomorphic 3-form $\Omega$ of $\CM$ as
\begin{equation}
\CR J=-J\ , \qquad \CR\Omega = \overline{\Omega}\, .
\label{iiaori}
\end{equation}
The fixed locus $\Pi_{\rm O6}$ of $\CR$ is one or several 3-cycles of $\CM_6$ in which O6-planes are located. The RR charge of such O6-planes can be cancelled by a combination of background fluxes and 4d space-time filling D6-branes wrapping three-cycles $\Pi_\alpha$ of $\CM_6$, together with their orientifold images.\footnote{For models that also consider D8-branes on coisotropic cycles see \cite{Font:2006na}.} More precisely, RR tadpole cancellation requires that the following equation in $H_3(\CM_6,\mathbb{Z})$ is satisfied 
\be
\sum_\a \left([\Pi_\a] + [\CR \Pi_\a]\right) - m [\Pi_{H}] - 4 [\Pi_{\rm O6}]\, =\, 0
\label{RRtadpole}
\ee
where $\Pi_{\rm O6}$ stands for the O6-plane loci, $[\Pi_H]$ is the Poincar\'e dual of the NS flux class $[H_3]$ and $m \in \IZ$ is the quantum of 0-form RR flux, see below for a precise definition. 

In the absence of internal background fluxes, dimensional reduction to 4d  \cite{Grimm:2004ua,Grimm:2011dx,Kerstan:2011dy} will yield a number of massless periodic scalars that are identified as axions.\footnote{These axions are in general subject to potentials generated by world-sheet and D-brane instantons. In this work we will consider a large volume regime of the compactification where such non-perturbative contributions can be neglected.} The axions arising from the closed string sector can be described by first specifying a basis of integer $p$-forms in which the K\"ahler two-form $J$ and holomorphic three-form $\Om$ of $\CM_6$ are expanded. Indeed, in general we have that
\be
e^{\phi/2} J = t^a \om_a
\ee
where $\phi$ is the 10d dilaton, $J$ is computed in the Einstein frame and $l_s^{-2} \omega_a$ are harmonic representatives of $H^2_-(\CM_6, \IZ)$, with $l_s^2  = 2\pi \sig =  4\pi^2 \a'$. The K\"ahler moduli $t^a$ are then understood as the saxionic partners of the $B$-field axions $b^a$, defined as
\be
B = b^a \om_a\, .
\ee
Similarly, we have the expansion of the holomorphic three-form $\Om = X^\lam \a_\lam - \CF_\lam\b^\lam$ in terms of a symplectic basis $l_s^{-3} (\a_\lam,\b^\lam) \in H^3 (\CM_6, \IZ)$ such that $l_s^{-6}\int_{\CM_6} \a_\rho \wedge \b^\sig = \d_\rho^\sig$. One then splits such decomposition into even $(\a_K, \b^\Lam) \in H_+^3(\CM_6)$ and odd $(\a_\Lam, \b^K) \in H_-^3(\CM_6)$ three-forms
\be
\re\, \Omega = X^K \a_K - \cf_\Lam \b^\Lam \qquad \qquad \im\, \Omega = X^\Lam \a_\Lam - \cf_K \b^K
\ee
and defines the RR axions of the compactification as
\be
C_3 = \xi^{\prime K} \a_K - \chi_\Lam' \b^\Lam + \dots
\label{C3axions}
\ee
which will pair up with the complex structure moduli above to form complex scalars. For simplicity in the following we will consider compactifications where the forms $(\a_\Lam, \b^\Lam)$ are absent. 

In addition there will be axions arising from the open string sector. In particular there will be $b_1(\Pi_\a)$ Wilson line axions $\th^i_\a$ for each D6-brane wrapping a three-cycle $\Pi_\a$. Such axions will combine with the worldvolume deformation moduli of BPS D6-branes to form complex scalars, and together they will redefine the notion of holomorphic variables in the closed string sector. We relegate the study of open string moduli to subsection \ref{ss:open}, and for now focus on the compactifications where they do not appear.

As shown in \cite{Louis:2002ny,Kachru:2004jr,Grimm:2004ua}, in the presence of background fluxes the above set of scalars develop an F-term scalar potential, which can be directly computed by dimensional reduction to 4d. More recently, it was pointed out  that such effective potential can also be entirely understood as arising from the 4d coupling of axions to four-forms \cite{Bielleman:2015ina}. In the following we will review this four-form formulation of the type IIA potential along the lines of \cite{Bielleman:2015ina,Carta:2016ynn}, and generalise the results therein.

\subsection{Closed string fluxes, axions and 4-forms}

Let us first rederive the closed string scalar potential by using the approach of \cite{Bielleman:2015ina,Carta:2016ynn}. For this we consider the type IIA  10d supergravity action in the string frame and the democratic formulation
\begin{equation}
S^{\rm 10d}_{\rm IIA} =\dfrac{1}{2\kappa_{10}^2} \int d^{10}x \sqrt{|g|} e^{-2\phi} \left( R+4 ( \partial \phi)^2  \right)  -   \frac{1}{4\kappa_{10}^2} \int e^{-2\phi} H \wedge \star_{10} H  +  \oh \sum_{n=0}^5  G_{2n}\wedge \star_{10} G_{2n}
\label{S10dIIA}
\end{equation}
where $\kappa_{10}^2 = l_s^8/4\pi$, and we have ignored for the moment
the contribution of localised sources. Of particular interest to us is the last term, which contains the dependence on the RR $p$-form potentials $C_p$ with $p=1,3,5,7,9$. It is useful to arrange such potentials in the polyforms 
\be
{\bf C} \, =\, C_1 + C_3 + C_5 + C_7 + C_9  \quad \quad {\rm or} \quad \quad {\bf A} \, =\, {\bf C} \wedge e^{-B}
\label{bfCA}
\ee
known as {C} and {A}-basis \cite{Bergshoeff:2001pv}. The corresponding gauge invariant field strengths are then given by
\be
{\bf G} \,=\, d{\bf C} - H \wedge {\bf C} + {\bf \bar{G}} \wedge e^B\, =\, e^B \wedge \left(d{\bf A} + {\bf \bar{G}}\right)
\label{bfG}
\ee
with ${\bf \bar{G}}$ a formal sum of closed $(p+1)$-forms of $\CM_6$ to be thought as the background value for the internal RR fluxes. The A-basis is quite useful in expressing the Bianchi identities and flux quantisation
\be
l_s^{2}\, d\left(d{\bf A} +  {\bf \bar{G}}\right) =- \sum_\a \d(\Pi_\a) \wedge e^{-\sig F_\a} \qquad {\rm and} \qquad \frac{1}{l_s^{p}} \int_{\pi_{p+1}} dA_{p} + \bar{G}_{p+1}  \in  \IZ
\label{BIG}
\ee
with $\Pi_\a \in \CM_6$ the internal cycle wrapped by a source and $\d(\Pi_\a)$ its bump-function with support on $\Pi_\a$ and indices transverse to it, such that $l_s^{p-9} \d(\Pi_\a)$ lies in the Poincar\'e dual class to $[\Pi_\a]$. In addition $F_\a$ is the quantised worldvolume flux threading $\Pi_\a$ and $\pi_{p+1} \in \CM_6$ is any $(p+1)$-cycle not intersecting the $\Pi_a$'s. In the absence of localised sources the $A_p$ are globally well-defined and the $\bar{G}_{p+1}$ are quantised, so one can define the flux quanta as the following integer numbers
\begin{equation}
m \, = \, l_s \bar{G}_0 ,  \quad  m^a\, =\, -\frac{1}{l_s^5} \int_{{\cal M}_6} \bar{G}_2 \wedge \tilde \om^a , \quad  e_a\, =\, \frac{1}{l_s^5} \int_{{\cal M}_6} \bar{G}_4 \wedge \om_a , \quad e_0 \, =\, -\frac{1}{l_s^5} \int_{{\cal M}_6} \bar{G}_6 
\label{RRfluxes}
\end{equation}
with  $\tilde \om^a$ the harmonic four-forms dual to $\om_a$ in the sense that $l_s^{-6} \int_{\CM_6} \om_a \wedge \tilde{\om}^b = \d_a^b$.

One important feature of the above 10d supergravity action is that the $p$-form degrees of freedom are doubled. In order to halve them one must impose the Hodge duality relations
\begin{equation}
G_{2n} = (-)^{n} \star_{10}  G_{10-2n}.
\label{10dH}
\end{equation}
either by hand or by adding a series of Lagrange multipliers to the action \cite{Bergshoeff:2001pv}. In the latter case one obtains a mother action which can be dimensionally reduced to 4d \cite{Carta:2016ynn}. As discussed in Appendix \ref{ap:dimred}, one then obtains a 4d effective action of the form
\be
S^{\rm 4d}=  - \frac{1}{16\kappa_4^2} \int_{\IR^{1,3}} Z_{AB} E_4^A \wedge *_4 E_4^B - \frac{1}{16\kappa_4^2}  \int_{\IR^{1,3}} Z^{AB} \varrho_A \varrho_B *_4 1  + \frac{1}{8\kappa_4^2}\int_{\IR^{1,3}} E_4^A \varrho_A
\label{S4formRR}
\ee
plus the kinetic terms for the RR axions (\ref{C3axions}). Here $\kappa_4^2 = \kappa_{10}^2/l_s^6$, $E_4^A$ are 4d four-forms and $\varrho_A$ are polynomials of fluxes and scalars, related to each other by 4d Hodge duality as
\be
 *_4 E_4^A= Z^{AB} \varrho_B\, ,
\label{4dH}
\ee
which can be deduced either from (\ref{S4formRR}) or by dimensionally reducing (\ref{10dH}). By plugging this relation back into the 4d action one can see that the first two terms of (\ref{S4formRR}) cancel each other. If we also use (\ref{4dH}) to eliminate the four-form dependence in the third term we obtain a scalar potential of the form
\be
V \, =\,  \frac{1}{\kappa_4^2} \frac{Z^{AB}}{8}   \varrho_A \varrho_B
\label{BilVRR}
\ee
which has a clear bilinear structure. 

Depending on the choice of four-form basis $E_4^A$ the quantities $\rho_A$ and $Z_{AB}$ will have one expression or the other. One obvious choice comes from  reducing the RR potentials in the C-basis to Minkowski three-forms. We have that
\be
C_3 = c_3^0 + \ldots \quad C_5 =  c_3^a \wedge \omega_a + \dots \quad C_7 = \tilde d_{3\, a} \wedge \tilde\omega^a +\dots \quad C_9 = \tilde d_{3} \wedge \om_6 + \dots
\label{C3form}
\ee
where $(c_3^0, c_3^a, \tilde d_{3\, a}, \tilde d_3)$  are three-forms with their indices in $\IR^{1,3}$, $\om_a$ and $\tilde \om^a$ are the harmonic forms of $\CM_6$ defined above and $\om_6$ is the harmonic six-form of $\CM_6$ such that $l_s^{-6} \int_{\CM_6} \om_6 =1$. Then, dimensional reduction of the RR 10d field strengths reads
 \be
 G_4 = F_4^0 + \ldots \quad G_6 =  F_4^a \wedge \omega_a + \dots \quad G_8 = \tilde F_{4\, a} \wedge \tilde\omega^a +\dots \quad G_{10} = \tilde F_{4} \wedge \om_6 + \dots
 \label{F4form}
 \ee
where the 4d four-forms are given by 
 \begin{align}
&F_4^0\, =\, dc_3^0 \,,\qquad F_4^a\, =\, dc_3^a -db^a \wedge c_3^0\,, \nonumber\\& \tilde F_{4\, a} \, =\, d\tilde d_{3\, a} -\CK_{abc}db^b \wedge c_3^c\,,\qquad \tilde F_4\, =\, d\tilde d_3 - db^a \wedge \tilde d_{3\, a}\,.
\label{M4forms}
\end{align}
For simplicity let us first assume a vanishing internal NS flux $H$. Then, if as in \cite{Bielleman:2015ina,Carta:2016ynn} we take $E_4^A = (F_4^0, F_4^a, \tilde F_{4\, a}, \tilde F_4)$ we have that $\varrho_A = (\rho_0, \rho_a, \tilde{\rho}^a, \tilde{\rho})$ with  
\be
 \begin{array}{rcl}
l_s \rho_0 & = &  e_0 - b^a e_a + \oh \CK_{abc}m^ab^bb^c - \frac{m}{6} \CK_{abc}b^ab^bb^c \\
l_s \rho_a & = & e_a - \CK_{abc}m^bb^c + \frac{m}{2} \CK_{abc}b^bb^c\\
l_s \tilde \rho^a & = & m^a - mb^a\\
l_s \tilde \rho & = & m 
 \end{array}
 \label{rhosRR}
 \ee
where $\CK_{abc} = l_s^{-6} \int_{\CM_6} \om_a \wedge \om_b \wedge \om_c$ are the triple intersection numbers of $\CM_6$. In addition $Z_{AB} = \frac{e^{-K}}{32} {\bf G}_{AB}$ with
\begin{equation}
{\bf G}= \left(
\begin{array}{c  c  c  c }
1 & & &  \\
 & 4g_{ab} & &  \\
& & \frac{9}{\CK^2}g^{ab} &  \\
& & & \frac{36}{\mathcal{K}^2}  
\end{array}
\right)
\label{Gmetric}
\end{equation}
where $\CK \equiv \CK_{abc} t^a t^b t^c$, and
\begin{equation}
g_{ab} = \frac{3e^{\phi/2}}{2 \CK l_s^6 } \int_{{\cal M}_6} \omega_a \wedge \star_6 \omega_b\,   \qquad g^{ab} = \frac{2\CK}{3e^{\phi/2}l_s^6} \int_{{\cal M}_6} \tilde \omega^a \wedge \star_6 \tilde \omega^b
\label{metrics}
\end{equation}
are 2 and 4-form metrics that only depend on the saxions $t^a$. Finally 
\be
e^{K}\, =\, \frac{e^{-\phi/2}}{8 {V}_6^3}
\label{eK}
\ee
with ${V}_6 = l_s^{-6} {\rm Vol} (\CM_6)$ the compactification volume in Einstein frame and string units. Putting all this together one finds that the scalar potential reads
\be
V_{\rm RR}\, =\,  \frac{1}{\kappa_4^2} e^K \left[4 \rho_0^2 + g^{ab} \rho_a \rho_b + \frac{4}{9} \CK^2 g_{ab} \tilde \rho^a \tilde \rho^b +  \frac{1}{9} \CK^2 \tilde \rho^2 \right]
\label{VRR}
\ee
which clearly has the bilinear structure of (\ref{BilVRR}). However, with this explicit expression we find a more specific structure, namely that 
\begin{itemize}

\item[{\it i)}] the $\rho$'s only depend on the fluxes (linearly) and on the axions $b^a$ (polynomially)

 \item[{\it ii)}] $Z_{AB}$ only depends on the saxions $t^a$

\end{itemize}

Alternatively, one may consider the choice of four-forms given by $E_4^A = (D_4^0, D_4^a, \tilde D_{4\, a}, \tilde D_4)$, which are related to the previous choice by the following change of basis
\begin{align}
\left(\begin{array}{c} F_4^0 \\[2mm] F_4^a \\[2mm] \tilde F_{4a} \\[2mm] \tilde F_4\end{array} \right) = \left(\begin{array}{cccc} 1 & 0 & 0 & 0 \\[2mm] b^a & \delta^a_b  & 0 & 0 \\[2mm] \frac{1}{2} \mathcal K_{abc} b^b b^c & \mathcal K_{abc} b^c & \delta^b_a  & 0 \\[2mm] \frac{1}{3!} \mathcal K_{abc} b^a b^b b^c & \frac{1}{2}\mathcal K_{abc} b^a b^c & b^b& 1\end{array}\right) \left(\begin{array}{c} D_4^0 \\[2mm] D_4^b \\[2mm] \tilde D_{4b} \\[2mm] \tilde D_4 \end{array}\right)\,.
\label{Rb}
\end{align}
One can check that this new set of four-forms are exact. More precisely, they are the field strengths of the Minkowski three-forms obtained from dimensionally reducing the RR potentials with an Ansatz similar to (\ref{C3form}) but now working in the A-basis. 

With this choice one finds that $l_s \varrho_A = (e_0, e_a, m^a, m)$ and 
\be
{\bf Z} = \frac{e^{-K}}{32} {\bf R}^{t} \cdot {\bf G} \cdot {\bf R}
\label{factorZ}
\ee
with {\bf R} the matrix in (\ref{Rb}). We may now plug in these expressions into (\ref{S4formRR}) and integrate out the four-forms in favour of $\varrho_A$ and $Z_{AB}$. Because now the four-forms are exact, this amounts to apply the procedure of Appendix E.2 of \cite{Louis:2002ny}, after which we again obtain (\ref{VRR}). Notice that with this description we obtain an even more precise description of the bilinear structure of the potential (\ref{BilVRR}). Namely
\begin{itemize}

\item[{\it i)}] $l_s \varrho_A$ are quantised fluxes

 \item[{\it ii)}] $Z_{AB}$ depends both on the saxions $t^a$ and axions $b^a$, but it factorises as (\ref{factorZ}), with ${\bf G} = {\bf G} (t)$ and ${\bf R} = {\bf R} (b)$

\end{itemize}
In the following sections, we will show that this is a quite general statement, even when we add more complicated ingredients to the compactification. 

Finally, let us recall that the same result for (\ref{VRR}) can be derived in the context of the standard $\CN=1$ supergravity formulation. Following the conventions in \cite{Grimm:2004ua}, one may do so by defining the complex K\"ahler variables
\be
T^a = b^a + i t^a
\ee
which enter the K\"ahler potential 
\be
K_K \,  = \, -{\rm log} \left(\frac{i}{6} \CK_{abc} (T^a - \bar{T}^a)(T^b - \bar{T}^b)(T^c - \bar{T}^c) \right) \, ,
\label{KK}
\ee
and the dilaton plus complex structure variables 
\be
N'^{K}\, =\, l_s^{-3} \int_{\CM_6} \Omega_c \wedge \b^K\, , \qquad \qquad \Omega_c = C_3 + i \re (C \Omega)
\label{defN}
\ee
where $C \equiv e^{-\phi} e^{\oh(K_{CS}-K_{K})}$ stands for a compensator term with $K_{CS} = - {\rm log}\left( \frac{i}{l_s^6} \int \Omega \wedge \bar{\Omega}\right)$.
These variables enter the K\"ahler potential 
\be
K_Q\,  = \, -2\,{\rm log} \left(-\frac{1}{4} \re(CX^K)\im(C\CF_K) \right) = -2\,{\rm log} \left(-\frac{1}{4} \im(\CF_{KL})\, n'^K n'^L \right)
\label{KQ}
\ee
where $\im (\CF_{KL})$ are zero order homogeneous functions of $n'^{K} \equiv \im \, N'^{K}$. Adding up both expressions we have that the full K\"ahler potential $K = K_K + K_Q$ indeed satisfies the relation (\ref{eK}). Finally, adding the RR flux superpotential
\be
l_s {W}_K  \, =\,  e_0 - e_a T^a + \frac{1}{2} {\cal K}_{abc} m^a T^b T^c - m \frac{1}{6} {\cal K}_{abc} T^a T^b T^c 
\label{WK}
\ee
one recovers (\ref{VRR}) as the F-term scalar potential, by applying the standard formula of 4d $\CN=1$ supergravity.

\subsubsection*{Adding $H$-flux}

Let us now consider adding a non-vanishing internal profile for the NS-flux $H$
\be
H \, =\, l_s^{-1} \sum_{K} h_K \beta^K \qquad \qquad h_K \in \IZ
\label{Hquanta}
\ee
and then expand the Hodge dual flux $H_7 =  e^{-2\phi} \star_{10} H$ in terms of even three-forms 
\be
H_7\, =\,  \sum_{K} H_4^K \wedge \a_K
\ee 
obtaining additional Minkowski 4-forms $H_4^K$ coming from the NS sector \cite{Bielleman:2015ina}. Then, on the one hand, dimensionally reducing the H-flux piece in (\ref{S10dIIA}) we obtain 
\be
S_{\rm NS}^{\rm 4d}=   \frac{1}{8\kappa_4^2 l_s}\int_{\IR^{1,3}} H_4^K h_K\, .
\label{S4formNS}
\ee
Here the 10d Hodge duality relation translates into
\be
 *_4 H_4^I = 32\,l_s^{-1} e^K  c^{IJ}  h_J
\label{4ddualH}
\ee
with $c^{IJ}$ the inverse of
\be
c_{IJ} = \frac{e^{K_Q/2}}{l_s^6} \int_{\CM_6} \a_I \wedge \star_6 \a_J\, .
\label{cIJ}
\ee
Hence, after imposing (\ref{4ddualH}) we recover a contribution to the potential of the form 
\be
V_{\rm NS} \, =\, \frac{1}{l_s^2 \kappa_4^2} 4 e^K {c^{IJ}} h_I h_J \, .
\ee

On the other hand, the RR piece of the action reads as in (\ref{S4formRR}) but with the replacement $e_0 \raw e_0 - h_K \xi^{\prime K}$, see Appendix \ref{ap:dimred}. That is, in the basis $E_4^A = (F_4^0, F_4^a, \tilde F_{4\, a}, \tilde F_4)$ we have that 
\be
l_s \rho_0  =   e_0 - b^a e_a + \oh \CK_{abc}m^ab^bb^c - \frac{m}{6} \CK_{abc}b^ab^bb^c - h_K \xi^{\prime K}
\label{rho0H}
\ee
and all the other $\rho$'s remain the same. Therefore one again obtains a contribution to the scalar potential of the form (\ref{VRR}) but with this new expression for $\rho_0$. 

Finally, due to the contribution of the NS flux $H$ and RR flux $m$ to the tadpole conditions (\ref{RRtadpole}) the total tension of the D6-branes will not cancel the negative tension of the O6-planes. This results in an extra contribution to the the scalar potential, that reads \cite{Villadoro:2005cu,Bielleman:2015ina}
\be
V_{\rm loc} \, =\,  \frac{4}{3\kappa_4^2l_s^2} e^K \CK\, m\, n'^K h_K \, .
\label{Vloc}
\ee
In supersymmetric vacua such contribution is negative, reflecting the corresponding D-brane deficit. To sum up, we end up with a full scalar potential of the form
\be
V\, =\, V_{\rm RR} + V_{\rm NS} + V_{\rm loc}
\label{VtotH}
\ee
which can again be derived as a standard 4d $\CN=1$ supergravity F-term potential \cite{Grimm:2004ua}. For this one again needs to consider the K\"ahler potential $K = K_K + K_Q$ with the previous expressions (\ref{KK}) and (\ref{KQ}), and the superpotential $W = W_K + W_Q$, with $W_K$ given by (\ref{WK}) and 
\be
l_s W_Q\, =\, - h_K N^K
\label{WQ}
\ee
with $N^K = N^{\prime K}$. 

It is interesting to notice that one can easily relate the first two pieces of (\ref{VtotH}) to an effective axion-four-form action of the form (\ref{S4formRR}), by using the duality relation (\ref{4ddualH}). As before, one can do it in different basis of 4d four-forms, of which two choices are particularly interesting. The most obvious one from the above discussion is to take
\be
E_4^A = (F_4^0, F_4^a, \tilde F_{4\, a}, \tilde F_4, H_4^K)
\label{EFwH}
\ee
so that
\be
\varrho_A = (\rho_0, \rho_a, \tilde{\rho}^a, \tilde{\rho}, \rho_K) \qquad \qquad l_s\rho_K = h_K
\label{varrhoH}
\ee
and 
\begin{equation}
{\bf Z}= \frac{e^{-K}}{32} \left(
\begin{array}{c  c}
{\bf G}  & \\
& {\bf C}
\end{array}
\right)
\label{GmetricH}
\end{equation}
with {\bf G} given by (\ref{Gmetric}) and the entries of {\bf C} given by (\ref{cIJ}). Alternatively one may consider the following rotated basis of four-forms
\begin{align}
\left(\begin{array}{c} F_4^0 \\  F_4^a \\  \tilde F_{4a} \\  \tilde F_4 \\  H_4^K\end{array} \right) = \left(\begin{array}{ccccc} 1 & 0 & 0 & 0 & 0\\  b^a & \delta^a_b  & 0 & 0 & 0 \\  \frac{1}{2} \mathcal K_{abc} b^b b^c & \mathcal K_{abc} b^c & \delta^b_a  & 0 & 0 \\  \frac{1}{3!} \mathcal K_{abc} b^a b^b b^c & \frac{1}{2}\mathcal K_{abc} b^a b^c & b^b& 1 & 0\\  \xi^{\prime K} & 0 & 0 & 0 & \d_L^K \end{array}\right) \left(\begin{array}{c} D_4^0 \\  D_4^b \\  \tilde D_{4b} \\  \tilde D_4 \\  S_4^L \end{array}\right)
\label{RbH}
\end{align}
where the above quantities read
\be
l_s \varrho_A = (e_0, e_a, m^a, m, h_K)
\label{rhoflux}
\ee
and
\begin{equation}
{\bf Z}= \frac{e^{-K}}{32} {\bf R}^{t} \left(
\begin{array}{c  c}
{\bf G}  & \\
& {\bf C}
\end{array}
\right) {\bf R}
\label{factorZH}
\end{equation}
with {\bf R} the axion-dependent rotation matrix in (\ref{RbH}). Again, when writing down these potential pieces as (\ref{BilVRR}), we recover a bilinear structure with factorised dependence on the saxions, axions and flux quanta. Remarkably, as we will discuss in the next section, this statement generalises for the full scalar potential (\ref{VtotH}), including the piece $V_{\rm loc}$. 

\subsection{Open string fluxes and moduli}
\label{ss:open}

Let us now consider the presence of D6-branes wrapping three-cycles $\Pi_\a$ of $\CM_6$. For such localised objects to preserve 4d $\CN=1$ supersymmetry they must wrap special Lagrangian three-cycles with vanishing worldvolume flux. That is they must satisfy 
\be
J_c|_{\Pi_\a} - \sig F_\a\, =\, \CF_\a + i J|_{\Pi_\a}\, =\, 0 \, ,
\label{D6Fterm}
\ee
where $\CF_\a = B|_{\Pi_\a} - \sig F_\a$ is the gauge invariant worldvolume flux, and
\be
\im\,\Om|_{\Pi_\a} \, =\, 0\, .
\label{D6Dterm}
\ee
Failure to satisfy (\ref{D6Fterm}) in some region of the closed string moduli space will be seen as a non-vanishing F-term in the 4d effective theory, and this will modify the above F-term scalar potential. This time the potential will also involve light open string fields of the compactification.

In general one may describe the open string moduli of a compactification in terms of a set of reference special Lagrangian three-cycles $\Pi_\a^0$ that, together with their orientifold images $\CR\Pi_\a^0$, satisfy the RR tadpole condition (\ref{RRtadpole}). Then one defines the space of light open string adjoint fields by considering the set of D6-brane Wilson lines and those three-cycle deformations that preserve the special Lagrangian condition. Due to McLean's theorem \cite{McLean} there is one field of each class per integer harmonic one-form $l_s^{-1} \zeta_i \in  \CH^1(\Pi_\a^0, \IZ)$ in each $\Pi_\a^0$, and so they can be paired up into $\sum_\a b_1(\Pi_\a^0)$ complex fields $\Phi_\a^i$, $i = 1, \dots, b_1(\Pi_\a^0)$. Equivalently, one may count such open string modes by a basis of integer harmonic two-forms  $l_s^{-2} \eta^j \in \CH^2(\Pi_\a^0, \IZ)$, defined such that $\int_{\Pi_\alpha^0} \zeta_i \wedge \eta^j = l_s^3 \d^j_i$. In particular one may define the open string moduli as \cite{Grimm:2011dx,Carta:2016ynn}
\be
\Phi^i_\a  \, =\, \frac{2}{l_s^{4}} \int_{\cc_4^\a} \left( J_c - \sig \tilde{F}_\a \right) \wedge \tilde{\eta}^i
\label{defPhi4ch}
\ee
where $\p\cc_4^\a = \Pi_\a - \Pi_\a^0$ is a four-chain that represents the homotopic deformation of $\Pi_\a^0$ to a new special Lagrangian $\Pi_\a$ and $\tilde{F}_\a$, $\tilde{\eta}^i$ are the extensions of the D6-brane worldvolume field strength $F=dA$ and of the  two-form $\eta^i$ to such a four-chain. In practice we may describe this open string field as
\be
\Phi_\a^i  \, =\,  T^a f_{a\, \a}^i - \theta_\a^i  
\label{defPhif}
\ee
with
\be
\theta_\a^i \, = \, \frac{2}{l_s^4} \int_{\Pi_\a^0} \sig A_\a \wedge \eta^i \qquad \qquad f_{a\, \a}^i \, = \,  \frac{2}{l_s^{4}} \int_{\cc_4^\a}\om_a \wedge \tilde{\eta}^i
\label{deffch}
\ee
see \cite{Carta:2016ynn} for more details.\footnote{Our definition of the open string fields differs by a global sign as compared to the one in \cite{Carta:2016ynn}, chosen like this for later convenience.}

If we neglect the effect of open string worldsheet instantons, there are two different mechanisms by which these open string fields may enter the type IIA scalar potential. The first one consists in adding a non-trivial profile for the worldvolume flux $F$ along the two-cycles $\pi_2^i$ of the special Lagrangian $\Pi_\a^0$, which are Poincar\'e dual to the quantised one-forms $l_s^{-1} \zeta^i$. That is, we consider the following worldvolume flux
\be
\sig F_\a\, =\, \sig d A_\a + n_{F\, i}^\a\, \eta^i \qquad n_{F\, i} \in \IZ
\ee
which clearly violates the F-term condition (\ref{D6Fterm}). The second mechanism is to consider that such two-cycles $\pi_2^i$ are non-trivial in the homology of the ambient space \cite{Marchesano:2014iea}, or in other words that some of the following integer numbers
\be
n_{a\, i}^\a  =  \frac{1}{l_s^{3}} \int_{\Pi_\a} \om_a \wedge \zeta_i
\label{nai}
\ee
are non-vanishing. As a result, when we move in the complexified K\"ahler moduli space the F-term condition (\ref{D6Fterm}) will be also generically violated. 

One may partially detect the effect of such F-term breaking by evaluating the DBI piece of the action of each D6-brane. Whenever the combined source of supersymmetry breaking is small in string units, the corresponding excess of energy is well-approximated at the two-derivative level by the following scalar potential\cite{Escobar:2015ckf}
\be
V_{\rm DBI}  \, =\,  \sum_\a \frac{e^K}{l_s^2 \kappa_4^2} G^{ij}_{\a} \left(n_{\CF\, i}^\a - n_{a\, i}^\a T^a\right)\left(n_{\CF\, j}^\a - n_{a\, j}^\a \bar{T}^a\right)   
\label{VDBI}
\ee
where, as in (\ref{RRtadpole}), $\a$ runs over pairs of D6-branes related by the orientifold action. Here $G^{ij}_{\a}$ is the inverse of
\be
G_{ij}^{\a} \, = \, \frac{3e^{-\phi/4}}{4 \CK l_s^{3}} \int_{\Pi_\a^0} \zeta_i \wedge *\, \zeta_j\, .
\label{gD6}
\ee
and we have defined
\be
n_{\CF\, i}^\a \, = \,  n_{F\, i}^\a - \oh g_{i\, \a}^K h_K \qquad \text{and} \qquad g_{i\, \a}^{K} \, =\, \frac{2}{l_s^{4}} \int_{\cc_4^\a}  \b^K \wedge \tilde{\zeta}_i \, .
\label{nCFg}
\ee
with $\tilde{\zeta}_i$ the extension of $\zeta_i$ to ${\cc_4^\a}$. 

Nevertheless, this is not the only effect of considering such D6-brane configurations. Indeed, one finds that the F-term scalar potential is further modified by terms that, unlike (\ref{VDBI}), depend on the open string moduli. The detection of such extra terms is not so obvious,  and one may do so by computing the increase in energy by the backreaction of such D6-branes \cite{Marchesano:2014iea} and by evaluating their Chern-Simons piece of their action \cite{Escobar:2015ckf,Carta:2016ynn}. The combined effect is however rather simple to describe. It amounts to again consider a 4d effective action of the form (\ref{4dH}) with the same four-forms as before, but where the $\varrho_A$ have been shifted by an open-string dependent term. Indeed, as discussed in Appendix \ref{ap:dimred} in the presence of $H$-flux we have that 
\be
\varrho_A = (\rho_0+ \upsilon_0, \rho_a+\upsilon_a, \tilde{\rho}^a+ \tilde{\upsilon}^a, \tilde{\rho}, \rho_K)
\label{rhosopen}
\ee
with $\rho_0$ given by (\ref{rho0H}), the other $\rho$'s by (\ref{rhosRR}) and the contribution to the $\upsilon$'s for each D6-brane $\a$ by
\be
 \begin{array}{rcl}
l_s \upsilon_0 & = & (b^c n_{c\, i}^\a - n_{F\, i}^\a) ( b^d f_{d\, \a}^i - \th^i_\a) + \oh h_K  b^a {\bf H}_{a\, \a}^{K}  - \oh h_K g^K_{i\, \a} \th^i_\a\, \\
l_s \upsilon_a & = & - n_{a\, i}^\a (b^c f_{c\, \a}^i - \theta^i_\a)  - (b^c n_{c\, i}^\a - n_{F\, i}^\a  ) f_{a\, \a}^i - \oh h_K  {\bf H}_{a\, \a}^{K} \\
l_s \tilde \upsilon^a & = &  q^a_\a \, =\,  \CK^{ab} \left(n_{b\, i}^\a f_{c\, \a}^i + n_{c\, i}^\a f_{b\, \a}^i\right)  t^c\, .
 \end{array}
 \label{varrhoHno}
 \ee
Here $q_\a^a$ and ${\bf H}_{a\, \a}^{K}$ are functions of the D6-brane position moduli defined by
\be
 q^a_\a  =  \frac{2}{l_s^{4}} \int_{\cc_4^\a} \tilde \om^a \qquad \qquad \text{and} \qquad \qquad \p_{{\rm Im} \Phi_\b^i} \left(t^a {\bf H}^K_{a\, \a}\right) =  g_{i\, \a}^{K} \d_{\a\b}\, ,
 \label{defH}
\ee
$\CK^{ab}$ is the inverse of $\CK_{ab} = \CK_{abc}t^c$  and we have used that $\CK_{abc} q_\a^c = n_{a\, i}^\a f_{b\, \a}^i + n_{b\, i}^\a f_{c\, \a}^i$ \cite{Carta:2016ynn}. As a result of these shifts, the combined contribution to the scalar potential of the 10d RR field strengths and the D6-brane Chern-Simons actions add up to the bilinear term (\ref{BilVRR}), with {\bf Z} as above (\ref{Gmetric}) and the $\varrho_A$ depending on the fluxes, closed string axions and open string moduli. 

To sum up, we find that the type IIA scalar potential in the presence of RR, NS and open string fluxes is given by
\be
V\, =\, V_{\rm RR+CS} + V_{\rm NS} + V_{\rm DBI} + V_{\rm loc}
\label{VtotD6}
\ee
As before, one may reproduce this whole expression in terms of a 4d $\CN=1$ supergravity F-term potential. Indeed, as shown in Appendix \ref{ap:sugra} the same expression follows if we consider the superpotential
\be
{W} = {W}_K + W_{Q} + {W}_{\rm D6}
\label{Wtotal}
\ee
with $W_K$ given by (\ref{WK}), $W_Q$ by (\ref{WQ}) and 
\be
l_s {W}_{\rm D6}(\Phi)  \, = \, - \Phi_\a^i( n_{F\, i}^\a - n_{a\, i}^\a T^a ) + l_s W_0\, .
\label{WD6}
\ee
where for simplicity we have suppressed the index $\a$ running over all D6-branes. Finally, $W_0$ is a constant piece defined in terms of the reference three-cycles $\{\Pi_\a^0\}$, see Appendix \ref{ap:sugra}. One important difference with the case without open strings is that the holomorphic variable $N^K$ that enters $W_Q$ is no longer the geometric variable $N^{\prime K}$ defined in (\ref{defN}), but instead it gets redefined by the open string moduli. More precisely from the discussion of Appendix \ref{ap:sugra} we find that
\be
N^K = N'^K  + \oh \sum_\a \left(g^K_{i\, \a} \th^i_\a -T^a  {\bf H}_{a \, \a}^K \right)  \, ,
\label{redefN}
\ee
which differs from previous identifications of the holomorphic variables in the literature, like those in \cite{Carta:2016ynn,Grimm:2011dx,Kerstan:2011dy}. Rewriting the K\"ahler potential $K = K_K + K_Q$ in terms of these new variables one indeeds reproduces the scalar potential (\ref{VtotD6}), as shown in Appendix \ref{ap:sugra}.

\section{The scalar potential as a bilinear form}\label{s:bilinear}

While not obvious, we will now show that one may also rewrite the full F-term potential (\ref{VtotD6}) in the bilinear form (\ref{BilVRR}). More precisely, one may take a choice of basis such that the $\varrho$'s are quantised open and closed string fluxes, and the matrix {\bf Z} takes the factorised form (\ref{factorZ}). As before, the matrix {\bf R} will only depend on the axionic components of the 4d sugra fields, which in this more general setup are given by
\be
b^a\, , \quad  \hat \th^i_\a = \re\, \Phi^i_\a =  b^a f^i_{a\, \a} - \th^i_\a  \, ,   \quad   \xi^K = \re\, N^K = \xi^{\prime K} -  \oh b^a \sum_\a {\bf H}_{a \, \a}^K +\oh g^K_{i\, \a}\th^i_\a \, .
\label{allaxions}
\ee

\subsubsection*{In the absence of open string moduli}

Let us first consider the case without open string moduli, so the axions of the compactification reduce to $b^a$, $\xi^{\prime K}$ and the potential takes the form (\ref{VtotH}). Notice that the negative definite term $V_{\rm loc}$ is bilinear in the fluxes $m$ and $h_K$, so one may easily incorporate it into the bilinear structure (\ref{BilVRR}) if one keeps the $\varrho$'s as in (\ref{varrhoH})  and takes
\begin{equation}
{\bf Z}^{-1}= 8\, e^{K} \left(
\begin{array}{c  c  c  c c c}
4 & & &  \\
 & g^{ab} & &  \\
& & \frac{4}{9}\CK^2g_{ab} &  \\
& & & \frac{1}{9} \CK^2 &  \frac{2}{3} \CK n^{\prime I}  \\
& & &   \frac{2}{3} \CK n^{\prime J} & 4 c^{IJ}
\end{array}
\right)\, ,
\label{ZinvHloc}
\end{equation}
or equivalently if one replaces (\ref{GmetricH}) by
\begin{equation}
{\bf Z}= \frac{e^{-K}}{32} \left(
\begin{array}{c  c  c  c c c}
1 & & &  \\
 & 4g_{ab} & &  \\
& & \frac{9}{\CK^2}g^{ab} &  \\
& & & -\frac{12}{\CK^2}  &  \frac{2}{\CK} c_{IJ} n^{\prime J}  \\
& & &  \frac{2}{\CK} c_{IJ} n^{\prime I} & c_{IJ} - \frac{1}{3}  c_{IK} n^{\prime K} c_{JL} n^{\prime L} 
\end{array}
\right)\, .
\label{ZHloc}
\end{equation}
where we have used that $c_{IJ}  n^{\prime I}  n^{\prime J} = 4$. In other words, one can absorb the potential piece $V_{\rm loc}$ into a modified metric for the 4d four-forms in the effective Lagrangian (\ref{S4formRR}). Notice that this new, modified metric is no longer definite positive, as needed to reflect the fact that the contribution from $V_{\rm loc}$ can be negative. In addition, the new metric has a non-trivial mixing between the four-forms in the RR and NS sector, which respectively couple to metrics of the K\"ahler and complex structure sectors of the compactification. This mixing seems to be a rather generic feature of massive type IIA string theory in Calabi-Yau compactifications. 

Finally, because through this modification the $\varrho$'s in (\ref{BilVRR}) remain unchanged, the effective potential still displays the triple factorisation into saxions, axions and flux quanta. More precisely in the basis of four-forms in the lhs of (\ref{RbH}), one again has that the $\varrho$'s are given by flux quanta as in (\ref{rhoflux}) and that (\ref{factorZH}) is replaced by
\be
{\bf Z} = {\bf R}^{t}\, {\bf M}\, {\bf R}
\label{factoRM}
\ee
with {\bf R} again the axion-dependent rotation matrix in (\ref{RbH}) and {\bf M} the saxion-dependent metric in the rhs of (\ref{ZHloc}).

\subsubsection*{In the presence of open string moduli}

Let us now consider the presence of open string moduli. As discussed above, this implies the shift of the closed string $\varrho$'s as in (\ref{rhosopen}) and the appearance of the new term (\ref{VDBI}) contributing to the potential. Now, because this extra term $V_{\rm DBI}$ is also quadratic on closed- and open-string fluxes, one may easily rewrite the full scalar potential in the form (\ref{BilVRR}). Indeed, for simplicity let us consider the case where open string moduli are present for a single D6-brane, so that we can suppress the index $\a$ in the following, and that $n_{a\, i} = 0$.  Then one may enlarge  the vector of $\varrho$'s (\ref{rhosopen}) to include the fluxes related to such D6-brane, as
\be
\varrho_A = (\rho_0+ \upsilon_0, \rho_a+\upsilon_a, \tilde{\rho}^a+ \tilde{\upsilon}^a, \tilde{\rho}, \rho_{\CF\, i}, \rho_{K})
\label{rhosopen+}
\ee
where
\be
l_s \rho_{K}\, =\, h_K \qquad \qquad l_s \rho_{\CF\, i}\, =\,  n_{F\, i} - \oh g_{i}^K h_K
\ee
and  rewrite the full scalar potential (\ref{VtotD6}) in the form (\ref{BilVRR}), where now
\begin{equation}
{\bf Z}^{-1}= 8\, e^{K} \left(
\begin{array}{c  c  c  c c c c }
4 & & &  \\
 & g^{ab} & &  \\
& & \frac{4}{9}\CK^2g_{ab} &  \\
& & & \frac{1}{9} \CK^2 &  0 & \frac{2}{3} \CK n^{\prime I} \\
& & &  0 &  G^{ij}   & 0 \\
& & & \frac{2}{3} \CK n^{\prime J} & 0 & 4 c^{IJ}\\
\end{array}
\right)\, ,
\label{ZinvHD6}
\end{equation}
whose inverse is given by
\begin{equation}
{\bf Z}= \frac{e^{-K}}{32} \left(
\begin{array}{c  c  c  c}
1 & & &  \\
 & 4g_{ab} & &  \\
& & \frac{9}{\CK^2}g^{ab} &  \\
& & & {\bf D}
\end{array}
\right)
\label{ZHD6}
\end{equation}
with
\begin{equation}
{\bf D}= 
\left(
\begin{array}{c c c}
1  & 0 & 0 \\
0 &  \d_{i}^j   &0 \\
-\frac{\CK}{6} c_{IK} n^{\prime K} & 0 &  \d_{I}^J
\end{array}
\right)
\left(
\begin{array}{c c c}
 -\frac{12}{\CK^2}   \\
 &4 G_{ij}    \\
 &  &  c_{IJ}
\end{array}
\right)
\left(
\begin{array}{c c c}
1  & 0 & - \frac{\CK}{6} c_{IJ} n^{\prime I} \\
0 & \d_{i}^j & 0  \\
0 & 0 &   \d_{I}^J 
\end{array}
\right)\, .
\label{D}
\end{equation} 

Therefore we can again relate the scalar potential to an effective action of the form (\ref{S4formRR}). Notice however that the $\varrho$'s in (\ref{rhosopen+}) not only depend on the axions of the compactification but also on the D6-brane position moduli, which are typically seen as 4d open string saxions.  It is then a priori not clear whether the triple factorisation of the potential into saxions, axions and fluxes holds for this case. Nevertheless, since one may rewrite the vector (\ref{rhosopen+}) as 
\begin{align}
\left(\begin{array}{c} l_s(\rho_0 + \upsilon_0) \\ l_s(\rho_a + \upsilon_a) \\  l_s (\tilde \rho^a + \tilde \upsilon^a) \\  l_s \tilde \rho \\ l_s  \rho_{{\CF\, i}} \\ l_s  \rho_{K}    \end{array} \right) \, =\,  {\bf S}^{t\, -1} {\bf R}^{t\, -1}  
\left(\begin{array}{c} e_0 \\ e_b \\  m^b \\  m \\  n_{F\, j}  \\ h_L   \end{array} \right) 
\label{SR}
\end{align}
where
\begin{align}
{\bf R} =  \left(\begin{array}{cccccc} 1 & 0 & 0 & 0 & 0& 0\\  b^b & \delta^b_a  & 0 & 0 & 0 &0 \\  \frac{1}{2} \mathcal K_{abc} b^a b^c & \mathcal K_{abc} b^c & \delta^a_b   & 0 & 0 & 0\\  \frac{1}{3!} \mathcal K_{abc} b^a b^b b^c & \frac{1}{2}\mathcal K_{abc} b^b b^c & b^a& 1 & 0 & 0 \\ \hat{\th}^j   & 0 & 0 & 0 & \d_i^j & 0  \\  \xi^{L} & 0 & 0 & 0 & 0 &  \d_K^L \end{array}\right), \
{\bf S}^{-1} =   \left(\begin{array}{cccccc} 1 & 0 & 0 & 0 & 0& 0\\  0 & \delta^b_a   & 0 & 0 & 0 &0 \\  0 & 0 & \delta^a_b  & 0 & 0 & 0\\ 0 &0 &0& 1 & 0 & 0 \\ 0 & f^j_a & 0 & 0 &  \d_i^j  & 0  \\  0 &  -\oh{\bf H}_a^L  & 0 & 0 & -\oh g_{i}^L  &  \d_K^L \end{array}\right)
\label{Ropen}
\end{align}
and $\hat \th^i_\a$ and $\xi^K$ are defined as in (\ref{allaxions}),  one can again factorise the dependence of the potential on the axionic and saxionic part of the moduli of the compactification.\footnote{In some cases like toroidal compactifications, the D6-brane position moduli can also take periodic values, and one should in principle be able to describe them on equal footing with the axions of the theory. We analyse this possibility in Appendix \ref{ap:micro}, where we show that the structure \eqref{SR} precisely allows to incorporate them in the rotation matrix {\bf R}.} More precisely, one recovers the previous triple factorisation structure, with a particular basis of four-forms in which the $\varrho$'s are given by the quantised fluxes
\be
l_s \varrho_A = (e_0, e_a, m^a, m, n_{F\, i}, h_K)
\label{rhopen}
\ee
and the four-form metric is of the factorised form (\ref{factoRM}), where {\bf R} is as in (\ref{Ropen}) and 
\be
{\bf M} \, =\, 
\frac{e^{-K}}{32} 
{\bf T}^t
\left(
\begin{array}{c  c  c  c c c}
1 & & &  \\
 & 4g_{ab} & &  \\
& & \frac{9}{\CK^2}g^{ab} &  \\
& & &  -\frac{12}{\CK^2}   \\
 & & &&  4 G_{ij}   \\
& & & &  & c_{IJ} 
\end{array}
\right)
{\bf T}
\label{MHD6}
\ee
with
\begin{align}
{\bf T} =  \left(\begin{array}{cccccc} 1 & 0 & 0 & 0 & 0& 0 \\  0 & \delta^a_b & 0 & 0 & 0 &0  \\  0 & 0 &   \delta^b_a & 0 & 0 & 0 \\ 0 & -\frac{\mathcal K}{12} (\mathbf{H}^I_b  - f^i_b g^K_i )c_{IJ} n'^J &0& 1 & -\frac{\mathcal K}{12} g^I_j c_{IJ} n'^J & -\frac{\CK}{6} c_{LJ} n^{\prime J}  \\ 0 & -f^i_b  & 0 & 0 & \d^j_i  & 0  \\  0 & \oh {\bf H}_b^K {-\frac{1}{2} f^i_b g^K_i} & 0 & 0 & \oh g_i^K & \d_L^K 
 \end{array}\right)\, .
\end{align}
It would be very interesting to interpret the final saxion-dependent matrix {\bf M} in terms of flux compactifications of M-theory on $G_2$ manifolds.

Let us turn to the case with several D6-branes with $n_{a\, i} \neq 0$. Then the natural choice is to further extend the vector \eqref{rhosopen+} to
\be
\varrho_A = (\rho_0+ \upsilon_0, \rho_a+\upsilon_a, \tilde{\rho}^a+ \tilde{\upsilon}^a, \tilde{\rho}, \rho_{\CF\, i}^\a,  \rho_{K},  \rho_{n_{a\, i}}^\a)
\label{rhosopen++}
\ee
where $\a$ runs over the D6-brane with open string moduli and 
\be
l_s \rho_{K}\, =\, h_K \qquad \qquad l_s \rho_{\CF\, i}^\a\, =\,  n_{F\, i}^\a - \oh g_{i\, \a}^K h_K - b^a n_{a\, i}^\a \qquad \qquad l_s \rho_{n_{a\, i}}^\a \, =\,  n_{a\, i}^\a
\ee
With this extension the saxion-dependent metric giving the potential is given by 
\begin{equation}
{\bf Z}^{-1}= 8\, e^{K} \left(
\begin{array}{c  c  c  c c c c  c}
4 & & &  \\
 & g^{ab} & &  \\
& & \frac{4}{9}\CK^2g_{ab} &  \\
& & & \frac{1}{9} \CK^2 &  0 & \frac{2}{3} \CK n^{\prime I} \\
& & &  0 &  G^{ij}_{\a\b}   & 0 \\
& & & \frac{2}{3} \CK n^{\prime J} & 0 & 4 c^{IJ}\\
&&&&&& G^{ij}_{\a\b} t^at^b
\end{array}
\right)\, ,
\label{ZinvHD6n}
\end{equation}
with $G_{\a\b}^{ij} \equiv G^{ij}_\a \d_{\a\b}$. In this case, interpreting ${\bf Z}^{-1}$ as the inverse of a four-form metric is problematic, because the lowest blocks of this matrix are not invertible. Hence, naively one cannot convert this bilinear expression for the scalar potential to an effective action the form (\ref{S4formRR}). 

Interestingly, one finds an analogous obstruction in the context of compactifications with metric fluxes, like the twisted tori analysed in Appendix \ref{ap:metric}. There, starting from the supergravity F-term potential, one only obtains an invertible matrix ${\bf Z}^{-1}$ after the Bianchi identities between metric fluxes have been taken into account. Since in principle both $\bar{G}_2$ and $n_{a\, i}^\a$ can be interpreted as metric fluxes in an M-theory uplift of our setup \cite{Hagen}, it is tempting to speculate that a similar kind of constraint should be imposed before attempting to invert \eqref{ZinvHD6n}. It would therefore be interesting to analyse the present setup from the viewpoint of  $G_2$-manifold with metric fluxes, a task that we leave for future work.

Instead of delving on the details of inverting ${\bf Z}^{-1}$, let us describe how to reproduce the previous factorised structure of the potential, now with the extended vector \eqref{rhosopen++}. For simplicity we will again consider a single D6-brane, the extension to several of them being trivial. As before, one may also reproduce the potential in terms of a vector of integer entries
\be
l_s \varrho_A = (e_0, e_a, m^a, m, n_{F\, i}, h_K , n_{a\, i}) \, ,
\label{rhopenn1}
\ee
which instead of \eqref{ZinvHD6n} is contracted with the metric ${\bf Z}^{-1} = {\bf R}^{-1} {\bf M}^{-1} {\bf R}^{t\, -1}$. Here the axion rotation matrix {\bf R} takes the form
\be
{\bf R} \, =\,\left(\begin{array}{ccccccc} 
1 & 0 & 0 & 0 & 0& 0 & 0 \\  b^b & \delta^b_a  & 0 & 0 & 0 &0  & 0 \\  \frac{1}{2} \mathcal K_{abc} b^a b^c & \mathcal K_{abc} b^c & \delta^a_b  & 0 & 0 & 0  & 0 \\  \frac{1}{3!} \mathcal K_{abc} b^a b^b b^c & \frac{1}{2}\mathcal K_{abc} b^b b^c & b^a & 1 & 0 & 0 & 0 \\ \hat{\th}^j   & 0 & 0 & 0 & \d_{i}^{j} & 0   & 0 \\  \xi^{L}  & 0 & 0 & 0 & 0 & \d_K^L   & 0 \\  \hat{\th}^i b^a &  \hat{\th}^i \delta^b_a& 0 &0&  b^a \d_{i}^{j} & 0 &  \d_{a\, i}^{b\, j} \end{array}\right),
\label{Ropenn1}
\ee
and the saxion-dependent matrix {\bf M}$^{-1}$ is 
\be
{\bf M}^{-1} \, =\, 
8 e^K 
{\bf T}^{-1}
\left(
\begin{array}{c  c  c  c c cc}
4 & & &  \\
 & g^{ab} & &  \\
& & \frac{4}{9}\CK^2g_{ab} &  \\
& & &  -\frac{\CK^2}{3}   \\
 & & &&  G^{ij}   \\
& & & &  & 4c^{IJ} \\
& & & &  & & G^{ij} t^at^b \\
\end{array}
\right)
{\bf T}^{t\, -1}
\label{MHD6n1}
\ee
with
\begin{align}
{\bf T}^{t\, -1} =  \left(\begin{array}{ccccccc} 1 & 0 & 0 & 0 & 0& 0 & 0 \\  0 & \delta^a_b & 0 & 0 & f^i_b  &- \oh {\bf H}_b^K & 0 \\  0 & 0 &   \delta^b_a & 0 & 0 & 0& \mathcal K^{ab} t^c f_{c}^i  + \mathcal K^{ac} f_c^i t^b \\ 0 & 0 &0& 1 & 0 & 0 & 0 \\ 0 & 0 & 0 & 0 & \d^j_i  & -\oh g_i^K  & 0 \\  0 &  0 & 0 & \frac{\CK}{6} c_{LJ} n^{\prime J} & 0 & \d_L^K& 0 \\
0 & 0 & 0 &0 &0&0 &  \d_{a\, i}^{b\, j} 
 \end{array}\right)\, ,
\end{align}

Given this description for the flux potential in Calabi-Yau compactifications, one may naturally wonder if the factorised bilinear structure also generalises to non-Calabi-Yau geometries. In Appendix \ref{ap:metric} we have worked out the case of toroidal orientifolds with metric fluxes, and found that the factorised structure indeed holds if we extend the flux vector with new entries $\varrho_{a_i}$, $\varrho_{b_{ij}}$ related to each additional metric flux. As mentioned before, we do also find that the metric for this extended flux vector is a priori not invertible, but that it becomes invertible once that 
 the Bianchi identities for metric fluxes are imposed. Presumably, an analogous constraint allows to invert the metric for the case of compactifications with $n_{a\, i} \neq 0$. These observations strongly suggest that the interpretation of the scalar potential in terms of four-forms can only be made after the whole set of Bianchi identities have been taken into account. In this sense, it would be very interesting to see if one can also generalise our results to compactification with more exotic non-geometric fluxes \cite{Shelton:2005cf,moredual}, and to reinterpret the structure of the potential analysed in \cite{Blumenhagen:2015lta,Gao:2017gxk} in this language. In particular, it would be interesting to further explore the connection between the invertibility of the saxion-dependent matrix {\bf M} and the Bianchi identities for these classes of vacua, where implementing the latter is oftentimes subtle. Finally, such a bilinear structure with factorised axion dependence also appears in the context of 4d $\CN=4$ gauged supergravity \cite{gerardo}, so one could apply the intuition drawn from our results to this case as well.

\section{4d strings and Freed-Witten anomalies}\label{s:FW}

As we have seen in the previous section, it is possible to understand the classical type IIA scalar potential in terms of a 4d effective action of the form (\ref{S4formRR}). Moreover one can choose a basis of four-forms such that the $l_s\varrho_A$ are given by flux quanta and the four-form metric {\bf Z} is of the factorised form (\ref{factoRM}), with {\bf M} purely saxion-dependent and {\bf R} depending on the axions and some topological numbers of the compactification manifold $\CM_6$.

It turns out that both the topological data and the precise axion dependence contained in {\bf R} have a neat microscopic description in terms of the Freed-Witten anomalies developed by branes in flux compactifications. In particular, in this section we will show that one can reconstruct {\bf R} in terms of the Freed-Witten anomalies of higher-dimensional branes that look like strings upon compactification to 4d. 

As discussed in \cite{BerasaluceGonzalez:2012zn}, 4d strings developing a Freed-Witten anomaly due to  an internal background flux is a ubiquitous effect in type II orientifold compactifications. The anomaly  is cured by adding further branes ending on the anomalous one, which in 4d are seen as domain walls ending on a string. Macroscopically, the presence of 4d domain-walls ending on certain strings is not surprising whenever the axions charged under such strings enter the scalar potential, and they are usually dubbed {\it axionic domain-walls}. Microscopically, the number $k$ of domain walls ending on a certain string depends on the topological details of the FW anomaly of the latter, and render the K-theory charge of such domain walls $\IZ_k$-valued \cite{BerasaluceGonzalez:2012zn}. 

\subsection{Reconstructing R}
 
As we have seen, in type IIA flux compactifications the axions that enter the classical scalar potential are given by (\ref{allaxions}). Let us first ignore the presence of D6-brane axions. Then we are left with the NS axions arising from $B = b^a \om_a$ and the RR axions from $C_3 = \xi^{\prime K} \a_K$. These axions are magnetically coupled to 4d strings which are NS5-branes wrapping four-cycles and D4-branes wrapping three-cycles of $\CM_6$, respectively. More precisely we have the following correspondence
\bea\nonumber
\text{B-field axion}\  b^a & \leftrightarrow & \text{NS5-brane wrapping} \ \pi_4^a \in \text{P.D.}[\omega_a] \\ \nonumber
\text{RR-axion}\ \xi^{\prime K} & \leftrightarrow & \text{D4-brane wrapping} \ \pi_3^K \in \text{P.D.}[\a_K]
\eea
where P.D. stands for Poincar\'e dual.

In general, D-branes develop FW anomalies in the presence of $H$-flux and NS5-branes in the presence of RR fluxes. The precise set of anomalies for the 4d strings above are summarised in table \ref{4dstFW}, adapted from table 3 of \cite{BerasaluceGonzalez:2012zn}. 
\begin{table}[!ht]
\begin{center}
\begin{tabular}{|c|c||c||c|c||c|}
\hline
\multicolumn{2}{|c||}{\bf String} & {\bf Flux} &\multicolumn{2}{|c||}{\bf Domain wall}  & {\bf Rank}\\
\hline
type & cycle & type & type & cycle &  \\
\hline\hline
NS5 & $[\pi_4^a]\in H_4(\CM_6,\IZ)$ & $F_{4} = e_b \tilde{\om}^b$  & D2 & $-$ &  $\int_{\pi_4^a} {F}_{4} = e_b$\\
\hline
NS5 & $[\pi_4^a]\in H_4(\CM_6,\IZ)$ & $F_2 = m^b \om_b$ & D4 & $\pi_{2} \in \text{P.D.}[F_2 \wedge \om_a]$ &  $\int_{\pi_2} \om_c = \CK_{abc}m^b$\\
\hline
NS5 & $[\pi_4^a]\in H_4(\CM_6,\IZ)$ & $F_0 = m$&  D6 & $[\pi_4^a]$ &  $m$\\
\hline
D4 & $[\pi_3^K] \in H_3(\CM_6,\IZ)$ & $H = h_K\b^K$ & D2 & $-$ &  $\int_{\pi_3^K} {H} = h_K $\\
\hline
\end{tabular}
\caption{4d strings that develop Freed-Witten anomalies in type IIA flux compactifications, together with the fluxes creating the anomaly and the domain walls curing it. The last column shows the amount of domain walls in terms of flux quanta. \label{4dstFW}}
\end{center}
\end{table}
The table also displays the kind of domain walls that are necessary to cure each anomaly, the internal cycle that they wrap and their multiplicity. 

Interestingly, one can encode the information of table \ref{4dstFW} in a set of square matrices. Indeed, notice that each 4d string can be seen as a linear map sending a quantised flux (the one creating the anomaly) to a 4d domain wall (the one curing the anomaly). Now, since the space of 4d domain walls is in one-to-one correspondence with the set of 4d four-forms that they couple to, and the latter is in one-to-one correspondence with the set of internal fluxes, we end up having an endomorphism in the lattice of quantised fluxes. If we represent the lattice of closed string fluxes by the vector $l_s \varrho_A = (e_0, e_a, m^a, m, h_K)$ then we can represent the endomorphisms for each of the above 4d strings as a set of square matrices
\be
P_a =\left(\begin{array}{ccccc}0 & \vec{\d}_a^{\, t} & 0 & 0 & 0 \\ 0 & 0 & \CK_{abc} & 0& 0 \\ 0 & 0 & 0 & \vec{\d}_a & 0\\0 & 0 & 0 &0& 0\\0 & 0 & 0 &0& 0\end{array}\right)\, ,\qquad \qquad
P_K =\left(\begin{array}{ccccc}0 & 0 & 0 & 0 & \vec{\d}_K^{\, t} \\ 0 & 0 & 0 & 0 & 0\\ 0 & 0 & 0 &0 & 0\\0 & 0 & 0 &0 & 0\\ 0 & 0 & 0 &0 & 0\end{array}\right)\,,
\ee
with $(\vec{\d}_a)^b = \d_a{}^b$. Here $P_a$ represents a NS5-brane wrapping the four-cycle class $[\pi_4^a] = \text{P.D.}[\omega_a]$ and $P_K$ a D4-brane wrapping the three-cycle class $[\pi_3^K] = \text{P.D.}[\a_K]$. For instance, $P_a$ maps a 0-flux $m$ to $m$ units of two-form flux $F_2$ along $\om_a$. This is precisely the flux jump  when crossing a 4d domain wall made of $m$ D6-branes wrapping $\pi_4^a$, which is the brane content needed to cancel the corresponding FW anomaly.

Notice that the matrices $P_a$, $P_K$ are strictly upper triangular and therefore nilpotent, and that they all commute with each other. Finally, one can easily check that
\be
e^{\phi^\a P_\a} \, =\,
e^{b^aP_a + \xi^{\prime K}P_K}\, =\, {\bf R}^{t}
\label{expR}
\ee
where $\phi^\a =\{b^a, \xi^{\prime K}\}$ runs over all axions, and {\bf R} is the rotation matrix in (\ref{RbH}). Therefore, we find that the axion dependence of the scalar potential is fully determined by simple topological data of the compactification, namely by the FW anomalies developed by 4d strings.

\subsection{Discrete shift symmetries}\label{ss:dss}

Four-dimensional effective theories of the form (\ref{S4formRR}), in which a set of axions couple to a set of four-forms, display a set of exact discrete shift symmetries. In those symmetries the shift of an axion by its period is compensated by a discrete shift of the four-form expectation values (or in other words by a shift of the quantised background fluxes) such that the scalar potential remains invariant. Since different choices of background fluxes corresponds to different scalar potentials, the presence of such discrete symmetries entails a potential structure with multiple identical branches, in which the action of a shift symmetry involves the jump from one branch to another. The simplest example of this setup is the minimal axion-four-form coupling considered in \cite{KS,KLS} to realise chaotic inflation, which results in the following quadratic potential
\be
V \, =\, \oh \left( q - m \phi \right)^2
\label{KSV}
\ee
with $\phi$ an axion of unit period and $q,m \in \IZ$. Here $q$ corresponds to a four-form flux that labels the different branches of the potential, that are connected to each other by means of the discrete shift symmetry
\be
\phi \raw \phi +1\, , \qquad \qquad q \raw q + m\, .
\label{simpleDSS}
\ee
Such simple 4d axion-four-form axion is recovered in certain subsectors of F-term axion monodromy models \cite{Marchesano:2014mla}. As pointed out in \cite{Bielleman:2015ina}, in more general setups like the type IIA flux compactifications at hand, the axion dependence of the potential is more involved than in (\ref{KSV}). This in turn translates into a more complicated multi-branched structure and discrete shift symmetries than in (\ref{simpleDSS}) \cite{Bielleman:2015ina,Carta:2016ynn}. 

It turns out that the expression (\ref{expR}) leads to a general description of the discrete shift symmetries of the potential. Indeed, the scalar potential resulting from (\ref{BilVRR}) and (\ref{factoRM}) is of the form
\be
V \, =\,  \frac{1}{\kappa_4^2} \frac{M^{AB}}{8}   \rho_A \rho_B
\label{BilVrho}
\ee
where, in the absence of open string moduli, the $\rho_A$ are the components of the following vector
\begin{align}
l_s \vec{\rho} \, =\,  {\bf R}^{t\, -1}  
\left(\begin{array}{c} e_0 \\ e_a \\  m^a \\  m \\  h_K   \end{array} \right) \, =\,  {\bf R}^{t\, -1} \vec{q}
\label{rhos}
\end{align}
with $q_A \in \IZ$ and  {\bf R} as in (\ref{expR}). These $\rho_A$ are precisely the axion polynomials identified in \cite{Bielleman:2015ina}, which are left invariant under the discrete shift symmetry of the effective theory. In terms of (\ref{expR}) periodic shifts of the axions $\phi^\lam$ corresponds to the transformation
\be
{\bf R}^{t\, -1} (\phi^\lam + k^\lam) \,=\, {\bf R}^{t\, -1} (\phi^\lam) \cdot e^{-k^\lambda P_\lam} \qquad \qquad k^\lam \in \IZ
\label{axionsh}
\ee
where we have used that $\{P_\a\}$ are commuting matrices. The vector (\ref{rhos}), and therefore the potential $V$, remain invariant if we perform the simultaneous shift
\be
\vec{q} \, \mapsto\, \vec{q}^{\, \prime} = e^{k^\lam P_\lam}\, \vec{q}
\label{fluxsh}
\ee
which we are always allowed to implement since $e^{k^\lam P_\lam}$ is a matrix of integers and so $q_A' \in \IZ$. The transformations (\ref{axionsh}) and (\ref{fluxsh}) are the generalisation of the discrete shift symmetry (\ref{simpleDSS}) to this more general setup. 

The vector entries $\rho_A$ can be seen as the basic building blocks of the potential. Indeed, any flux dependence of the potential or any axion dependence which is not periodic must come through a function of the $\rho_A$, or else it will not respect the underlying discrete shift symmetry of the theory. This explains why the dimensional reduction of a two-derivative action must yield an effective potential of the form (\ref{BilVrho}). However, since the symmetry is exact,  the statement also applies when we consider arbitrary corrections to the potential. As pointed out in \cite{Carta:2016ynn} these symmetries must also leave invariant the superpotential of the 4d effective theory. In fact, as we will discuss in the next section, they allow to reconstruct $W$ from the $\rho_A$, and vice versa.

\subsubsection*{Further discrete symmetries}

The bilinear form of the scalar potential (\ref{BilVrho}) is also useful in order to identify classical symmetries under transformations of fields and fluxes, beyond the shift symmetries discussed above. In particular one may identify duality symmetries involving the saxions. Indeed, notice that the potential is invariant under orthogonal $O(N, \IZ)$ transformations
\beq
\vec{\rho}\ \rightarrow \ O \vec{\rho} \quad \quad \quad M\ \rightarrow \ OMO^{t} \ ,
\eeq
where $\vec{\rho} = {\bf R}^{t\, -1} \vec{q}$ is defined similarly to (\ref{rhos}) and $N$ is its number of entries. Consider in particular the RR sector of the potential and the following transformation on $\vec{\rho}$
\beq 
(\rho_0,\rho_a,{\tilde \rho}^a,{\tilde \rho})\ \rightarrow \ ({\tilde \rho},{\tilde \rho}^a,\rho_a,\rho_0) .
\eeq
This is a symmetry as long as we transform 
\beq
\CK\ \rightarrow \  \frac {36}{\CK}\, , \quad \quad  g^{ab}\ \rightarrow \ 16g_{ab} \, .
\eeq
This is a duality symmetry which relates large and small volumes in the classical potential. Further choices for transformations
of the flux-axion polynomials contained in $\vec\rho$ come along with different duality actions on the saxion metrics and volumes.
These symmetries together with the shift symmetries belong to the full duality group of the compactifications. In the case of toroidal
compactifications one can explicitly verify that both types of symmetries correspond to the action of the modular duality groups. 
In particular, the above volume-duality transformation corresponds to the action of the ${\cal S}$ modular generators of the K\"ahler moduli group $SL(2, \IZ)^3$. The  RR fluxes transform like a $(2,2,2)$ representation of $SL(2, \IZ)^3$ and the shift symmetries correspond to the action of the shift generators ${\cal T}$  in these modular groups. This interesting structure is described in detail in Appendix \ref{ap:sym}.

\subsection{Adding open strings}\label{ss:FWopen}

Let us now reinstate the presence of open string moduli and let us see how the previous prescription to obtain the axion rotation matrix {\bf R} generalises to this case. In general, having open string moduli will not only increase the number of axions (i.e., the D6-brane Wilson lines) and therefore of 4d strings of the compactification. It will also increase the number of 4d domain walls. In order to classify the latter, recall that one can locally describe the open string moduli space of the compactification in terms of a set of reference special Lagrangian three-cycles $\{\Pi_\a^0\}$ with no worldvolume flux on them and satisfying the RR tadpole condition (\ref{RRtadpole}). Each three-cycle $\Pi_\a^0$ is homotopic to the actual three-cycle $\Pi_\a$ where the D6-brane sits, and which corresponds to the point in open string moduli space that we are looking at. 

In this description, there are two new kinds of 4d domain walls. The first one is made of a D6-brane stretched along the four-chain $\cc_{4}^{\a}$ such that $\p\cc_{4}^{\a} = \Pi_\a - \Pi_\a^0$, and that connects the vacuum that corresponds with the reference D6-brane configuration $\{\Pi_\a^0\}$ with the one that corresponds with the actual configuration $\{\Pi_\a\}$. The second one is made of D4-branes stretched along two-chains $\cc_{2}^{\a\, i}$ defined by $\p \cc_{2}^{\a\, i} = \g_\a^i -  \g_\a^{i\, 0}$, where $[\g_\a^i]$ is the class of one-cycles Poincar\'e dual to the two-form class $[\eta^i]$ in $\Pi_\a$, and $\g_\a^{i\, 0}$ is its homotopic relative in $\Pi_\a^0$. Having $k$ of such domain walls corresponds to switching on $n_{F\, i}^\a = k$ units of worldvolume flux on the D6-brane wrapping $\Pi_\a$. 

\begin{figure}[t]
\begin{center}
\includegraphics[width=140mm]{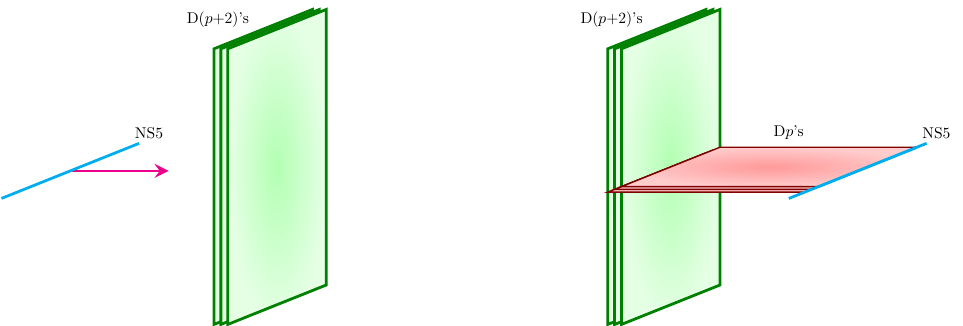}
\caption{Hanany-Witten effect \cite{Hanany:1996ie}: when an NS5-brane crosses $M$ D($p$+2)-branes, $M$ D$p$-branes will appear stretching between the two.}\label{fig1}
\end{center}
\end{figure}

Given this new set of 4d domain walls one may consider how they interact with the previously discussed 4d strings. Instead of describing this interaction in terms of the Freed-Witten anomaly, we will now do so in terms of the Hanany-Witten effect. More precisely, we will use the fact that when an NS5-brane crosses a D($p$+2)-brane, a D$p$-brane extended along the two and wrapping their common directions is created, see figure \ref{fig1} for a cartoon of the process. Since crossing a D-brane corresponds to switching on a background flux, this brane creation effect is in one-to-one correspondence with the Freed-Witten anomaly cancellation, as explained e.g. in Appendix B of \cite{BerasaluceGonzalez:2012zn}. Therefore, we can use it to reconstruct the generators $P_\a$ and the axion rotation matrix {\bf R}.

Indeed, in our setup a HW brane creation effect will occur whenever a 4d string made up of a NS5-brane wrapping a four-cycle $\pi_4^\a$ crosses one of the above 4d domain walls made up of D($p$+2)-branes stretching along $p$-chains. If the three-cycle $\Pi_\a$ contains a two-cycle $\pi_2^i$ that is non-trivial in the bulk, that is if $n_{a \, i}^\a \neq 0$, then an NS5-brane wrapping $\pi_4^a$ will intersect $\Pi_\a$ on the dual one-cycle $\g_\a^i$, and so the four-chain $\cc_{4}^{\a}$ along $\cc_{2}^{\a\, i}$. Therefore, due to the HW effect, when a 4d NS5-string crosses a 4d domain wall made up of a D6-brane along $\cc_{4}^{\a}$, then $n_{a \, i}^\a$ 4d domain walls made up of D4's along $\cc_{2}^{\a\, i}$ are created. Finally, as mentioned above the latter domain walls are dual to the worldvolume flux $n_{F\, i}^\a$. In practice all this implies that the entries of (\ref{rhosopen}) must depend on the following combination of axions and fluxes
\be
n_{F\,i}^\a - b^a n_{a \, i}^\a
\ee
as is indeed the case. The same NS5-brane will also intersect $\cc_{4}^{\a}$ along $\cc_{2}^{\a\, i}$ if the chain integral $f_{a\, \a}^i$ in (\ref{deffch}) does not vanish. The intersection will now be along the two-cycle $\pi_2^i$, whose decomposition in terms of bulk two-cycles is given by $[\pi_2^i] = n_{a \, i}^\a [\pi_2^a]$. Therefore when a 4d NS5-string crosses the same D6-domain wall, $n_{b \, i}^\a f^i_{a\, \a}$ D4-branes wrapping $\pi^b_2$ will be created. This implies that the following combination of axions and fluxes appears in (\ref{rhosopen})
\be
e_b - n_{b \, i}^\a b^a f^i_{a\, \a} \qquad \raw \qquad e_b + n_{b \, i}^\a \hat{\th}^i_\a
\ee
where have promoted the combination $b^a f^i_{a\, \a}$ to the full open string axion $\hat{\th}^i_\a$. 
Finally, a NS5-brane wrapping $\pi_4^a$ will intersect a D4-brane along $\cc_{2}^{\a\, i}$ if $f^i_{a\, \a} \neq 0$. When such a 4d NS5-string crosses the 4d D4-domain wall, D2-branes along 4d will be created, that is the objects dual to the flux $e_0$. this implies the following combination of axions and fluxes in (\ref{rhosopen})
\be
e_0 - n_{F\,i}^\a b^a f^i_{a\, \a} \qquad \raw \qquad e_0 -  n_{F\,i}^\a \hat{\th}^i_\a
\ee
where we have again completed the combination to include the full $\hat{\th}^i_\a$. 

Putting all these result together, we obtain the generalisation of (\ref{expR}) that includes open string domain walls and axions. First we define the flux vector 
\be
l_s \varrho_A = (e_0, e_a, m^a, m, n_{F\, i}^\a, h_K,  n_{a\, i}^{\a})\, .
\label{qopen}
\ee
Here the entries $n_{F\, i}^\a$ are worldvolume fluxes whose corresponding 4d domain walls are D4-branes wrapping $\cc_{2}^{\a\, i}$. The 4d domain walls corresponding to the entries $n_{a\, i}^{\a}$ and $p_{\a'}$ are more obscure, and are presumably related to those made up of D6-branes wrapping $\cc_{4}^{\a}$. As pointed out in \cite{Marchesano:2014mla}, the 4d four-form that corresponds to $n_{F\, i}^\a$ can be obtained from dimensional reduction of the four-form $A_4$ dual to the D6-brane gauge vector potential. There is however no obvious description for the 4d four-forms that correspond to the last two entries of \eqref{qopen} and it would be interesting to develop one. In particular, it would be interesting to describe their interplay with the presence of RR two-form fluxes since then, as mentioned before, one must satisfy the consistency condition $m^an_{a\, i}^\a$, $\forall \, \a, i$. 

Acting on this flux vector, the FW and HW relations translate into the following matrices 
\be
P_a =\left(\begin{array}{ccccccc}0 & \vec{\d}_a^{\, t} & 0 & 0 & 0 &0& 0 \\ 0 & 0 & \CK_{abc} & 0& 0 &0& 0\\ 0 & 0 & 0 & \vec{\d}_a & 0&0& 0\\0 & 0 & 0 &0& 0&0& 0\\0 & 0 & 0 &0& 0&0& \Delta_a\\0 & 0 & 0 &0& 0&0 & 0 \\0 & 0 & 0 &0& 0&0& 0\end{array}\right)\quad
P_K =\left(\begin{array}{ccccccc}0 & 0 & 0 & 0 & 0&  \vec{\d}_K^{\, t} & 0 \\0 & 0 & 0 &0& 0&0& 0\\0 & 0 & 0 &0& 0&0& 0\\0 & 0 & 0 &0& 0&0& 0\\0 & 0 & 0 &0& 0&0& 0\\0 & 0 & 0 &0& 0&0& 0\\0 & 0 & 0 &0& 0&0& 0\end{array}\right)\,
\label{PaKopen}
\ee
with $(\Delta_{a})^j_{k \, b}  = \d_a^b \d^j_k$,  and
\be
P_i^{\, \a} =\left(\begin{array}{ccccccc}0 & 0& 0 & 0 &  \vec{\d}_{i}^{\, t} & 0 & 0 \\ 0 & 0 &0 & 0& 0 & 0 &\Delta_a\\ 0 & 0 & 0 & 0 & 0&0& 0\\0 & 0 & 0 &0& 0&0& 0\\0 & 0 & 0 &0& 0&0& 0 \\0 & 0 & 0 &0& 0& 0 & 0 \\0 & 0 & 0 &0& 0&0& 0\end{array}\right)\, ,
\label{Piaopen}
\ee
where the last set corresponds to the Wilson line axions $\hat{\th}^i_\a$. For simplicity we have considered the case of a single D6-brane with moduli. Exponentiating we arrive to the axion rotation matrix
\be
{\bf R}  = \left(e^{b^aP_a + \xi^KP_K +  \hat \theta^i_\a P_i^{\a}}\right)^{t} = 
\left(\begin{array}{ccccccc} 1 & 0 & 0 & 0 & 0& 0 & 0\\  b^a & \delta^a_b  & 0 & 0 & 0 &0  & 0\\  \frac{1}{2} \mathcal K_{abc} b^b b^c & \mathcal K_{abc} b^c & \delta^b_a  & 0 & 0 & 0  & 0\\  \frac{1}{3!} \mathcal K_{abc} b^a b^b b^c & \frac{1}{2}\mathcal K_{bac} b^a b^c & b^b& 1 & 0 & 0 & 0 \\  \hat{\th}^i & 0 & 0 & 0 & \d^i_j   & 0   & 0\\ \xi^{K}   & 0 & 0 & 0 & 0 & \d_L^K & 0 \\   \hat{\th}^i b^a &  \hat{\th}^i \delta^b_a& 0 &0&  b^a \d_{i}^{j} & 0 &  \d_{a\, i}^{b\, j}\end{array}\right)
\label{Rotopen}
\ee
that reproduces the results in section \ref{s:bilinear}. Needless to say, the description of discrete shift symmetries made in subsection \ref{ss:dss} readily generalises to this more general case, with the flux vector $\vec{q}$ replaced by (\ref{qopen}) and the axion rotation matrix {\bf R} by (\ref{Rotopen}).

\section{The superpotential and the master axion polynomial}\label{s:master}

As already mentioned, the discrete shift symmetries of section \ref{ss:dss} are in fact discrete gauge symmetries, and as such they must also be respected by the superpotential of the effective 4d theory, which in the classical regime in which we are working is given by the polynomial (\ref{Wtotal}). Explicitly, we have that up to the constant piece $W_0$ such polynomial reads
\be         
l_s W = e_0 - e_a T^a + \frac{1}{2} {\cal K}_{abc} m^a T^b T^c - m \frac{1}{6} {\cal K}_{abc} T^a T^b T^c - h_K N^K - \Phi_\a^i( n_{F\, i}^\a - n_{a\, i}^\a T^a ) \, .
\ee
One can check explicitly that the combination of axion plus flux shifts given by (\ref{axionsh}) and (\ref{fluxsh}) leaves this expression invariant, just like it leaves invariant the axion polynomials defined by
\be
l_s \vec{\rho} \, =\,  {\bf R}^{t\, -1} \vec{q} \, =\, {\bf R}^{t\, -1}  
\left(\begin{array}{c} e_0 \\ e_a \\  m^a \\  m \\ n_{F\, i}^\a  \\   h_K \\ n_{a\, i}^\a  \end{array} \right)
\label{vecrho}
\ee
where {\bf R} is given by (\ref{Rotopen}). This already suggests that the axion polynomial vector $\vec{\rho}$ and the superpotential above are intimately related. In fact, we will show that they contain the same information, and there is a one-to-one dictionary between both.

Let us first show how to obtain the superpotential from $\vec{\rho}$. For this, recall that in general a flux superpotential can be written down as \cite{Gukov:1999ya,Taylor:1999ii} 
\be
l_s W\, =\, \Pi^{\, t}(\psi) \cdot \vec q
\ee
where $\Pi$ is a matrix of periods from where one can construct the tree-level K\"ahler potential, with one entry normalised to the unity and the rest depending holomorphically on the complex fields $\psi^\lam = \phi^\lam + i s^\lam$. In the case at hand we have
\be
\Pi^{\, t}\, =\, (1, -T^a, \oh \CK_{abc} T^aT^b, -\frac{1}{6} \CK_{abc} T^aT^bT^c, - \Phi^i_\a, -N^K, n_{a\, i}^\a T^a \Phi_\a^i) \, ,
\ee
although its precise form will not be essential in the following. Indeed, in general one can rewrite the superpotential as
\be
W\, =\, [{\bf R}(\phi) \Pi(\psi)]^{\, t} \cdot \vec \rho\, ,
\label{Wprod}
\ee
where we recall that {\bf R} only depends on the axionic part $\phi^\lam$ of each complex field $\psi^\lam$. Now, both $W$ and $\vec{\rho}$ are  invariant under the discrete shifts (\ref{axionsh}) and (\ref{fluxsh}), so the product {\bf R}$\, \Pi$ must be invariant as well. Because this product does not depend on the fluxes,\footnote{{\bf R} does not depend on the fluxes by construction, and $\Pi$ does not depend on the fluxes whenever the K\"ahler potential does not depend either.}  this means that it must be invariant under periodic axion shifts alone, so it can only depend on the saxions $s_\lam$ or on periodic terms of the form $e^{2\pi i \phi^\lam}$. Finally, the latter are absent whenever we neglect world-sheet and D-brane instanton contributions, so in this regime we necessarily have that
\be
{\bf R}(\phi) \Pi(\psi) \, =\, \Pi(s) \quad \Raw \quad W\, =\, [\Pi(s)]^{\, t} \cdot \vec \rho \, . 
\label{Wpirho}
\ee
Moreover, we can always express a holomorphic function as $W(\psi) =  e^{is^\lambda  \frac{\p}{\p \phi^\lam}} W(s =0)$, which applied to the above expression gives
\be
W\, =\, e^{is^\lam \frac{\p}{\p \phi^\lam}} \left[ \Pi^{\, t} (0) \cdot \vec{\rho} \right] \, .
\ee
In general, $\Pi(0)$ only contains one non-vanishing entry, the one that is normalised to unity. Therefore, the product $\Pi^{\, t} (0) \cdot \vec{\rho}$ selects one of the axion polynomials contained in $\vec \rho$, which in the following we dub {\em master axion polynomial} and denote by $\rho_0$. To sum up, we have that the dictionary that takes us from $\vec\rho$ to $W$ reads
\begin{equation}
\label{superpo}
\boxed{
W=e^{is^\lam \frac{\partial}{\partial \phi^\lam}} \rho_0
	}
\end{equation}
with $\rho_0$ the particular component of $\vec{\rho}$ selected by $\Pi(0)$. In the case at hand, such master axion polynomial is nothing but the first entry in (\ref{vecrho}) and it reads
\be
l_s \rho_0 \, =\, e_0 - b^a e_a +\frac {1}{2} \mathcal{K}_{abc} m^ab^bb^c- \frac {m}{6}\mathcal{K}_{abc}b^ab^bb^c  - h_K \xi^{K} - \hat \theta_\a^i \left(n_{F\, i}^\a  - n_{a\, i}^\a b^a  \right) \, ,
\ee
from where it is obvious that (\ref{superpo}) holds. This axion polynomial is associated to the four-form $F_4^0$ defined in (\ref{F4form}), that arises from the direct dimensional reduction of $dC_3$ with all indices in Minkowski. Notice that this 4-form is universal, compactification independent and present in any Type IIA orientifold of the class here considered. 

Let us now discuss how to derive $\vec{\rho}$ in (\ref{vecrho}) from $W$. Obviously, the entry given by $\rho_0$ is easily recovered from $W$ by definition, since $\rho_0 \equiv W|_{s=0}$. Now, it turns out that we can also recover the remaining components of $\vec{\rho}$ by taking successive derivatives of $\rho_0$ with respect to all the different axions that it depends on. Schematically, we have that all the components of $\vec{\rho}$ are linear combination of the following axion polynomials 
\begin{equation}
\label{generatorrhos}
\boxed{
\rho_{\lam_1...\lam_n}=\dfrac{\partial^n \rho_0}{\partial\phi^{\lam_1}...\partial\phi^{\lam_n}}
}
\end{equation}
and vice versa. To see this let us first characterise all the entries of $\vec{\rho}$ as 
\be
 \rho_A\, =\,  \vec{\d}_A^{\, t} \cdot \vec{\rho}\, =\, l_s^{-1} \vec{\d}_A^{\, t} \cdot {\bf R}^{t\, -1} \cdot \vec{q}\, =\, l_s^{-1}  \vec{\d}_A^{\, t} \cdot e^{-\phi^\lam P_\lam} \cdot \vec{q}
\ee
where $\vec{\d}_A$ is a vector with the $A^{th}$ entry equal to one and all the other vanishing, and in particular we have that $\vec{\d}_0^t = (1, 0 ,0, \dots)$. Also, in the last equality we have used the expression of {\bf R} in terms of the nilpotent generators $P_\lam$, as in (\ref{Rotopen}). We then have that
\be
\p_{\lam_1} \dots \p_{\lam_n} \rho_0\, =\, (-1)^{n} \vec{\d}_0^{\, t} \cdot P_{\lam_1} \dots P_{\lam_n} \cdot \vec{\rho}\, =\, (-1)^{n} \left(P_{\lam_1}^t \dots P_{\lam_n}^t \vec{\d}_0\right)^t \cdot \vec{\rho} \, .
\ee
Since the generators $P_\lam$ do not depend on axions or fluxes, one recovers linear combinations of the axion polynomials in $\vec{\rho}$, whose coefficients at most depend on the topological numbers $\CK_{abc}$. Also, since the matrices $P^t_\lam$ are strictly lower triangular, their action lowers the position of the non-vanishing entry of any vector $\vec{\d}_A$. Finally, the different combinations of generators acting on $\vec{\d}_0$ scan all possible entries of this vector, as one can check with the expressions (\ref{PaKopen}) and (\ref{Piaopen}). 

For completeness, let us show explicitly how these axion polynomials are obtained. By successive derivation we find
\begin{equation}
\begin{array}{r c l}\vspace*{.2cm}
 l_s \dfrac{\partial \rho_0}{\partial b^a} & = &  - e_a+ \mathcal{K}_{abc} m^b b^c - \dfrac{m}{2} \mathcal{K}_{abc} b^b b^c + n_{a\, i}^\a \hat\theta^i \, =\,-  l_s\rho_a,  \\ \vspace*{.2cm}
  l_s\dfrac{\partial^2 \rho_0}{\partial b^a \partial b^b} & = &  \mathcal{K}_{abc} \left(m^c - m b^c \right)\,  = \,  l_s\mathcal{K}_{abc} \tilde{\rho}^c, \\ \vspace*{.2cm} 
 \dfrac{\partial^3 \rho_0}{\partial b^a \partial b^b \partial b^c} & = & - \mathcal{K}_{abc} m \, =\, -  \mathcal{K}_{abc} \rho_m,  \\ \vspace*{.2cm}
  l_s\dfrac{\partial \rho_0}{\partial \hat\theta^i_\a}  & = &  -n_{F\, i}^\a+ n_{a\, i}^{\a} b^a \, =\, -  l_s\rho_{{F\, i}^\a},  \\ \vspace*{.2cm} 
  l_s\dfrac{\partial \rho_0}{\partial \xi^{K}}& = &  - h_K \, =\, -  l_s\rho_{K}, \\
  l_s\dfrac{\partial^2 \rho_0}{\partial b^a \partial \hat\theta^i_\a}  & = &   n_{a\, i}^\a \, =\,  l_s \rho_{n_{a\, i}^{\a}}\, .
\end{array}
\label{derirhos}
\end{equation}

To summarise, one may consider the master axion polynomial $\rho_0$ as the generator of all other axion polynomials associated to all possible Minkowski 4-forms. As a consequence of this and the results of section \ref{s:bilinear}, the scalar potential can be written in the general form
\beq
V\, =\, G^{\lam_1\dots\lam_n}_{\mu_1\dots \mu_n} \left(\frac {\partial \rho_0}{\partial \phi_{\lam_1}..\partial \phi_{\lam_n}}\right)
\left(\frac {\partial \rho_0}{\partial\phi_{\mu_1}..\partial \phi_{\mu_n}}\right)
\eeq
with $G$ encoding all the saxion-dependent geometric data, and $\rho_0$ all the information about axions and fluxes. Since $\rho_0$ keeps the information about the rest of the $\rho_A$'s, one can consider this formulation as an alternative recipe to the Cremmer et al. potential, computed in terms of the K\"ahler potential and the superpotential. Here the full scalar potential may be constructed in terms of the matrix of metrics {\bf M} and $\rho_0$. In the first formulation supersymmetry is explicit whereas in the second it is not but, as follows from the discussion of the last section, the duality symmetries (like shift symmetries) are more transparent.

\section{4-forms and sugra auxiliary fields}\label{s:aux}

Since 4-forms do not propagate, it is interesting to investigate the possible connection between them and the moduli auxiliary fields 
of the minimal 4d $\mathcal{N}=1$ sugra formulation. In the closed string sector, the off-shell action includes $2(h_{11}^++h_{21}^++1)$  real auxiliary scalars, corresponding to one complex field per modulus. To those one has to add two more from the supergravity multiplet. On the other hand  there are $2(h_{11}^+ +1)$ RR 4-forms and $(h_{21}^++1)$ NS ones. If open string moduli are added, there is one complex auxiliary field per D6-brane, whereas $b_1(\Pi_\alpha^0)$  additional 4-forms per brane appear. It is clear that the most naive expectation of a matching between sugra scalar auxiliary fields and 4-forms does not work, at least for Type IIA orientifolds. Still, it is interesting to express the sugra auxiliary fields in terms  of the $\rho$ polynomials given by eq. \eqref{generatorrhos}. Hopefully this may give us hints about a possible new  off-shell formulation with 4-forms acting as auxiliary fields (notice that, using eq. \eqref{4dH} it is straightforward to obtain the expressions in terms of the 4-forms from those in terms of the $\rho$'s) . 

The minimal $\mathcal{N}=1$ supergravity F-term auxiliary fields are given by
\beq
{\overline F^{\bar \beta} = e^{\frac{K}{2}} K^{\bar \beta \alpha } D_\alpha W \,.
}
\eeq
Decomposing them as $\psi^a = \phi^\a + i s^\a$ one can evaluate them by computing
\beq
{ \p_\alpha W = \frac{1}{2} (\p_{\phi^\alpha} - i \p_{s^\alpha}) \left[e^{i s^\beta \p_{\phi^\beta} } \rho_0 \right] = e^{i s^\beta \p_{\phi^\beta}} \p_{\phi^\alpha} \rho_0 =e^{i s^\beta \p_{\phi^\beta}}\rho_\alpha\,,
}
\eeq
where we have used the expression \eqref{superpo}.  Using
\beq
{ K^{\bar \beta \alpha} K_\alpha = 2 i s^\alpha\,,
}
\eeq
one finally obtains
\beq
{ \overline F^{\bar \beta} = e^{\frac{K}{2}} e^{i s^\gamma \p_{\phi^\gamma}} \left[ K^{\bar \beta \alpha} \rho_\alpha + 2 i s^\beta \rho_0\right]\,,
}
\eeq
for all the auxiliary fields. Let us be more explicit and provide the detailed  dependence on each of the different $\rho$-polynomials.

\subsubsection*{In the absence of open string moduli}

Let us first restrict ourselves to closed string moduli in the presence of RR and NS fluxes. In this case, the K\"ahler potential can be separated in two pieces in the following way:
\begin{equation}
\label{Kclosed}
K=K_K (t^a)+K_{Q} (n^{'K}),
\end{equation}
where the first term only depends on the K\"ahler moduli and is given by the expression \eqref{KK} and the second on the complex structure and is shown in eq. \eqref{KQ}. From this K\"ahler potential we can obtain the metric in the K\"ahler moduli space and its inverse  as  \cite{philip,Grimm:2004ua} 
\begin{equation}
\label{kahlermetricclosed}
K_{a \bar{b}}=-\dfrac{3}{2\mathcal{K}}\left( \mathcal{K}_{ab}-\dfrac{3}{2}\dfrac{\mathcal{K}_a \mathcal{K}_b}{\mathcal{K}}\right) \ ;\ 
K^{a  \bar{b}}=-\dfrac{2\mathcal{K}}{3} \left( \mathcal{K}^{ab}-\dfrac{3t^a t^b}{\mathcal{K}}\right),
\end{equation}
where we have defined the following contractions of the triple intersection numbers with the imaginary parts of the K\"ahler  moduli
\begin{equation}
\begin{array}{c c }
\mathcal{K}_{ab}=\mathcal{K}_{abc}t^c, & \mathcal{K}_{a}=\mathcal{K}_{abc} t^b t^c,
\end{array}
\end{equation}
with $\mathcal{K}^{ab}$ being the inverse of $\mathcal{K}_{ab}$. Notice that, due to this separation in the K\"ahler potential, each piece separately satisfies a no-scale type condition, that is
\begin{equation}
\begin{array}{l  c r}
\label{noscale}
K_{a}K^{a \bar{b}}K_{\bar{b}}=3\, , & & K_{I}K^{I\bar{J}}K_{\bar{J}}=4\, .
\end{array}
\end{equation} 
By applying eq.(\ref{Wpirho}) one finds
\begin{equation}
W=\rho_0 -\dfrac{1}{2}\mathcal{K}_a \tilde{\rho}^a- i \left[  t^a \rho_a-n^{'I} \rho_{h_I} +\dfrac{ \mathcal{K}}{6} \rho_m\right],
\end{equation}
and it may be easily checked  that this agrees with eqs. \eqref{WK} and \eqref{WQ}, that is,  the RR+NS flux superpotential \cite{Grimm:2004ua}. The auxiliary fields can then be computed yielding
\begin{equation}
\begin{array}{r c l}
\label{auxfieldsct}
\bar{F}^{T^a} &=& e^{K/2}\left\{  \left[ \left(2 t^a t^b - K^{\bar{a} b}  \right) \rho_b + 2 t^a n'^{I} \rho_I+ \left(\dfrac{1}{2} K^{\bar{a}b}\mathcal{K}_b -\dfrac{1}{3}t^a \mathcal{K}\right) \rho_m\right] \right.\\
& & \qquad \quad  \left. + i \left[ 2 t^a \rho_0 + \left( K^{\bar{a}b}\mathcal{K}_{bc} - t^ a \mathcal{K}_c \right)\tilde{\rho}^c  \right]  \right\}\\ \\
 				    &=& e^{K/2}\left\{  \left[\dfrac{2}{3}\mathcal{K} \mathcal{K}^{ab} \rho_b + 2 t^a n'^{I} \rho_I+ \dfrac{1}{3}\mathcal{K}t^a \rho_m\right]+ i \left[ 2 t^a \rho_0 + \left( t^a \mathcal{K}_c-\dfrac{2}{3}\mathcal{K} \delta_c^a \right)\tilde{\rho}^c  \right] \right\}, \\ 
\end{array}
\end{equation}
\begin{equation}
\begin{array}{r c l}
\label{auxfieldscn}
\bar{F}^{N^I}& = &  e^{K/2} \left\{ \left[  \left( 2 n'^{I} n'^{J}- K^{\bar{I}J}\right)\rho_J + 2 n'^{I} t^a \rho_a -\dfrac{1}{3}n'^{I} \CK \rho_m \right] + i \left[ 2 n'^{I} \rho_0 - n'^{I} \CK_a \tilde{\rho}^a  \right]  \right\}.
\end{array}
\end{equation}
where in the second line of $\bar{F}^{T^a}$ we have expressed the inverse K\"ahler metric as in eq. \eqref{kahlermetricclosed}. In addition, we have used  $K^{\bar{a} b}K_b=2 i t^{a}$ and $K^{\bar{I} J}K_{J}=2 i n'^{I}$, which is only valid thanks to the block diagonal form of the K\"ahler metric coming from the fact t hat  $n'^{I}$ in eq. \eqref{Kclosed} is the field that enters the holomorphic variable $N^{I}$ (i.e. $n'^{I}=n^{I}=\im\, N^I$), as opposed to the case in which open strings are taken into account.  As a remark, in the absence of NS fluxes, there is a one to one correspondence between 4-forms and sugra auxiliary fields. This is the case treated in \cite{Farakos:2017jme}  and, as expected, our expressions for the auxiliary fields $F^{T^a}$ and $W$ match the ones in there after performing the corresponding substitutions in the 4-forms and K\"ahler metric. 

However, a point to consider in the more general case with NS fluxes is that the imaginary part of all the complex structure auxiliary fields $\im\, F^{N^I}$ is the same for all the fields (modulo the metrics) and is directly given by the real part of the superpotential. This is in agreement with the fact that the number of NS 4-forms is half the number of complex structure fields, and hence there should be half the number of independent auxiliary fields. 

An interesting question is how this structure changes in the presence of additional closed string fluxes. One could conceive that the presence of extra e.g. NS fluxes could lead to a match of sugra auxiliary fields and 4-forms or some modification of the bilinear structure that we have found. This does not seem to be the case.  As an example  we have worked out  in Appendix \ref{ap:metric} a toroidal Type IIA $\IZ_2\times \IZ_2$ orientifold example in the presence of additional NS metric fluxes, beyond the standard RR and NS ones. The structure we have found in the above sections goes through with the addition of new 4-forms associated to the metric fluxes. In this case the total number of 4-forms is larger than the number of complex structure fields and no matching seems possible.  As already pointed out in \cite{Bielleman:2015ina}, this suggests the necessity of a $\mathcal{N}=1$ sugra formulation in which the 4-forms are contained as auxiliary fields of new $\mathcal{N}=1$ multiplets. This has recently been explored in \cite{Farakos:2017jme} .

Finally, recall that the scalar potential cam be written in terms of the auxiliary fields and the superpotential as
\begin{equation}
V= F^{T^a}K_{a \bar{b}} F^{T^{*b}}+F^{u^I}  K_{I\bar{J}} F^{u^{*J}}- 3 e^K | W |^2\ ,
\end{equation}
so substituting \eqref{auxfieldsct})-\eqref{auxfieldscn} and using (\ref{noscale}) we get the general expression for the scalar potential as a function of the $\rho$'s and the derivatives of the K\"ahler potential
\begin{equation}
\begin{array}{l c l}
V&=&  e^K\left[ 4 \rho_0^2 + g^{a b} \rho_a \rho_b + \dfrac{4}{9} \mathcal{K}^2 g_{ab} \tilde{\rho}^a\tilde{\rho}^b+\dfrac{1}{9} \mathcal{K}^2 \rho_m^2+  K^{LJ}  \rho_{h_L}\rho_{h_J} +\dfrac{1}{3} \mathcal{K} n^{'L}\rho_{h_L} \rho_m \right].
\end{array}
\end{equation}
The interest of an expression like this is that the dependence on the axions of the system is encapsulated inside the $\rho$ polynomials. 
The invariance of the potential under the monodromy shift symmetries is explicit, since the $\rho$ polynomials are invariant.  It also facilitates the
study  of the extrema of the potential with respect to the axions. For example, we know that only  $\rho_0$ depends on the axions of the complex structure $\xi^ I$. Thus minimising with respect to them one finds immediately the condition $\rho_0=0$ at the extrema, so that a linear combination of axions is fixed at the minima.  It would be interesting to study minima of this kind of potentials in the presence of open string fields, as discussed in the previous sections. We leave this task for future work.

\subsubsection*{In the presence of open string moduli}

The same exercise of relating the 4d $\mathcal{N}=1$ sugra auxiliary fields with the $\rho$'s (or the 4-forms) can be performed in the presence of open string moduli and the corresponding fluxes. In this case, the K\"ahler metric is not block diagonal due to the redefinition of the holomorphic variables explained in section \ref{ss:open} (see Appendix \ref{app:kahlermetrics} for more details on the K\"ahler metrics) and the computation of the auxiliary fields becomes more cumbersome. The expression \eqref{noscale} no longer separates into two pieces and is replaced by 
\begin{equation}
K^{\alpha \bar \beta} K_{\alpha} K_{\bar \beta}=7.
\end{equation}
We can use eq. \eqref{superpo} again to obtain the superpotential as a function of the $\rho$'s, which reads
\begin{equation}
W=\rho_0 - i t^a \rho_a - i n^I \rho_I - i t^a f_a^i \rho_{{F\, i}}-\dfrac{1}{2}\CK_a \tilde{\rho}^a-\dfrac{1}{2}n_{a \, i} t^a t^b f_b^i+\dfrac{i}{6}\CK \rho_m
\end{equation}
Once more, one can check that this superpotential is the same as the one given by eq. \eqref{Wtotal}. With this expression plus the inverse K\"ahler metric given by eqs. \eqref{Kinverse}-\eqref{eq:im3}, the auxiliary fields can be computed and they have the following form:
\begin{equation}
\begin{array}{l}
\label{auxfieldsot}
\bar{F}^{T^a} =\ e^{K/2}\left\{  i \left[ 2 t^a \rho_0 + \left( K^{\bar{a}b}\mathcal{K}_{bc} - t^ a \mathcal{K}_c \right)\tilde{\rho}^c  + f_c^i \left(K^{\bar{a}b} t^c-K^{\bar{a}c} t^b - t^a t^b t^c \right)n_{b\, i} \right] \right.\\ \\
     \left.  +\left[ \left(2 t^a t^b - K^{\bar{a} b}  \right) \rho_b +\left(\dfrac{1}{2}K^{\bar{a}c} \textbf{H}_c^I + 2 t^a n^I\right) \rho_I+ \left(2 t^a t^b f_ b^i+ f_c^i K^{\bar{c}a} \right)\rho_{{F\, i}}+ \left(\dfrac{1}{2}K^{\bar{a}b} \CK_b -\dfrac{1}{3}t^a\right) \rho_m \right]\right\},\\ \\ 
\end{array}
\end{equation}
\begin{equation}
\begin{array}{l}
\label{auxfieldson}				  
\bar{F}^{N^I}=\  e^{K/2} \left\{  i \left[ 2 n^{'I} \rho_0 - \left(\dfrac{1}{2}K^{\bar{d}a}\textbf{H}_d^I \CK_{ac}+ n^I \CK_c \right)\tilde{\rho}^c - f_d^i t^d \left(\dfrac{1}{2}K^{\bar{c}a} \textbf{H}_c^I+ n^I t^a \right)n_{a\, i}  \right] \right. \\ \\
\qquad \qquad \qquad \quad \left . +\left[ \left(\dfrac{1}{2}K^{\bar{c}a}\textbf{H}_c^I+2 n^{I} t^a\right) \rho_a + \left( K^{\bar{I}J}+2 n^{I} n^{J} \right)\rho_J \right. \right. \\ \\
\qquad \qquad \quad \quad \quad \quad \left. \left.+ \left( \dfrac{1}{2}D^{lj} g_l^I + K^{\bar{c}d} f_d^j \textbf{H}_c^I + 2 t^a f_a^i n^I \right) \rho_{{F\, j}} -\left( \dfrac{1}{2}K^{\bar{c}a}\textbf{H}_c^I \CK_a + \dfrac{\CK}{3}n^I  \right)\rho_m \right]  \right\}, \\ \\ \\
\end{array}
\end{equation}
\begin{equation}
\begin{array}{l}
\label{auxfieldsophi}
\bar{F}^{\Phi^i}=\  e^{K/2} \left\{  i \left[ 2 t^a f_a^i \rho_0 - \left(f_d^i K^{\bar{d}a} \CK_{ac}+ t^a f_a^i \CK_c \right)\tilde{\rho}^c +\left(D^{ij} t^a + K^{\bar{d}b}f_d^i f_b^j t^a - f_c^i K^{\bar{c}a}t^d f_d^j  \right)n_{a\, j}  \right] \right. \\ \\
\qquad \qquad \qquad \quad \left . +\left[ \left(f_c^i K^{\bar{c}a}+ 2 t^a t^b f_b^i \right) \rho_a + \left( \dfrac{1}{2}D^{li}g_l^J+\dfrac{1}{2}K^{\bar{c}d}f_d^i \textbf{H}_c^J+2 t^d f_d^i n^J \right)\rho_J \right. \right. \\ \\
\qquad \qquad \quad \quad \quad \quad \left. \left.+ \left( 2 t^a f_a^i t^d f_d^j - D^{ij}-K^{\bar{a}b}f_a^i f_b^j \right) \rho_{{F\, j}} -\left( \dfrac{1}{2} f_c^i K^{\bar{c}a}+\dfrac{\CK}{3}t^a f_ a^i \right)\rho_m \right]  \right\},
\end{array}
\end{equation}
where the $K^{\bar{I}J}$ in the expression for  $\bar{F}^{N^I}$ is given by eq. \eqref{eq:im3}. As in the previous case, the scalar potential can be expressed in terms of the auxiliary fields and it is computed in Appendix \ref{app:Fterm}.

\section{Outlook}\label{s:outlook}

In this paper we have analysed the scalar potential of flux compactifications from a different perspective, that makes manifest the discrete shift symmetries involving fluxes and periodic scalars such as axions. We have focused our analysis in 4d type IIA Calabi-Yau orientifold compactifications with RR and NS fluxes and D6-branes, for which we have computed the full scalar potential at tree-level and in the large volume regime. This class of compactifications is particularly interesting for our purposes, because each complex scalar $\psi^\lam = \phi^\lam + i s^\lam$ entering the potential consists of a periodic scalar $\phi^\lam$, which in above regime is an axion, and a saxionic partner $s^\lam$.  Moreover, as pointed out in \cite{Bielleman:2015ina,Carta:2016ynn} the 4d effective theory of these vacua can to great extent be understood in terms of couplings of axions to four-forms, which in turn makes manifest the discrete shift symmetries of the potential and its multi-branched structure.

In the present work we have extended the analysis of \cite{Bielleman:2015ina,Carta:2016ynn}, by considering the F-term scalar potential that results from combining the presence of RR and NS fluxes and localised sources like D6-branes and O6-planes, and that involves both closed- and open-string axions. As shown in Appendix \ref{ap:dimred}, an important part of the potential can be computed by considering the three-form fields $C_3^A$ that appear in 4d, computing the couplings of scalars to their four-form field strengths $F_4^A$ and then dualising the latter. By construction, this piece of the potential is of the form
\be
V \, =\, Z^{AB} \varrho_A \varrho_B
\label{Vout}
\ee
where $Z^{AB}$ is the inverse metric of the four-forms, and $\varrho_A$ the different combination of scalars coupling to each of them. One also finds that each of the entries of $\vec{\varrho}$ is linear in the quantised fluxes of the compactification, and therefore the potential is quadratic in them. This is probably not so surprising, given that we are obtaining this piece of the potential by dimensionally reducing a two-derivative 10d action. What it is perhaps more remarkable is that one can manage to write the full F-term scalar potential in the form \eqref{Vout}, including the contribution from localised sources. Again, one finds that each of the entries $\varrho_A$ is linear in the different RR and NS fluxes and discrete D-brane data. Therefore, one may postulate a 4d effective theory describing the scalar potential purely in terms of 4d four-forms. 

Given the same potential, the specific form of {\bf Z} and $\vec{\varrho}$ in \eqref{Vout} depends on the basis of 4d four-forms that one considers. In general we find that there is always a particular basis in where $\varrho_A = q_A$, with $\vec{q}$ the vector of quantised fluxes of the theory. In this description, the full potential is encoded in the kinetic terms {\bf Z} of the four-forms, and the flux quanta appear as integration constants of their dualisation, as in \cite{Farakos:2017jme}. There is however an alternative, more interesting choice in which $\vec{\varrho}$ equals
\be
\vec{\rho} \, =\,  {\bf R}^{t -1}  \vec{q}
\label{Rout}
\ee
where $\vec{q}$ is the previous vector of flux quanta and ${\bf R}(\phi)$ a rotation matrix that only depends on the axionic components $\phi^\lam$ of the scalars fields and topological data. All the dependence on the saxions $s^\lam$ is kept in the metric for this choice of four-forms. One then finds a interesting factorisation of the scalar potential in terms of its dependence on axions, saxions and fluxes, namely
\be
V\, =\, \vec{q}^{\, t} \left({\bf R}^t\, {\bf M}\, {\bf R} \right)^{-1} \vec{q}
\label{factout}
\ee
with {\bf R} only depending on the axions $\phi^\lam$ and {\bf M} only on the saxions $s^\lam$.\footnote{To be precise, this factorisation only holds when $\phi^\lam$ have unit period. When normalised canonically the factorisation will be generically lost. } In fact, one can be more precise, and write the axion rotation matrix {\bf R} as
\be
{\bf R}^t\, =\, e^{\phi^\lam P_\lam}
\ee
with the generators $P_\lam$ integer-valued, nilpotent matrices that only depend on the topological data of the compactification. More precisely, one constructs $P_\lam$ by considering the 4d string that is associated with the axion $\phi^\lam$ and looking at how it interacts with the 4d domain walls of the effective theory: when the 4d string crosses a certain domain wall, a second domain wall may need to appear stretching between the two. At the microscopic level, this rules are encoded in the Freed-Witten and Hanany-Witten effects between the different branes of the compactification that appear as 4d defects. 

Since the FW and HW effects are ultimately related to gauge invariance, one may interpret the two previous choices of basis, leading to $\varrho_A = q_A$ vs. $\varrho_A = \rho_A$, in terms of a well-known effect in 10d supergravity \cite{Marolf:2000cb}. Indeed, in general one may describe the notion of charge in terms of quantised, non-gauge invariant quantities (the $q_A$) or in terms of non-quantised, gauge invariant quantities (the $\rho_A$). The higher dimensional gauge transformations correspond in the effective 4d theory in the combined discrete shifts of axions and fluxes that leave invariant the $\rho_A$ but not the $q_A$. As the $\rho_A$ are basic invariants of this discrete shift symmetry, any 4d quantity that depends on the fluxes must do it as a function of the $\rho_A$, or else it would result into a violation of a microscopic gauge invariance. This applies, in particular, to the corrections to the scalar potential of any sort. Notice that this generalises in a very concrete way the protection mechanism used in \cite{KS} to propose models of large field inflation. 

The gauge invariance of $\vec{\rho}$ is also manifest in its close relationship to another gauge invariant quantity, the flux superpotential $W$. Neglecting contributions from world-sheet and D-brane instantons, this superpotential is a polynomial on the complex fields $\psi^\lam$, whose coefficients are given by the quantised fluxes and the topological  compactification data. We find that $\vec{\rho}$ and $W$ contain the same information, and one can build a dictionary between the two quantities. More precisely, there is one of the components of $\vec{\rho}$, dubbed master axion polynomial $\rho_0$, that becomes $W$ upon the replacement $\phi^\lam \raw \psi^\lam$. In addition one can recover all the other components of $\vec{\rho}$ by taking all the possible derivatives with respect to the axions present in $\rho_0$, completing the correspondence $\vec{\rho} \lraw W$. As a result one may give concrete expression for the standard $\CN=1$ auxiliary fields in terms of the $\rho_A$ and the saxion-dependent metrics contained in {\bf M}.

Notice that most of our results can be purely formulated in 4d language, and as such one could hope to extend them to other classes of string compactifications. Indeed, to obtain the key element of our analysis, the vector $\vec{\rho}$, one just needs to know the spectrum of 4d domain walls and string defects plus their interactions. This topological information is typically much easier to extract than performing the full dimensional reduction. Reversing the logic, one may reinterpret the scalar potentials already present in the literature in terms of gauge invariant quantities like $\vec{\rho}$ and from there the physics of strings and domain walls in the 4d effective theory. One interesting generalisation would be to consider type IIA backgrounds with metric fluxes and D-branes, which give more general superpotentials \cite{Marchesano:2006ns}, or compactifications that do not have a standard geometric interpretation, see e.g. \cite{Shelton:2005cf,moredual,Blumenhagen:2015lta,Gao:2017gxk}. In this context, it would be particularly interesting to develop the connection that we have found in Appendix \ref{ap:metric} between the invertibility of the bilinear form of the potential and the implementation of the Bianchi identities for the background fluxes. In general, a relation between the two is to be expected, since the Bianchi identities constrain the naive lattice of flux quanta to the subset of those that correspond to truly independent fluxes. As the effective Lagrangian \eqref{S4formRR} describes independent 4d four-forms, it seems reasonable that one has to restrict the lattice of fluxes to independent ones before attempting to translate their effect in the potential into a Lagrangian of this form, that is, before attempting to invert the bilinear form $Z^{AB}$. We find quite amusing that consistency conditions of the microscopic theory such as Bianchi identities are encoded in this manner in the structure of the scalar potential, and it would be interesting to use this property to improve our understanding of non-geometric compactifications. In some cases, the correct application of Bianchi identities in this class of compactifications can be rather subtle and have no clear prescription, so demanding that the F-term potential can be obtained from a four-form Lagrangian could set a criterium to implement them.

In any event, it would be interesting to generalise our result, perhaps in combination with those in \cite{Farakos:2017jme}, to give a complete, model-independent description of the 4d four-form Lagrangian and the related scalar potential. In addition, it would be interesting to sharpen the dictionary that relates the couplings in the former to the standard $\CN=1$ quantities $K$ and $W$. This would in particular allow to reformulate the more developed type IIB scenarios of moduli stabilisation from the vantage point of 4d four-form Lagrangians and, as mentioned in the introduction, to connect with the proposal of Bousso and Polchinski \cite{BP}. 

Finally, this alternative description of the scalar potential may be useful in understanding the structure of vacua in flux compactifications. In particular, the bilinear form of the potential could be used to reanalyse the set of supersymmetric and non-supersymmetric vacua found in the type IIA flux literature, see e.g. \cite{DeWolfe:2005uu,Camara:2005dc,Villadoro:2005cu,Shelton:2005cf,moredual,Grana:2006kf,Palti:2008mg,Lust:2008zd}. Moreover, most of the search for vacua has neglected the presence of open strings fields in the scalar potential, while in our approach they appear on equal footing with the closed string fields. One could then incorporate open strings in the analysis of type IIA moduli stabilisation to construct new and more general classes of vacua. At any rate, we find remarkable that something as intricate as an F-term scalar potential can have a general factorised dependence between axions and saxions, and that such factorisation is ultimately due to the consistency conditions of the microscopic underlying theory. Hopefully, developments along these lines will give us a better idea of the class of scalar potentials that one may obtain out of string compactifications, and eventually a better characterisation of the String Landscape.

\bigskip

\bigskip

\centerline{\bf \large Acknowledgments}

\bigskip

\noindent We would like to thank G. Aldazabal, A. Font, E.~Palti, R.~Savelli, W.~Staessens,  A.~Uranga and I.~Valenzuela for useful comments and discussions. This work is supported by the Spanish Research Agency (Agencia Estatal de Investigaci\'on) through the grant IFT Centro de Excelencia Severo Ochoa SEV-2016-0597, by the grant FPA2015-65480-P from MINECO/FEDER EU, by the ERC Advanced Grant SPLE under contract ERC-2012-ADG-20120216-320421 and by the HKRGC grants HUKST4/CRF/13G and 16304414. A.H. is supported by the Spanish FPU Grant No. FPU15/05012.

\newpage

\appendix

\section{Dimensional reduction and 4d four-forms}
\label{ap:dimred}

In this appendix we compute the 4d scalar potential from the dimensional reduction of the massive type IIA 10d supergravity Lagrangian, in the presence of fluxes and localised sources. We will do so by obtaining a 4d effective four-form Lagrangian (\ref{S4formRR}) from the democratic action of \cite{Bergshoeff:2001pv}, along the lines of Appendix B of \cite{Carta:2016ynn}. We will extend the computation therein by adding the necessary ingredients for the discussion of section \ref{s:4forms}, like the simultaneous presence of D-branes and NS fluxes. To streamline the presentation we will divide the computation into two, first focusing on the contribution coming from the closed string sector and adding afterwards the open string sector. 

\subsection{Closed string sector}

Let us first compute the closed string sector contribution to the 4d Lagrangian (\ref{S4formRR}). Following the structure of section \ref{s:4forms} we will first do so in the {\bf C}-basis, then in the {\bf A}-basis and then add the contribution of the NS sector. 

\subsubsection*{The C-basis}

To perform the dimensional reduction of the closed string flux sector one needs to consider the terms in the action \eqref{S10dIIA} that involve the RR $G_{2n}$ and NSNS $H$ field strengths. We will first focus on the former, supplying the action with the Lagrange multiplier terms to enforce the adequate Bianchi identities in the {\bf C}-basis
\begin{align}\label{eq:RRdem} 
S_{\text{RR}} = - \frac{1}{4 \kappa_{10}^2} \int  \left[ \oh {\bf G} \wedge \star_{10}\, {\bf G} - \sig'({\bf C}) \wedge\left( d \mathbf G - H\wedge \mathbf G\right) \right]_\text{top}\,,
\end{align}
where we have used the polyform notation of eqs.(\ref{bfCA})  and (\ref{bfG}), $\sig'$ reverses the indices of a $p$-form and {\em top} indicates that we extract the 10-form from the wedge product. Defining the flux quanta as in (\ref{RRfluxes}) and (\ref{Hquanta}), a solution to the Bianchi identities is
 \begin{align}
 G_0 & = l_s^{-1} m\,,\nonumber \\
 G_2 &= l_s^{-1} ( b^a m-m^a )\, \omega_a +\dots\,,\nonumber\\
 G_4 &= l_s^{-1} \left[e_a - \mathcal K_{abc} m^b b^c +\frac{m}{2} \mathcal K_{abc} b^b b^c\right] \tilde \omega^a  +F_4^0+ d\xi'^K \wedge \alpha_K +\dots \,,\nonumber\\
 G_6 &= l_s^{-1} \left[-e_0 + e_a b^a -\frac{1}{2} \mathcal K_{abc} m^a b^b b^c + \frac{m}{3!} \mathcal K_{abc} b^a b^b b^c + \xi'^K h_K\right] \omega_6 + F_4^a  \wedge \omega_a \nonumber  \\ & + \left[d\Xi_K+ h_K c^0_3\right] \wedge \beta^K +\dots \nonumber\\
 G_8 &=\tilde F_{4\,a}\wedge \tilde \omega^a\,,\nonumber\\
\label{eq:Gred} G_{10} &= \tilde F_4\wedge \omega_6 \,,
 \end{align}
 where we have used the basis of quantised harmonic $p$-forms $\{\om_a, \tilde{\om}^a, \om_6,  \a_K, \b^K\}$ and the definition for the closed string axions $\{b^a,  \xi'^K\}$ of section \ref{s:4forms}. As in there, we have defined the 4d RR three-forms in the {\bf C}-basis as
\be
C_3 = c_3^0 + \ldots \quad C_5 =  c_3^a \wedge \omega_a + \dots \quad C_7 = \tilde d_{3\, a} \wedge \tilde\omega^a +\dots \quad C_9 = \tilde d_{3} \wedge \om_6 + \dots
\label{ap:C3form}
\ee
 and the corresponding four-form field strengths by
 \begin{align}
&F_4^0\, =\, dc_3^0 \,,\qquad F_4^a\, =\, dc_3^a -db^a \wedge c_3^0\,, \nonumber\\& \tilde F_{4\, a} \, =\, d\tilde d_{3\, a} -\CK_{abc}db^b \wedge c_3^c\,,\qquad \tilde F_4\, =\, d\tilde d_3 - db^a \wedge \tilde d_{3\, a}\,.
\label{ap:M4forms}
\end{align}
Moreover $\Xi_K$ stand for the magnetic duals of the axions $\xi'^K$, which appear in the dimensional reduction of the RR potential $C_5$ as follows
\begin{align}
C_5 = \Xi_K \wedge \beta^K + \dots\,.
\end{align}
Finally, the dots stand for non-closed, co-closed pieces of the background fluxes, that appear due to the presence of localised sources and non-vanishing $m$ and $H$. Such pieces do contribute to the potential when plugged into the first term of the 10d Lagrangian  (\ref{eq:RRdem}), but they essentially correspond to the backreaction of localised sources. Because of that, their contribution is already taken into account by the potential terms $V_{\rm DBI}$ and $V_{\rm loc}$ defined in Section \ref{s:4forms}, and will not be considered in the following.\footnote{It is possible to see \cite{Koerber:2007hd} that the $d_H$ non-closed pieces give contributions localised at the sources, and that in supersymmetric vacua these contributions coincide with the DBI potential energy.}

Plugging these expressions into (\ref{eq:Gred}) and integrating over the compactification manifold $\CM_6$ we will recover a 4d action with the structure \eqref{S4formRR}. The first two terms in \eqref{S4formRR} will come from the kinetic terms of the RR field strengths and the last term coming from the Lagrange multipliers. More precisely, from the reduction of the kinetic terms we obtain
\begin{align}\label{eq:RRk4d}
S_{\text{RR,}k} = -\frac{1}{16 \kappa_4^2} \int_{\mathbb R^{1,3}}& \left[\frac{4 e^{\frac{5\phi}{2}}}{V_6} \tilde \rho^2  + \frac{16 e^{\frac{5\phi}{2}}}{V_6} g_{ab}  \tilde \rho^a \tilde \rho^b+ \frac{e^{-\frac{\phi}{2}}}{V_6^3} g^{ab} \rho_a \rho_b+ \frac{4 e^{-\frac{\phi}{2}}}{V_6^3}\rho_0^2\right] *_4 1\nonumber\\
+&\frac{V_6 e^{-\frac{5\phi}{2}}}{4} \tilde F_4 \wedge *_4 \, \tilde F_4  + \frac{V_6 e^{-\frac{5\phi}{2}}}{16} g^{ab} \tilde F_{4\, a} \wedge *_4\, \tilde F_{4\, b}\nonumber\\
+& V_6^3\, e^{\frac{\phi}{2}}  g_{ab} F_4^a \wedge *_4\, F_4^b + \frac{V_6^3 e^{\frac{\phi}{2}}} {4} F_4^0 \wedge *_4\, F_4^0\nonumber\\
+& 2 c_{KL} d\xi^K \wedge *_4\, d\xi^L+\frac{1}{2} c^{KL} \left[d\Xi_K +h_K C_3^0\right]\wedge *_4 \left[d\Xi_L +h_L C_3^0\right]\,.
\end{align}
with the definition of the $\rho_A$ given in \eqref{rhosRR}. Here $\phi$ is the 10d dilaton and $V_6$ the volume of $\CM_6$ in string units and 10d Einstein frame (obtained via the replacement $g_{MN} \rightarrow e^{\frac{\phi}{2}}g_{MN}$), where also the different $p$-form metrics are computed
\begin{equation}
g_{ab} = \frac{e^{-\phi}}{4 V_6 l_s^6 } \int_{{\cal M}_6} \omega_a \wedge \star_6 \omega_b\,    \qquad g^{ab} = \frac{4 V_6}{e^{-\phi}l_s^6} \int_{{\cal M}_6} \tilde \omega^a \wedge \star_6 \tilde \omega^b\, 
\label{ap:metrics}
\end{equation}
with $c_{IJ}$ given by \eqref{cIJ}. Also, the result is obtained after performing the 4d Weyl rescaling $g_{\mu \nu} \rightarrow \frac{g_{\mu\nu}}{V_6/2}$ and integrating over the orientifold quotient space. Finally, using the relations (\ref{eK}) and $\CK = 6 e^{3\phi/2} V_6$ one finds
\begin{align}\label{eq:RRk4d2}
S_{\text{RR,}k} = -\frac{1}{16 \kappa_4^2} \int_{\mathbb R^{1,3}} &32e^{K} \left[\frac{\CK^2}{36} \tilde \rho^2  + \frac{\CK^2}{9} g_{ab}  \tilde \rho^a \tilde \rho^b+  \frac{g^{ab}}{4} \rho_a \rho_b+ \rho_0^2\right] *_4 1\nonumber\\
-\frac{1}{16 \kappa_4^2} \int_{\mathbb R^{1,3}} & \frac{1}{32e^K} \left[ \frac{36}{\CK^2} \tilde F_4 \wedge *_4 \, \tilde F_4  + \frac{9}{\CK^2} g^{ab} \tilde F_{4\, a} \wedge *_4\, \tilde F_{4\, b} +  4g_{ab} F_4^a \wedge *_4\, F_4^b + F_4^0 \wedge *_4\, F_4^0\right] 
\nonumber\\
-\frac{1}{16 \kappa_4^2} \int_{\mathbb R^{1,3}} & 2 c_{KL} d\xi'^K \wedge *_4\, d\xi'^L+\frac{1}{2} c^{KL} \left[d\Xi_K +h_K c_3^0\right]\wedge *_4 \left[d\Xi_L +h_L c_3^0\right]\, .
\end{align}
It is easy to see that the first two lines correspond to the first two terms of \eqref{S4formRR} with the metric corresponding to (\ref{Gmetric}). 

To perform the dimensional reduction of the Lagrange multipliers notice that up to boundary terms
\be
 - \sig'({\bf C}) \wedge d_H \mathbf G \, \simeq \,  \sig'(d_H {\bf C}) \wedge \mathbf G \, =\, \sig'(d_H {\bf C}) \wedge e^B \wedge \bar{\bf G} = - d_H {\bf C} \wedge e^{-B} \wedge  \sig'(\bar{\bf G})
 \label{idLagC}
 \ee
where $d_H = d - H \wedge$ and we have used (\ref{bfG}) and the fact that $\sig'({\bf P}) \wedge {\bf P}|_\text{top} \equiv 0 $ for an even polyform {\bf P} in 10d. Integrating over the quotient space then gives 
\begin{align}
S_{\text{RR,L}} = \frac{1}{8 \kappa_{4}^2} \int_{\mathbb R^{1,3}}  \tilde F_4 \, \tilde \rho + \tilde F_{4\, a}  \, \tilde \rho^a 
+ F_4^a \, \rho_a + F_4^0 \, \rho_0
\label{eq:topC}\,,
\end{align}
reproducing the last term in \eqref{S4formRR}. 

In fact, the second equality in \eqref{idLagC} is not entirely correct, because it assumes that the RR potential polyform {\bf C} in the Lagrange multipliers and the one contained in {\bf G} are identical. However, as discussed in \cite{Jockers:2004yj}, in order to be able to dualise the 4d two-forms $\Xi_K$ into the axions $\xi'^K$ the former should not appear in the Lagrange multipliers, while they do appear in \eqref{eq:Gred}. This mismatch gives an extra term in the above reduction, which finally translates into
\begin{align}
S_{\text{RR,L}} = \frac{1}{8 \kappa_{4}^2} \int_{\mathbb R^{1,3}} \tilde F_4 \, \tilde \rho + \tilde F_{4\, a}  \, \tilde \rho^a 
+ F_4^a \, \rho_a + F_4^0 \, \rho_0 +d\xi'^K\wedge \left[d\Xi_K +h_K c_3^0\right]
\label{eq:topC2}\,.
\end{align}
Notice that \eqref{10dH} and \eqref{eq:Gred} will imply the following 4d duality relation
\begin{align}
d\Xi_K + h_K c_3^0 = -2c_{KL} *_4 \,d\xi'^L\, .
\end{align}
Eliminating the two-forms $\Xi_K$ by imposing this relation, the last line in \eqref{eq:RRk4d2} will vanish and the last term in \eqref{eq:topC} will transform into the kinetic term for the RR-axions $\xi'^K$.

\subsubsection*{The A-basis}

For completeness, let us redo the computation in the {\bf A}-basis, following more closely Appendix B of \cite{Carta:2016ynn}.\footnote{Notice that the definition of flux quanta made in \eqref{RRfluxes} differ from those in \cite{Carta:2016ynn} by some signs.} Here the relevant piece of the 10d action reads
\begin{align}\label{eq:RRdemA} 
S_{\text{RR}} = - \frac{1}{4 \kappa_{10}^2} \int  \left[ \oh {\bf G} \wedge \star_{10}\, {\bf G} - \sig'({\bf A}) \wedge d \left( e^{-B} \wedge \mathbf G \right) \right]_\text{top}\, .
\end{align}
In this basis 4d three-forms are defined as
 \begin{align}
a_3^0 =  c_3^0 , \qquad a_3^a = c_3^a - b^a c_3^0 , \qquad \tilde a_{4\, a}  = \tilde d_{3\, a} -\CK_{abc}b^b c_3^c , \qquad \tilde a_3 =  \tilde d_3 - b^a  \tilde d_{3\, a}\, ,
\label{ap:A3form}
\end{align}
and the corresponding field strength four-forms as
 \begin{align}
D_4^0\, =\, da_3^0 \,,\quad D_4^a\, =\, d a_3^a \,, \qquad \tilde D_{4\, a} \, =\, d\tilde a_{3\, a} \,,\qquad \tilde D_4\, =\, d\tilde a_3 \, ,
\label{ap:A4forms}
\end{align}
which are related to the four-forms \eqref{ap:M4forms} by the rotation \eqref{Rb}. In terms of them, the solution to the Bianchi identities is
\begin{align}
 G_0 & = l_s^{-1} m\,,\nonumber \\
 G_2 &= l_s^{-1} ( b^a m-m^a )\, \omega_a +\dots\nonumber\\
 G_4 &= l_s^{-1} \left[e_a - \mathcal K_{abc} m^b b^c +\frac{m}{2} \mathcal K_{abc} b^b b^c\right] \tilde \omega^a  +D_4^0 + d\xi'^K \wedge \alpha_K\, +\dots \nonumber\\
 G_6 &= l_s^{-1} \left[-e_0 + e_a b^a -\frac{1}{2} \mathcal K_{abc} m^a b^b b^c + \frac{m}{3!} \mathcal K_{abc} b^a b^b b^c + \xi'^K h_K\right] \omega_6  \nonumber  \\ & + (D_4^a + b^a D_4^0)  \wedge \omega_a + \left[d\Xi_K+ h_K c^0_3\right] \wedge \beta^K +\dots\nonumber\\
 G_8 &=\left[\tilde D_{4\, a} + \mathcal K_{abc} b^b D_4^c +\frac{1}{2} \mathcal K_{abc} b^b b^c D_4^0\right]\wedge \tilde \omega^a\,,\nonumber\\
\label{eq:GredA} G_{10} &= \left[\tilde D_4 + \tilde D_{4\, a} b^a + \frac{1}{2} \mathcal K_{abc} b^a b^b D_4^c + \frac{1}{3!} \mathcal K_{abc} b^a b^b b^c D_4^0\right]\wedge \omega_6 \,.
 \end{align}
When plugged into the 10d RR kinetic terms, one again obtains the result \eqref{eq:RRk4d2}, but now written in terms of the four-forms \eqref{ap:A4forms}. As discussed in Section \ref{s:4forms}, this translates into the first two terms of \eqref{RRfluxes} with $E_4^A = (D_4^0, D_4^a, \tilde D_{4\, a}, \tilde D_4)$, $l_s \varrho_A = (e_0, e_a, m^a, m)$ and the four-form metric given by \eqref{factorZ}. To evaluate the contribution from the Lagrange multipliers one notices that
\be
- \sig'({\bf A}) \wedge d \left( e^{-B} \wedge \mathbf G \right) \, \simeq \,  \sig'( d {\bf A}) \wedge e^{-B} \wedge \mathbf G \, =\, \sig'(d {\bf A}) \wedge \bar{\bf G} = - d {\bf A} \wedge \sig'( \bar{\bf G})
 \ee
that translates into
\begin{align}
S_{\text{RR,L}} = \frac{1}{8 \kappa_{4}^2 l_s} \int_{\mathbb R^{1,3}} \tilde D_4 \, m + \tilde D_{4\, a}  \, m^a 
+ D_4^a \, e_a + D_4^0 \, \left[e_0-h_K \xi'^K\right] 
\label{eq:topA}\,.
\end{align}
Again, taking into account the absence of 4d two-forms in the Lagrange multipliers will give the extra term $d\xi'^K\wedge \left[d\Xi_K +h_K c_3^0\right]$, and eliminating the two-forms $\Xi_K$ in favour of their axions $\xi'^K$ gives the same result as before.

\subsubsection*{The NS sector}

The treatment of the NSNS flux sector is quite similar to the discussion carried above. In 10d we have the action
\begin{align}
S_{\text{NSNS}} = -\frac{1}{4\kappa_{10}^2} \int \frac{1}{2} e^{-2\phi}H \wedge \star H +\frac{1}{2}e^{2\phi} H_7 \wedge \star H_7 + H \wedge H_7\,.
\end{align}
The ansatz for the dimensional reduction in this case is 
\begin{align}
H &= l_s^{-1} \, h_K \beta^K\, , \qquad  H_7 = H_4^K \wedge \alpha_K\,,
\end{align}
and the resulting action in 4d is
\begin{align}
S_{\text{NSNS}} &= -\frac{1}{16\kappa_4^2 }\int_{\mathbb R^{1,3}}  \frac{32 e^{K}}{l_s^2 }c^{KL} h_K h_L *_4 1 +\frac{ e^{-K}}{32} c_{KL} H_4^K \wedge *_4\, H_4^L +\frac{1}{8 l_s\kappa_4^2 } \int_{\mathbb R^{1,3}} H_4^K h_K\,,
\end{align}
where the 4d four-form metric $c_{KL}$ is given by \eqref{cIJ}. Added to the contribution from the RR sector in the {\bf C}-basis, it reproduces a 4d Lagrangian of the form \eqref{RRfluxes} with the choices \eqref{EFwH}-\eqref{GmetricH}.

\subsection{Open string sector}

The presence of D-branes and O-planes modifies the above computation in a two-fold manner. On the one hand, they act as sources for the RR field strengths and modify {\bf G} through their backreaction. On the other hand, they contribute to the axion-four-form couplings through their Chern-Simons action. 

In the case of D6-branes the sum of Chern-Simons actions reads
\begin{align}
S_{\text{CS}} = -\sum_{\alpha}  \mu_6 \int_{\IR^{1,3} \times \Pi_\alpha} \mathbf C \wedge e^{\sigma F - B}\,,
\label{CSD6}
\end{align}
where $\mu_6 = 2\pi l_s^{-7}$ and $\Pi_\a$ are the three-cycles wrapped by the D6-branes. This piece of the action is related to how D6-branes enter the Bianchi identity \eqref{BIG}, which is implemented in the 10d democratic action as
\begin{align}\label{eq:RRdemD6} 
S_{\text{RR+CS}} = &- \frac{1}{4 \kappa_{10}^2} \int  \left[ \oh {\bf G} \wedge \star_{10}\, {\bf G} - \sig'({\bf C}) \wedge\left( d \mathbf G - H\wedge \mathbf G +  \frac{1}{l _s^2} \sum_\a q_\a \d(\Pi_\a) \wedge e^{B-\sig F_\a}  \right) \right]_\text{top}\nonumber \\
 = &- \frac{1}{4 \kappa_{10}^2} \int  \left[ \oh {\bf G} \wedge \star_{10}\, {\bf G} - \sig'({\bf A}) \wedge\left( d (e^{-B} \wedge \mathbf G)  +  \frac{1}{l _s^2} \sum_\a q_\a \d(\Pi_\a) \wedge e^{-\sig F_\a}  \right) \right]_\text{top} \, .
\end{align}
Two comments are in order regarding this expression. First, unlike in the main text $\a$ runs over D6-branes and their orientifold images and we have also included in the sum the three-cycles wrapped by the O6-planes. For those one should take $\CF = B - \sig F \equiv 0$ and $q_\a = -4$, while for D6-branes we have $q_\a=1$. Second, the new terms in \eqref{eq:RRdemD6} induced by D6-branes amount to one-half \eqref{CSD6}. This is because to embed the Chern-Simons action into the democratic formulation one must split the RR potentials into two equal parts made of electric and magnetic components (see \cite{Koerber:2007hd}), and only the former are relevant for the computations to follow. Notice that with these choices and from demanding that there is a solution for the RR field strength $G_2$ one recovers the RR tadpole condition \eqref{RRtadpole}.

As in the main text, one may explore the dependence of the type IIA flux potential on the open string moduli by first considering a D6-brane configuration $\{\Pi_\a^0\}$ that satisfies the tadpole condition \eqref{RRtadpole} and preserves supersymmetry. Notice that the latter require that $\Pi_\a^0$ is a special Lagrangian with $\CF \equiv 0$, which in particular implies that $H|_{\Pi_\a^0} = 0$.\footnote{The discussion below can be generalised  to relax the condition $\CF\equiv0$, which in some cases may be too restrictive. We will however impose it in this appendix for the sake of simplicity.} In addition we will impose that the D6-brane Wilson lines vanish for each $\Pi_\a^0$. One may then consider homotopic deformations of the D6-brane embeddings $\Pi_\a^0 \raw \Pi_\a$ as well as changes in their worldvolume fluxes and Wilson lines. In general the RR background fluxes will depend on such deformations of the open string sector, and so will the dimensional reduction of the 10d action \eqref{eq:RRdemD6}. 

At the reference configuration $\{\Pi_\a^0\}$ the Bianchi identities (\ref{BIG}) read
\be
l_s^{2}\, d\left(d{\bf A} +  {\bf \bar{G}}\right) =- \sum_\a q_\a \d(\Pi_\a^0)
\label{ap:BI}
\ee
and so all the $A_{p}$ except $A_1$ are globally well-defined. Given the quantisation condition in (\ref{BIG}), this implies that all the forms $\bar{G}_{p+1}$ are quantised except $\bar{G}_2$. In practice this means that if we consider the entries of the flux vector $q_A = (e_0, e_a, m^a, m)$ defined by \eqref{RRfluxes}, all of them are integer number except the $m^a$, whose precise value depends on the choice of reference D6-brane configuration. Of course if we fix $\{\Pi_\a^0\}$ these entries also take discrete values, in the sense that one can shift them by integer numbers without spoiling the quantisation condition in (\ref{BIG}). Other than this subtlety for the values of $m^a$, the dimensional reduction of the 10d democratic action parallels the discussion made for the closed string sector, and gives a similar 4d effective action. 

Given a solution for the RR fluxes {\bf G} at the reference configuration $\{\Pi_\a^0\}$, let us see how they change as we deform the embeddings $\Pi_\a^0 \raw \Pi_\a$ homotopically, and we switch on Wilson lines and worldvolume fluxes on the D6-branes. Let us first consider deforming a single D6-brane $\a$ and call the corresponding change in RR background as $\Delta_\a {\bf G}$. Following the discussion in appendix \ref{sap:supo}, one can characterise such change by
\be
\Delta_\a {\bf G} \, =\,  \frac{1}{l_s^2} \d(\Pi_\a) \wedge \left( \sig A - \oh \sig^2 A \wedge F\right)  \wedge e^{B}  - \frac{1}{l_s}  \d({\cc_4^\a}) \wedge \left( e^{B} - \varpi_4\right) 
\label{shiftGT}
\ee
Here $\p\cc_4^\a = \Pi_\a - \Pi_\a^0$ is a four-chain that represents the homotopic deformation of $\Pi_\a^0$ to a new special Lagrangian $\Pi_\a$, and $\d({\cc_4^\a})$ is the bump delta-function or current associated to it, with $l_s d\, \d({\cc_4^\a}) = \d(\Pi_\a) - \d(\Pi_\a^0)$.  Finally $\varpi_4$ is a co-exact form such that $d\varpi_4 = H \wedge B$. 

Due to the shift (\ref{shiftG}) the dimensional reduction of the 10d Lagrange multiplier will be modified. Instead of the rhs of (\ref{idLagC}) we will have
\be
- d_H {\bf C} \wedge \left(e^{-B} \wedge \left[ \sig'(\bar{\bf G}) + \sig'(\Delta_\a {\bf G})  \right]\right)
\ee
where $\bar{\bf G}$ corresponds to the reference D6-brane configuration $\{\Pi_\a^0\}$. Upon integration one finds that \eqref{eq:topC} is replaced by
\begin{align}
S_{\text{RR+CS,L}} = \frac{1}{8 \kappa_{4}^2} \int_{\mathbb R^{1,3}} \tilde F_4 \, \tilde \rho + \tilde F_{4\, a}  \, ( \tilde \rho^a +  \tilde\upsilon^a)
+ F_4^a \, (\rho_a + \upsilon_a) + F_4^0 \, (\rho_0 + \upsilon_0) 
\label{eq:topCD}\,,
\end{align}
where
\bea\nonumber
l_s \tilde \upsilon^a  &=& \frac{2}{l_s^4} \int_{\cc_4^\a} \tilde \om^a\, , \quad l_s \upsilon_a  =  \frac{2}{l_s^4} \int_{\Pi_\a} \om_a \wedge \sig A -  \frac{2}{l_s^4} \int_{\cc_4^\a} \om_a \wedge B \, , \\ l_s \upsilon_0  &=& - \frac{2}{l_s^4} \int_{\Pi_\a}  \sig A  \wedge  (B -\sig F)+ \frac{1}{l_s^4} \int_{\cc_4^\a} B^2 - \varpi_4\, ,
\label{ap:upsilons}
\eea
and one may then rewrite these chain integrals in terms of the D6-brane moduli, which as in section \ref{s:4forms} can be defined in terms of the reference configuration $\{\Pi_\a^0\}$. 

One can check that these expressions are related to the ones obtained in section 3 of \cite{Carta:2016ynn}, where the shift (\ref{eq:topCD}) was computed directly from the variation in the D6-brane CS action. The main difference with respect to that computation is that now we are in the presence of a non-vanishing NS flux $H$, and so the B-field is not globally well-defined. However, imposing the absence of Freed-Witten anomalies for each D6-branes we have that $\int_{\Pi_\a} H = 0$ and so $B$ is well-defined in their worldvolume. Therefore it can also be made well-defined on the four-chain ${\cc_4^\a}$ that represents the homotopic deformation from $\Pi_\a^0$. We are then able to perform the split
\be
B|_{\cc_4^\a}\, =\, b^a \om_a + \tilde{B}
\label{splitB}
\ee
with $\tilde{B}$ the co-exact piece of the B-field satisfying $d\tilde{B} = H|_{\cc_4^\a}$. Given this split one can see that $\varpi_4|_{\cc_4^\a} = \oh \tilde{B} \wedge \tilde{B}|_{\cc_4^\a}$. 

With these conventions, one may easily evaluate (\ref{ap:upsilons}) in terms of open string moduli and fluxes by using the definitions \eqref{defPhi4ch}-\eqref{nai}. One finds 
\begin{align}
l_s \upsilon_0 & =  \left(b^c n_{c\, i}^\a - n_{F\, i}^\a + \oh h_K g^K_{i\, \a}\right) ( b^d f_{d\, \a}^i - \th^i_\a) +  b^a \aleph_{a\, \a}  \,, \nonumber\\
l_s \upsilon_a & =  - n_{a\, i}^\a (b^c f_{c\, \a}^i - \theta^i_\a)  - (b^c n_{c\, i}^\a - n_{F\, i}^\a  ) f_{a\, \a}^i - \aleph_{a\, \a}\,, \label{eq:upsA}\\
l_s \tilde \upsilon^a & =  q^a\,, \nonumber
\end{align}
where we have defined
\begin{align}
\aleph_{a \, \alpha} = \frac{2}{l_s^4} \int_{\mathcal C_4^\alpha} \tilde B \wedge \omega_a
\end{align}
which is independent of the choice of four-chain ${\mathcal C_4^\alpha}$ provided that $H \wedge \om_a \equiv 0$.\footnote{This is true at the vacua of the theory, like the supersymmetric vacua where $H \in H^{(3,0)+(0,3)}_-(\mathcal M_6)$.} Notice that the shift (\ref{shiftG}) will also affect the dimensional reduction of the 10d kinetic terms, resulting in the above shift $\rho_A \raw \rho_A + \upsilon_A$ also in the second term of  \eqref{RRfluxes}.

Eqs.\eqref{eq:upsA} reproduce \eqref{varrhoHno} provided that 
\be
\aleph_{a \, \alpha} \, =\, \oh h_K {\bf H}^K_{a\,\alpha}
\label{realepH}
\ee
where ${\bf H}^K_{a\,\alpha}$ are functions of the D6-brane geometrical deformations satisfying (\ref{defH}), and used by Hitchin in order to in order to describe the metric on the moduli space of special Lagrangian submanifolds \cite{Hitchin:1997ti}. A more precise definition for our purposes is \cite{Carta:2016ynn}
\be
\p_{\varphi_\b^j} {\bf H}^K_{a\, \a}\, =\, (\eta_{a\, \b})^i{}_j g_{i\, \a}^K\d_{\a\b}\, .
\label{ap:defH}
\ee
where $g_{i\, \a}^K$ is defined as in \eqref{nCFg} and 
\be
 (\eta_{a\, \a})^i{}_j\, \equiv\, l_s^{-3} \int_{\Pi_\a} \iota_{X_j} \omega_a \wedge \eta^i = \frac{\p f_{a\, \a}^i}{\p \varphi^j_\a}
\ee
measures the deformation of the special Lagrangian three-cycle $\Pi_\a$ with respect to a normal vector $X = \frac{1}{2} l_s \varphi_\a^j X_j$ in terms of its cohomology  (recall that $l_s^{-2}\eta^i$ is a basis of integer harmonic two-forms of $\Pi_\a$). See \cite{Carta:2016ynn} for a more detailed discussion of these quantities. With these definitions at hand one can see that
\begin{align}
\p_{\varphi^i_\a} (\aleph_{a \, \alpha})&= \frac{1}{l_s^3} \int_{\mathcal C_4^\alpha}\mathcal L _{X_i} \left(\tilde B \wedge \omega_a\right)=\frac{1}{l_s^3} \int_{\Pi_\alpha-\Pi_\alpha^0}\iota_{X_i} \left(\tilde B \wedge \omega_a \right)+ \frac{1}{l_s^3}\int_{\mathcal C_4^\alpha} \iota_{X_i} \left(H\wedge \omega_a\right)\nonumber\\ &= \frac{1}{l_s^3} \int_{\Pi_\alpha-\Pi_\alpha^0}\iota_{X_i} \left(\tilde B \wedge \omega_a \right) = \frac{1}{l_s^3} \int_{\Pi_\alpha-\Pi_\alpha^0}\tilde B \wedge \iota_{X_i}\omega_a = \frac{1}{2} h_K\, g^K_j \left(\eta_{a \, \alpha}\right)^j_i
\end{align}
where we have used $H \wedge \omega_a=0$ and that $\tilde B$ can be chosen to have an even number of legs along $\Pi_\alpha$. This determines the desired relation up to a constant and, since both $\aleph_{a\, \a}$ and ${\bf H}^K_{a \, \a}$ vanish at the reference cycle $\Pi_\a^0$, \eqref{realepH} follows. 

Before concluding let us mention that there are some additional couplings in the 4d action between the two-forms $\Xi_K$ and the open strings moduli. Specifically we find that the last term in \eqref{eq:topC} has now the form
\begin{align}
S_{\text{two-forms}} = \frac{1}{8 \kappa_{4}^2} \int_{\mathbb R^{1,3}} \left[d\xi'^K+\frac{1}{2} \theta^i dg_i^K\right]\wedge \left[d\Xi_K +h_K c_3^0\right]\,.
\end{align}
Integrating out the two-forms proceeds exactly in the same way as before taking into account the modification of the duality relation due to the presence of open string degrees of freedom.

\section{The potential from standard 4d supergravity}
\label{ap:sugra}

In this appendix we recover the scalar potential (\ref{VtotD6}) from the Cremmer et al. F-term potential of standard 4d $\CN=1$ supergravity, using the tree-level superpotential and K\"ahler potentials given in the main text. We follow the same approach as in Section 6 of \cite{Carta:2016ynn}, but now to the presence of D6-brane moduli we add a non-trivial NS $H$-flux. This complicates the computation because then several no-scale identities are no longer valid, and in addition the superpotential depends on the 4d holomorphic variables  $N^K$, which are  complex structure moduli redefined by the open string moduli, as in \eqref{redefN}. We first discuss this dependence from the viewpoint of holomorphicity of the superpotential, and then discuss several properties of the K\"ahler metrics. As in \cite{Carta:2016ynn}, under certain simplifying assumptions for such metrics one is able to carry the computation of the F-term potential analytically, and we show that the result matches the potential obtained from the dimensional reduction of Appendix \ref{ap:dimred}.

\subsection{The superpotential}\label{sap:supo}

The type IIA flux superpotential can be written in the form \cite{Koerber:2007xk}
\be
iW\, =\,\frac{1}{l_s^{6}} \int_{\CM_6}  e^{-\phi} \re\,  \Omega \wedge H - i {\bf G} \wedge e^{iJ} 
\label{supoflux}
\ee
which is manifestly gauge invariant and globally well-defined. Notice that in a Calabi-Yau $J$ is harmonic, so only the harmonic pieces of {\bf G} contribute to the integral (\ref{supoflux}). As pointed out in \cite{Koerber:2007xk,Carta:2016ynn}, this expression not only contains the closed string superpotential but also the open string one. To see how both are contained, one may proceed as in section \ref{s:4forms} and consider a reference configuration of D6-branes wrapping special Lagrangian three-cycles $\{\Pi_\a^0\}$, with vanishing worldvolume flux $F$ and satisfying the tadpole condition \eqref{RRtadpole}. The local open string moduli space can then be parametrised through homotopic deformations $\Pi_\a^0 \raw \Pi_\a$. Then one can split the RR flux background {\bf G} into two pieces 
\be
{\bf G} \, =\, {\bf G}^0 +  \sum_\a \Delta_\a {\bf G}
\label{splitG}
\ee
with ${\bf G}^0$ satisfying the Bianchi identities and quantisations conditions for the reference configuration, and $\Delta_\a {\bf G}$ representing the change in {\bf G} as we replace the D6-brane at $\Pi_\a^0$ with the one at $\Pi_\a$. More precisely we have that
\be
d_H {\bf G}^0 \, =\, - \sum_\a q_\a \d(\Pi_\a^0) \wedge e^B\, , \quad \qquad d_H \Delta_\a{\bf G} \, =\, \d(\Pi_\a^0) \wedge e^B -  \d(\Pi_\a) \wedge e^{B - \sig F}
\label{splitBI}
\ee

In compactifications with vanishing $H$-flux it is quite simple to describe suitable solutions for ${\bf G}^0$ and $\Delta_\a {\bf G}$. For instance one has that 
\be
\Delta_\a {\bf G} \, \simeq \, - j_{({\cc_4^\a},\CF)} \, =\,  - \frac{1}{l_s} \d({\cc_4^\a}) \wedge e^{B -\sig \tilde F}\, .
\label{shiftG}
\ee
Here $j_{({\cc_4^\a},\CF)}$ is a generalised current in the sense of \cite{Koerber:2010bx,Koerber:2006hh}, that connects the D6-brane wrapping $\Pi_\a$ to the reference three-cycle $\Pi_\a^0$. As before $\d({\cc_4^\a})$ is the bump delta-function associated to a connecting four-chain ${\cc_4^\a}$, such that $l_s d\, \d({\cc_4^\a}) = \d(\Pi_\a) - \d(\Pi_\a^0)$. In general, the rhs of \eqref{shiftG} will not be the same polyform as $\Delta_\a{\bf G}$, but they will only differ by an exact piece, and so for the purposes of evaluating the integral \eqref{supoflux} they will be equivalent. To see this, one may proceed as in \cite{Hitchin:1997ti,Marchesano:2014bia} and perform a Hodge decomposition of both polyforms into coexact, harmonic and exact pieces. Because the rhs of (\ref{shiftG}) satisfies the same Bianchi identity as $\Delta_\a{\bf G}$ and $H=0$, the coexact piece of both polyforms is equal. Now, because of the quantisation condition for RR fluxes, the harmonic piece of $\Delta_\a {\bf G}$ is determined by its coexact piece, just like it happens for $j_{({\cc_4^\a},\CF)}$. Therefore only the exact pieces of these functions can differ. 

These observations are also useful in finding a suitable solution for ${\bf G}^0$ which, in general, can be separated into a $d_H$ closed and non-closed piece, namely
\be
{\bf G}^0 = - j_0- H \wedge C_3 + e^{B} \wedge \bar{\bf G} + \dots
\ee
where the dots stand for terms that will not contribute to the integral in (\ref{supoflux}). Here $C_3\, =\, \xi'^K\a_K$ stands for the harmonic piece of $C_3$, $\bar{\bf G}$ is a harmonic polyform of integer fluxes and $l_s^2 d_H(j_0 + \bar{G}_0)\, =\,  \sum_\a q_\a \d(\Pi_\a^0)$. 
In compactifications where $H \equiv 0$, the RR tadpole condition \eqref{RRtadpole} implies that there is a four-chain 
 connecting the whole set of D6-branes and O6-planes. In particular for the reference configuration we have that $\p \cc_4^0 = \sum_\a \Pi_\a^0 + \CR\Pi_\a^0 - 4\Pi_{\rm O6}$. One may then define the current $j_0$ in terms of such a four-chain \cite{Marchesano:2014iea,Carta:2016ynn}
 \be
 j_0 \, =\, \d(\cc_4^0) \wedge e^B \, .
 \label{j04ch}
 \ee
Notice that a choice of four-chain $\cc_4^0$ with a fixed boundary is only determined up to the choice of a four-cycle $\Lam_4$. Nevertheless, choosing different four-cycles can be interpreted as taking different flux quanta $m^a$ for $\bar{G}_2$, as their contribution to the superpotential is the same. In general, one can interpret the contribution of $j_0$ as shifting the lattice of integer two-form fluxes $m^a \in \IZ$ to $\tilde m^a = m^a + \eps_0^a$, with $\vec{\eps}_0$ a fixed vector that depends on the choice of reference D6-brane configuration $\{\Pi_\a^0\}$. 

Plugging these expressions into (\ref{supoflux}) we obtain
\be
W\, =\,  \frac{1}{l_s^6} \int_{\CM_6} H \wedge \Omega_c  - \bar{\bf G} \wedge e^{J_c} + \sum_\a \frac{1}{l_s^5}  \int_{\cc_4^\a} \left(J_c - \sig \tilde{F} \right)^2 + W_0
\label{Wmodu}
\ee
where $\Om_c$ and $J_c$ are defined as in \eqref{defN} and \eqref{D6Fterm}, respectively, $\a$ runs over only half of the D6-branes of the compactification and not over their orientifold images and
\be
W_0 \, = \, \frac{1}{2l_s^5}  \int_{\cc_4^0} (J_c - \sig \tilde{F})^2\,.
\ee
Notice that the first two terms of \eqref{Wmodu} reproduce the standard form of the Calabi-Yau closed string flux superpotential \cite{Gukov:1999ya,Taylor:1999ii,Gurrieri:2002wz,Villadoro:2005cu}, while the third corresponds to the D6-brane superpotential \cite{Thomas:2001ve,Martucci:2006ij}.

When the $H$-flux is non-vanishing, the equivalence (\ref{shiftG}) between polyforms is no longer valid. Indeed, both forms still satisfy the same Bianchi identity but, because it implies the operator $d_H$ and now $H$ is non-trivial, their coexact piece is now different. Since the coexact piece determines their harmonic part, when plugged into \eqref{supoflux} these two polyforms will give different results. In fact, one can easily check that if we consider \eqref{Wmodu} in the presence of non-vanishing $H$ the integral will depend continuously on the choice of four-chain $\cc_4^\a$, which is unacceptable. 

So instead of the rhs of \eqref{shiftG} one should consider replacing $\Delta_\a G$ by a polyform with the same coexact piece. A suitable choice seems to be
\be
\Delta_\a {\bf G} \, \simeq\,  \frac{1}{l_s^2} \d(\Pi_\a) \wedge \left( \sig A - \oh \sig^2 A \wedge F\right)  \wedge e^{B}  - \frac{1}{l_s}  \d({\cc_4^\a}) \wedge \left( e^{B} - \varpi_4\right) 
\label{shiftG2}
\ee
where $\varpi_4$ is the co-exact form such that $d\varpi_4 = H \wedge B$. Notice that when $H=0$ this is equivalent to the previous Ansatz. Replacing this into \eqref{supoflux} we obtain
\be
- \frac{2}{l_s^4} \int_{\Pi_\a} \sig A \wedge \left( J_c - \sig F\right) + \frac{1}{l_s^5} \int_{\cc_4^\a} J_c^2 - \varpi_4
\label{newchain}
\ee
instead of the four-chains in \eqref{Wmodu}. Notice that now these chain-integral are invariant under continuous deformations of $\cc_4^\a$ as long as $d_H J = 0$.

 Interestingly, armed with this last expression we can determine the definition of 4d holomorphic variable of the complex structure sector used in the main text. For that we need to extract the dependence of the full superpotential \eqref{supoflux} with respect to the quanta of background NS flux $H$.  Indeed, performing the  split (\ref{splitB}) in $\Pi_\a$ and $\cc_4^\a$ with the same gauge choice for $\tilde{B}$, we obtain that \eqref{newchain} gives
\be\nonumber
 \frac{1}{l_s^4}  \int_{\cc_4^\a} J_c \wedge  \tilde{B}   -  \frac{2}{l_s^4} \int_{\Pi_\a} \sig A \wedge \tilde{B} + \dots \, =\, \oh \aleph_{a\, \a} T^a - \frac{1}{2} h_K g^K_{i\, \a}  \th^i_\a \, =\, \oh h_K \left( {\bf H}^K_{a\,\alpha} T^a - g^K_{i\, \a}  \th^i_\a\right) + \dots 
\ee
where we have used \eqref{realepH}. Adding this contribution to the one from the first term of \eqref{Wmodu} one obtains that the full superpotential depends on
\be
- h_K N^K\, =\, - h_K \left[N'^K + \oh \sum_\a \left(g^K_{i\, \a} \th^i_\a -T^a  {\bf H}_{a \, \a}^K \right)\right]
\ee
which indeed corresponds to the 4d holomorphic variable \eqref{redefN}. 

In general, evaluating the expression \eqref{Wmodu} for a Calabi-Yau flux compactification with D6-branes we will obtain the following polynomial superpotential 
\begin{align}
l_s (W - W_0) = e_0 - e_a T^a + \frac{1}{2} \mathcal K_{abc} m^a T^b T^c - \frac{1}{3!} m T^a T^b T^c- h_K N^K - \Phi^i_\a \left(n_{F\, i}^\a -n_{a\, i}^\a T^a\right)\,.
\end{align}
which indeed matches the expression (\ref{Wtotal}) used in the main text. In the following we will use this same expression to evaluate type IIA the flux potential via the standard 4d $\CN=1$ supergravity formula. 

\subsection{The K\"ahler metrics}
\label{app:kahlermetrics}
Before computing the F-term scalar potential it is useful to discuss the structure of the K\"ahler metric that arises from the K\"ahler potential $K= K_K + K_Q$, obtained from rewriting \eqref{KK} and \eqref{KQ} in terms of the holomorphic variables. Here the discussion is rather similar to the one in Appendix A of \cite{Carta:2016ynn}, since the metrics are the same. First we have the general relations that come from the fact that $e^{-K}$ is a homogeneous function of degree 7
\begin{align}
K^{\alpha \bar \beta} K_{\alpha} K_{\bar \beta}=7\,, \qquad K^{\alpha \bar \beta} K_{\bar \beta} =-2i \text{Im}\, \Psi^\alpha\,,
\label{noscale7}
\end{align}
where $\a, \b$ run over all the fields $\Psi^\alpha$ in the effective theory.  

Then, one may consider the simplifying assumption that the chain integrals $f^i_{a\, \a}$ and $g^K_{i \a}$ defined in \eqref{deffch} and \eqref{nCFg}, respectively, do not depend on the complex structure of the compactification, and in particular that they only depend on the K\"ahler moduli $t^a$ and on $\im\, \Phi^i_\a$ through the D6-brane position $\varphi^i_\a$ defined as in \eqref{ap:defH} (see \cite{Carta:2016ynn} for a justification of this approximation).  Then, following Appendix A of \cite{Carta:2016ynn}, we have that the K\"ahler metric has the form
\begin{align}\mathbf K = \left(\begin{array}{cc} \mathbf N & \mathbf N . \Xi^\dag\\ \Xi. \mathbf N&\mathbf \Omega + \Xi.\mathbf N . \Xi^\dag\end{array}\right)\,,
\end{align}
where $\mathbf N_{I \bar J} = \frac{1}{4} \p_{n'^I}\p_{n'^J} K_{Q}$ and $\Xi^K_{\hat \alpha} =  \p_{\psi^{\hat \alpha}} n'^K$, with $\psi^{\hat \alpha}$ the imaginary part of the K\"ahler and brane moduli and $n'^K = \im\, N'^K$ of the complex structure moduli. Finally 
\begin{align}
\Omega = \left(\begin{array}{cc}\mathbf A & \mathbf B \\ \mathbf C & \mathbf D\end{array}\right)\,,
\end{align}
with
\begin{align}
\mathbf A_{a\bar b} &= \p_a \p_{\bar b} \mathbf K_{\mathbf K}+ (\p_{n'^K}\mathbf K_{\mathbf Q}) \p_a \p_{\bar b} n'^K\,,\\
\mathbf B_{a\bar \jmath} &= (\p_{n'^K} \mathbf K_{\mathbf Q})\p_a \p_{\bar \jmath} n'^K\,,\\
\mathbf D_{i\bar \jmath} &= (\p_{n'^K} \mathbf K_{\mathbf Q}) \p_i \p_{\bar \jmath} n'^K\,,
\end{align}
and $\mathbf C = \mathbf B^\dag$. The inverse of the full K\"ahler metric $\mathbf K$ is then easily computed to be
\begin{align}
\label{Kinverse}
\mathbf K^{-1} = \left(\begin{array}{cc} \mathbf N^{-1}+\Xi^{\dag}.\Omega^{-1}.\Xi & -\Xi^{\dag}.\Omega^{-1}\\ - \Omega^{-1}.\Xi&\Omega^{-1}\end{array}\right)\,.
\end{align}
From the definition of the 4d holomorphic variable defined in (\ref{redefN}) we have that
\begin{align} n'^K = n^K +\frac{1}{2} t^a \mathbf H_a^K\,,
\end{align}
where we have dubbed $n^K = \im\, N^K$. We can use this relation to compute the various components of the inverse K\"ahler metric, obtaining
\begin{align} K^{\bar I a } &= -\p_\alpha n'^I (\mathbf \Omega^{-1})^{a\alpha} = -\p_i n'^I (\mathbf \Omega^{-1})^{a i}-\p_b n'^I (\mathbf \Omega^{-1})^{a b} =\nonumber\\&= -\frac{1}{2}\left[K^{ab} f^{i}_b  g^I_i +K^{ab}(\mathbf H_b^I - f^i_b g_i^I)\right]= -\frac{1}{2} K^{ab} \mathbf H^I_b\,,\\
K^{\bar I i } &= -\p_\alpha n'^I (\mathbf \Omega^{-1})^{i\alpha} = -\p_j n'^I (\mathbf \Omega^{-1})^{ij}-\p_a n'^I (\mathbf \Omega^{-1})^{ia}= \nonumber\\&= -\frac{1}{2}\left[g^I_j \left(\mathbf D^{ij}+ f^i_af^j_b K^{ab}\right)+K^{ab}f_b^i (\mathbf H_a^I - f^j_a g^I_j)\right]=-\frac{1}{2} \mathbf D^{ij} g_j^I- \frac{1}{2} K^{ab}f_b^i \mathbf H^I_a\,,
\end{align}
where we used that $\p_i n'^I = \frac{1}{2} g_i^I$ and $\p_a n'^I = \frac{1}{2}(\mathbf H_a^I - f_a^i g_i^I)$.  A similar computation gives $K^{\bar J I}$. Summarising we find that
\begin{align}\label{eq:im1}K^{\bar I a}& = -\frac{1}{2} K^{\bar b a}\mathbf{H}^I_b\,,\\
\label{eq:im2}K^{\bar I i} &= -\frac{1}{2}\left[\mathbf D^{ji} \,g_j^I+K^{\bar b a }f_a^i\,\mathbf{H}_b^I \right]\,,\\
\label{eq:im3}K^{\bar J I} &=\mathbf N^{\bar J I}+ \frac{1}{4} \left[K^{\bar b a}\,\mathbf H_b^J\mathbf H_a^I+ \mathbf D^{ij}\, g_i^I g_j^J\right]\,.
\end{align}

\subsection{The F-term potential}
\label{app:Fterm}

Let us now compute the F-term scalar potential through the standard formula
\begin{align}
V_F = \frac{1}{\kappa_4^2} \,e^K \left(K^{\alpha \bar \beta} D_\alpha W D_{\bar \beta} \overline W -3|W|^2\right)\,,
\end{align}
where the index $\alpha$ runs over the entire set of the fields in the 4d theory. 

First the relations \eqref{noscale7} allow to rewrite the above expression as
\begin{align}
\kappa_4^2 V = e^K \left(K^{\alpha \bar \beta}\p_\alpha W \p_{\bar \beta} \overline W +4 \text{Im} \left(\text{Im} \Psi^\alpha \p_{\alpha} W \overline W\right)+4 |W|^2\right)\, ,
\label{V3terms}
\end{align}
and then we can proceed now to the computation of the individual terms. For simplicity we will merge the two indices of the open string moduli $\Phi_\a^i$ into a single one $\Phi^i$. 

The computation of the last two terms of \eqref{V3terms} follows closely the one in \cite{Carta:2016ynn}:
\be
\label{eq:1}
\begin{split}
4 l_s^2 &(|W|^2+ \text{Im} \left[\Psi^\alpha \p_\alpha W \overline W\right] ) = \\
=&\bigg[2 e_0 - 2 e_a b^a + \mathcal K_{abc} m^a b^b b^c  - \frac{1}{3} m \mathcal K_{abc} b^a b^b b^c -2 \text{Re} \, \Phi^i (n_{F\, i} - n_{a\, i} b^a) {- 2 h_I \xi^I}\bigg]^2 \\
- &\bigg[\mathcal K_a m^a - m \mathcal K_a b^a + 2 \text{Im} \, \Phi^i n_{a\, i} t^a\bigg]^2  + \frac{4}{3} m \,\mathcal K \,\text{Im}\, W
\end{split}
\ee
In addition we can employ the relations \eqref{eq:im1}-\eqref{eq:im3} for the K\"ahler metrics to simplify the remaining terms in the scalar potential. We find that
\bea\label{Kabterm}
K^{\alpha \bar \beta} \p_\alpha W \p_{\bar \beta} \overline W & = & K^{a \bar b}  \hat \p_a W \hat \p_{\bar b } \overline W  +{\bf N}^{I\bar{J}} \p_I W \p_{\bar J} \overline W \\
&+& G^{ij}_{\rm D6} \left[ \p_i W -\frac{1}{2}  g_i^K \p_K W \right] \left[ \p_{\bar \jmath} \overline W - \frac{1}{2} g_j^L \p_{\bar L} \overline W \right]
\label{Gijterm}
\eea
where the modified derivative is $\hat \p_a = \p_a +f_a^i \p_i -\frac{1}{2} {\bf H}_a^K \p_K$. The second term in \eqref{Kabterm} is familiar from compactifications with H-flux without D6-branes. It is of the form
\be
 {\bf N}^{I\bar{J}} h_I h_J = \frac{1}{2} e^{-2 \phi} \int H_3 \wedge * H_3\,,
\ee
and it is expected to arise from integrating out the three-form corresponding to the NS-flux. The term proportional to the D6-brane inverse metric can, as in \cite{Carta:2016ynn}, be identified with the DBI piece of the potential. Notice that now we have
\bea
\label{ap:VDBI}
V_{\rm DBI}  & = & \frac{e^K}{\kappa_4^2} G^{ij}_{\rm D6} \left[ \p_i W - \frac{1}{2} g_i^K \p_K W \right] \left[ \p_{\bar \jmath} \overline W - \frac{1}{2} g_j^L \p_{\bar L} \overline W \right] \\
& = &  \frac{e^K}{ l_s^2 \kappa_4^2} G^{ij}_{\rm D6}  \left(n_{\CF\, i} - n_{a\, i} T^a\right)\left(n_{\CF\, j} - n_{a\, j} \bar{T}^a\right)
\nonumber   
\eea
where we have defined $n_{\CF\, i}$ as in \eqref{nCFg}. 

Finally, let us look at the first term in the rhs of (\ref{Kabterm}), which is to be combined with \eqref{eq:1}. The computation parallels again the one in \cite{Carta:2016ynn}. We have that
\be\label{eq:2}\begin{split}
&l_s^2\, \text{r.h.s.} \eqref{Kabterm} = \\
&K^{a \bar b} \bigg[e_a - \mathcal K_{acd } m^c b^d + \frac{1}{2} m \mathcal K_{acd} b^c b^d -\frac{1}{2} m\mathcal K_{a} - \text{Re} \,\Phi^i n_{a\, i} + f_a^i (n_{F\, i} - n_{c\, i} b^c) {- \frac{1}{2} {\bf H}_a^K h_K} \bigg]\\
&\quad \times \bigg[e_b - \mathcal K_{bcd } m^c b^d + \frac{1}{2} m \mathcal K_{bcd} b^c b^d -\frac{1}{2}m \mathcal K_{b} - \text{Re} \,\Phi^i n_{b\, i} + f_b^i (n_{F\, i} - n_{c\, i} b^c) {-\frac{1}{2} {\bf H}_b^K h_K} \bigg]\\
&+K^{a \bar b} \bigg[\mathcal K_{ac} m^c - m \mathcal K_{ac} b^c + \text{Im} \,\Phi^i n_{a\, i} + f_a^i  n_{c\, i} t^c \bigg]\\
&\quad \times \bigg[\mathcal K_{bc} m^c - m \mathcal K_{bc} b^c + \text{Im} \,\Phi^i n_{b\, i} + f_b^i  n_{c\, i} t^c \bigg]\,.
\end{split}\ee
To proceed we add up the first term in the second line in \eqref{eq:1} and the last two lines of \eqref{eq:2}. Just like in \cite{Carta:2016ynn} we obtain
\be
\frac{4}{9} \mathcal K^2 K_{a \bar b} ( m^a -m b^a +q^a) ( m^b - m b^b +q^b)\,.
\ee
Next we take the first two lines of \eqref{eq:2} and rewrite them as
\be\begin{split}
& \,
K^{a \bar b} \bigg[e_a - \mathcal K_{acd } m^c b^d + \frac{1}{2} m \mathcal K_{acd} b^c b^d -\frac{1}{2} m\mathcal K_{a} - \text{Re} \,\Phi^i n_{a\, i} + f_a^i (n_{F\, i} - n_{c\, i} b^c) {- \frac{1}{2} {\bf H}_a^K h_K} \bigg]\\
&\quad \times  \bigg[e_b - \mathcal K_{bcd } m^c b^d + \frac{1}{2} m \mathcal K_{bcd} b^c b^d -\frac{1}{2}m \mathcal K_{b} - \text{Re} \,\Phi^i n_{b\, i} + f_b^i (n_{F\, i} - n_{c\, i} b^c) {- \frac{1}{2} {\bf H}_b^K h_K} \bigg]  \\
& = \, K^{a \bar b} \bigg[e_a - \mathcal K_{acd } m^c b^d + \frac{1}{2} m \mathcal K_{acd} b^c b^d  - \text{Re} \,\Phi^i n_{a\, i} + f_a^i (n_{F\, i} - n_{c\, i} b^c){- \frac{1}{2} {\bf H}_a^K h_K} \bigg]\\
&\quad \times  \bigg[e_b - \mathcal K_{bcd } m^c b^d + \frac{1}{2} m \mathcal K_{bcd} b^c b^d - \text{Re} \,\Phi^i n_{b\, i} + f_b^i (n_{F\, i} - n_{c\, i} b^c) {- \frac{1}{2} {\bf H}_b^K h_K} \bigg]  \\
& - \frac{4}{3} \mathcal K m t^a \bigg[e_a - \mathcal K_{acd } m^c b^d + \frac{1}{2} m \mathcal K_{acd} b^c b^d  - \text{Re} \,\Phi^i n_{a\, i} + f_a^i (n_{F\, i} - n_{c\, i} b^c){- \frac{1}{2} {\bf H}_a^K h_K} \bigg] \\ & +\frac{1}{4} K^{a \bar b} \mathcal K_a \mathcal K_b m^2
\end{split}\ee
and we combine the last line of this equation with the last term in \eqref{eq:1} to obtain
\be\begin{split}
& \frac{4}{3} m \,\mathcal K \,\text{Im}\, W - \frac{4}{3} \mathcal K m t^a \bigg[e_a - \mathcal K_{acd } m^c b^d + \frac{1}{2} m \mathcal K_{acd} b^c b^d  - \text{Re} \,\Phi^i n_{a\, i} + f_a^i (n_{F\, i} - n_{c\, i} b^c){- \frac{1}{2} {\bf H}_a^K h_K} \bigg]   \\
&  +\frac{1}{3} \mathcal K^2  m^2 = \left( \frac{1}{3}- \frac{2}{9}\right)  \mathcal K^2  m^2 + \frac{4}{3} m \,\mathcal K \left( n^I  + \frac{1}{2} t^a {\bf H}_a^I\right) h_I\, =\, \frac{1}{9}  \mathcal K^2  m^2 + \frac{4}{3} m \,\mathcal K  n'^Ih_I \, .
\end{split}\ee
To sum up, we find that the F-term scalar potential reads
\be
V_{\rm total}   = V_{\rm DBI} + V_{\rm loc} + \frac{1}{\kappa_4^2} e^K \left[4 \varrho_0^2 + g^{ab} \varrho_a \varrho_b + \frac{4}{9} e^K  \CK^2 g_{ab} \tilde \varrho^a \tilde \varrho^b + \frac{1}{9} e^K \CK^2 \tilde \rho^2  + {\bf N}^{IJ} h_Ih_J\right]
\label{Vtotal}
\ee
where
 \be
 \begin{array}{rcl}
l_s \varrho_0 & = &  e_0 - b^a e_a + \oh \CK_{abc}m^ab^bb^c - \frac{m}{6} \CK_{abc}b^ab^bb^c -h_I \xi^I - (b^a f_a^i-\theta^i  ) (n_{F\, i} - n_{a\, i} b^a)\,, \\
 l_s \varrho_a & = & e_a -\CK_{abc}m^bb^c + \frac{m}{2} \CK_{abc}b^bb^c - ( b^a f_a^i-\theta^i ) n_{a\, i} + f_a^i (n_{F\, i} - n_{c\, i} b^c) {- \frac{1}{2} {\bf H}_a^K h_K} \,, \\
l_s \tilde \varrho^a & = & m^a - mb^a + q^a\,, \\
 l_s \tilde \varrho & = & m\,,
 \end{array}
 \label{varrhos}
 \ee
the term $V_{\rm DBI}$ is given by \eqref{ap:VDBI} and 
\be
V_{\rm loc}\, =\,  \frac{4}{3} m \,\mathcal K  n'^Ih_I \, .
\ee
Notice that everywhere in the potential appear the geometric variables of the complex structure moduli $N'^K$. On the other hand, the term $\xi^I$ within $\varrho_0$ stands for the axionic component of the 4d holomorphic field $N^K$. When expressed in term of the geometric axions, this term will contain further dependence in the open string moduli, namely given (\ref{redefN}) we have that 
\be
\xi^I\, =\, \xi'^I - \frac{1}{2} b^a {\bf H}^I_a + \oh g_i^I \th^i \, ,
\ee
and therefore
 \be
\begin{array}{rcl}
l_s \varrho_0 & = &  e_0  - h_K \xi'^K - (n_{F\, i} - \oh g_i^K h_K) \theta^i   -  (e_a + \theta^i  n_{a\, i} + f_a^i n_{F\, i} - \frac{1}{2} {\bf H}_a^K h_K)b^a \\
& & + \oh \CK_{abc}(m^a +q^a) b^bb^c - \frac{m}{6} \CK_{abc}b^ab^bb^c  \\
 l_s \varrho_a & = & (e_a + \theta^i  n_{a\, i} + f_a^i n_{F\, i} - \frac{1}{2} {\bf H}_a^K h_K) -\CK_{abc}(m^b + q^b)b^c + \frac{m}{2} \CK_{abc}b^bb^c   \\
l_s \tilde \varrho^a & = & m^a + q^a - mb^a \\
 l_s \tilde \varrho & = & m\, ,
 \end{array}
 \label{varrhos2}
 \ee
matching the results of Appendix \ref{ap:dimred}.

\section{Periodic D6-brane positions}\label{ap:micro}

In some particular compactifications, the position of the three-cycles wrapped by the D6-branes is of periodic nature. One then expects that such directions in open string moduli space can be described in terms of periodic scalars that enter the monodromic structure of the potential, just like any of the multiple axions. One familiar class of models where this occurs are toroidal and orbifold compactifications, where typically D-brane positions can be understood as Wilson line scalars in dual descriptions of the theory. In those cases, the functions $f_{a\, \a}^i$ and $g^K_{i\, \a}$ defined in \eqref{deffch} and \eqref{nCFg} are linear in the microscopic parameter $\varphi_\a^i$ describing the positions of the three-cycle $\Pi_\a$ wrapped by the D6-brane. In general one can write them in the form
\be
f_{a\, \a}^i = (\eta_{a\, \a})^i{}_j \varphi^j_\a\, , \qquad g^K_{i\, \a} = (\CQ_\a^K)_{ij}\varphi^j_\a\, , \qquad {\bf H}_{a\, \a}^K = \oh (\eta_{a\, \a})^k{}_i (\CQ^K_\a)_{kj} \varphi^i_\a \varphi^j_\a
\ee
where $(\eta_{a\, \a})^i{}_j$ and $(\CQ_\a^K)_{ij}\varphi^j_\a$ are constant tensors whose precise value is not relevant for the present discussion (see section 2 of \cite{Carta:2016ynn} for their precise definition). What is important is that then these quantities satisfy the following relation
\be
{\bf H}_{a\, \a}^K = \oh f_{a\, \a}^i g^K_{i\, \b} \d_{\a\b}\, .
\label{Hid}
\ee
When plugged into the matrix {\bf S} in \eqref{Ropen}, this implies that we can write {\bf S} as the exponential of a nilpotent matrix. In particular for the toroidal case we have that
\be
{\bf S}^t = e^{\varphi^i_\a Q_i^\a}
\ee
with
\be
Q_i^\a =\left(\begin{array}{cccccc}0 & 0 & 0 & 0 & 0&  0  \\0 & 0 & 0 &0& (\eta_a)^j_i &0\\0 & 0 & 0 &0& 0&0\\0 & 0 & 0 &0& 0&0\\0 & 0 & 0 &0& 0&\oh Q^K_{ij}\\0 & 0 & 0 &0& 0&0\\ \end{array}\right)\, .
\label{Qopen}
\ee
Therefore, we can treat $Q_i^\a$ as one of the nilpotent generators of section \ref{s:FW}, describing the interplay of 4d axions and fluxes. Finally, because of the structure of eq.\eqref{SR}, one can incorporate the D6-brane periodic positions $\varphi$ into the definition of the axion rotation matrix {\bf R}. 

Now, and interesting point is that the generator matrices $Q_i^\a$ do not commute with the generators $P$ of section \ref{s:FW}. In particular, they do not commute with the Wilson line matrices of \eqref{Piaopen}. This fact is not that surprising, since when one describes discrete gauge symmetries involving two periodic scalars of the same complex field one often finds non-commutativity, see \cite{BerasaluceGonzalez:2012vb} for other examples. In our case this translates into the fact that the matrices {\bf R} and {\bf S} above do not commute.

The way that {\bf R} and {\bf S} do not commute is quite interesting. To see this let us define
\be
\hat{\bf R}
=  \left(\begin{array}{cccccc} 1 & 0 & 0 & 0 & 0& 0\\  b^b & \delta^b_a  & 0 & 0 & 0 &0 \\  \frac{1}{2} \mathcal K_{abc} b^a b^c & \mathcal K_{abc} b^c & \delta^a_b   & 0 & 0 & 0\\  \frac{1}{3!} \mathcal K_{abc} b^a b^b b^c & \frac{1}{2}\mathcal K_{abc} b^b b^c & b^a& 1 & 0 & 0 \\ -{\th}^j   & 0 & 0 & 0 & \d_i^j & 0  \\  \xi'^{L} & 0 & 0 & 0 & 0 &  \d_K^L \end{array}\right), \
\label{Ropenhat}
\ee
That is, $\hat{\bf R}$ the axion rotation matrix but with the 4d supergravity axions replaced by the microscopic, geometric axions of the compactification. Now one can check that
\be
\hat{\bf R} \,= \, {\bf S} {\bf R} {\bf S}^{-1}
\label{dict}
\ee
and so the non-commutativity of the above generators translates into the dictionary between the notion of microscopic geometric axions and macroscopic 4d axions. Notice that this observation relies on the precise definition of holomorphic variable \eqref{redefN} in terms of open string moduli, and provides a cross-check of the latter. It is quite remarkable that the matrices {\bf S} give us the dictionary between geometric and 4d supergravity axions. It would be interesting to see if this can be related to the fact that 4d supergravity variables have to transform holomorphically when performing closed loops in open string moduli space, which is one of the criteria used to find the 4d redefinitions of closed string moduli with open string moduli, see e.g. \cite{Camara:2009uv}.

Moreover, \eqref{dict} implies that  the description of all the flux-axion polynomials from the axion polynomial $\rho_0$ in \eqref{generatorrhos} can be made with both macroscopic and microscopic axions. Indeed, one can define
\begin{align}
l_s \vec{\rho}\, =\,  {\bf S}^{t} 
\left(\begin{array}{c} l_s(\rho_0 + \upsilon_0) \\ l_s(\rho_a + \upsilon_a) \\  l_s (\tilde \rho^a + \tilde \upsilon^a) \\  l_s \tilde \rho \\ l_s  \rho_{{\CF\, i}} \\ l_s  \rho_{K}    \end{array} \right)  \, =\,
{\bf R}^{t\, -1}  
\left(\begin{array}{c} e_0 \\ e_b \\  m^b \\  m \\  n_{F\, j}  \\ h_L   \end{array} \right)
\end{align}
as done in \eqref{vecrho}. The components of this vector then satisfy eqs.\eqref{derirhos}. Using \eqref{dict} one can also write
\begin{align}
l_s \vec{\rho}^{\, \prime} \, =\, \left(\begin{array}{c} l_s(\rho_0 + \upsilon_0) \\ l_s(\rho_a + \upsilon_a) \\  l_s (\tilde \rho^a + \tilde \upsilon^a) \\  l_s \tilde \rho \\ l_s  \rho_{{\CF\, i}} \\ l_s  \rho_{K}    \end{array} \right) \, =\,
\hat{\bf R}^{t\, -1}   {\bf S}^{t\, -1}
\left(\begin{array}{c} e_0 \\ e_b \\  m^b \\  m \\  n_{F\, j}  \\ h_L   \end{array} \right) \, .
\end{align}
By following similar arguments to those below \eqref{generatorrhos}, one can see that the components of the vector $\vec{\rho}^{\, \prime}$ satisfy equations similar to \eqref{derirhos}, but now instead deriving with respect to the microscopic axions.

\section{Simple type IIA toroidal orientifold with metric fluxes}\label{ap:metric}

In this appendix we calculate the type IIA scalar potential in the toroidal orientifold  $\mathcal{T}^6/(\IZ_2\times \IZ_2)$ presented in \cite{Villadoro:2005cu}, that is, considering only the closed string moduli and in the presence of the usual  RR and NS  fluxes plus  metric fluxes. As we will see, the bilinear structure of the potential (including the triple factorisation) still holds when we define the new $\rho$'s according to eq. \eqref{generatorrhos}.

The complete 4d scalar potential has the following contributions:
\begin{equation}
\label{eq:potentialmetric}
V \ = \ V_{\mathrm{RR}}+V_{\mathrm{NS}}+V_{\mathrm{loc}}+V_{\mathrm{SS}},
\end{equation}
where the first three pieces are the contributions from RR, NS fluxes and localised sources, respectively. The last piece is the Scherk-Schwarz potential, which appears in the presence of metric fluxes when one performs the dimensional reduction of  the purely gravitational part in the 10d action. Before computing the potential, let us recall  how the Bianchi identities get modified in the presence of metric fluxes $\omega$ \cite{Villadoro:2005cu} 
\begin{equation}
\begin{array}{l c l}
\label{BImetric}
H_3&=&dB_2+ \omega \cdot B_2+\bar{H}_3,\\
G_p&=&dC_{p-1}+ \omega \cdot C_{p-1}-H\wedge C_{p-3} + (\bar{\textbf{G}} e^{-B}).\\
\end{array}
\end{equation}
Recall that these can be obtained from the expression without metric fluxes by making the substitution $dX \rightarrow dX + \omega \cdot X$. In the following, we will use the notation from \cite{Camara:2005dc} for the metric fluxes, that is
\beq 
\bmat{c} a_1 \\ a_2 \\
a_3 \emat = \bmat{c} \om^1_{56} \\ \om^2_{64} \\ \om^3_{45} \emat
\quad ; \quad \bmat{ccc} b_{11} & b_{12} & b_{13}ß \\ b_{21} & b_{22}
& b_{23} \\ b_{31} & b_{32} & b_{33} \emat = \bmat{ccc} \! \! \!
-\om^1_{23} & \, \om^4_{53} & \, \om^4_{26} \\ \, \om^5_{34} & \! \!
\! -\om^2_{31} & \, \om^5_{61} \\ \, \om^6_{42} & \, \om^6_{15} & \!
\! \! -\om^3_{12} \emat \ .
\label{abmatrix}
\eeq 
In addition, let us recall that in the toroidal compactification the K\"ahler and complex structure moduli are $T^i= b^i + i t^i$ and $N^I=\xi^I+ i n^I$, with $i=1,\ 2,\ 3$ and $I=0,\ 1,\ 2, \ 3$. From eq. \eqref{BImetric} one can compute the explicit expressions for the field strengths along the compact dimensions in terms of the fluxes and the axions by expanding them in the usual basis of harmonic forms.
After integrating  upon the compact dimensions in the 10d action, one can identify $\rho_0$ and from there calculate the rest of the $\rho$'s and the superpotential by applying eqs. \eqref{generatorrhos} and \eqref{superpo}, respectively. The $\rho$'s are then 
\begin{equation}
\label{ap:rhos}
\begin{array}{l}\vspace*{.2cm}
l_s \rho_0 \ = \  e_0 - e_i b^i+ m^1 b^2 b^3+ m^2 b^1 b^3 + m^3 b^1 b^2- m b^1 b^2 b^3 - h_I \xi^I +  b_{ij} b^i \xi^j -  a_i  b^i\xi^0, \\ 	  \vspace*{.2cm}
l_s\dfrac{\partial \rho_0}{\partial b^i}  \ = \  \ - l_s \rho_i \ = - \left( \ e_i - m_j b_k - m_k b_j + m b^j b^k - b_{ij} \xi^j - a_i  \xi^0\right),  \qquad (i \neq j \neq k \neq i),\\ \vspace*{.2cm}
l_s\dfrac{\partial ^2 \rho_0}{\partial b^i \partial b^j} \ = \  l_s \tilde{\rho}^k \ = \ \left( m^k - m b^k \right), \qquad (i \neq j \neq k \neq i),\\ \vspace*{.2cm}
l_s\dfrac{\partial ^3 \rho_0}{\partial b^i \partial b^j \partial b^k} \ = \ -l_s \rho_m \ = \ - m, \qquad (i \neq j \neq k \neq i),\\ \vspace*{.2cm}
l_s\dfrac{\partial \rho_0}{\partial \xi^0}  \ = \ - l_s \rho_{h_0} \ = \ - \left( h_0+a_i b^i\right), \\ \vspace*{.2cm}
l_s\dfrac{\partial \rho_0}{\partial \xi^i}  \ = \ - l_s \rho_{h_i} \ = \ - \left( h_i -b_{ij} b^j \right),\\ \vspace*{.2cm}
l_s\dfrac{\partial ^2 \rho_0}{ \partial b^i \partial \xi^j} \ = \  l_s \rho_{b_{ij}}\ = \ b_{ij}\, , \\ \vspace*{.2cm}
l_s\dfrac{\partial ^2 \rho_0}{\partial b^i \partial \xi^0 } \ = \  l_s \rho_{a_i}\ = \ -a_{i}\, , 
\end{array}
\end{equation}
and the superpotential reads
\begin{equation}\nonumber
l_sW=e_0 - e_i T^i+ m^1 T^2 T^3+ m^2 T^1 T^3 + m^3 T^1 T^2- m T^1 T^2 T^3 - h_I N^I +  b_{ij} T^i N^j-  a_i  T^i N^0, 
\end{equation}
which matches the superpotential in \cite{Camara:2005dc} up to the different conventions used here.  Let us now compute the different pieces of the scalar potential in \eqref{eq:potentialmetric}  in terms of the $\rho$'s and the geometric moduli, which enter the K\"ahler potential as $e^K=-1/(2^7 n_0  t_1 t_2 t_3 n_1 n_2 n_3)$.  The RR piece takes the form 
\begin{equation}
V_{\mathrm{RR}}  =  \frac{4 e^K}{\kappa_4^2 l_s^2}  t_1 t_2 t_3 \left\{ \dfrac{1}{t_1 t_2 t_3}\rho_0^2+\dfrac{t_1}{t_2 t_3}  \rho_1^2 +\dfrac{t_2}{t_1 t_3}  \rho_2^2+\dfrac{t_3}{t_1 t_2}  \rho_3^2 + \dfrac{t_2 t_3}{t_1}  \tilde{\rho}_1^2+ \dfrac{t_1 t_3}{t_2}  \tilde{\rho}_2^2+ \dfrac{t_1 t_2}{t_3}  \tilde{\rho}_3^2 + t_1 t_2 t_3  \rho_m^2 \right\}.
\end{equation}
The piece coming from the NSNS part of the action is
\begin{equation}
V_{\mathrm{NS}} = \frac{4 e^K}{\kappa_4^2 l_s^2} \left(n_0^2 \rho_{h_0}^2+n_1^2 \rho_{h_1}^2+n_2^2 \rho_{h_2}^2 +n_3^2 \rho_{h_3}^2 \right).
\end{equation}
The contribution from localised sources as D6-branes or O6-planes that preserve $\mathcal{N}=1$ supersymmetry gets an extra piece from the metric fluxes  with respect to the usual one. The general expression for the whole localised term is given in \cite{Villadoro:2005cu} and using the Bianchi identities for $G_2$ (which include the source terms from the D6-branes and O6-planes) one can express it in terms of the geometric moduli and the fluxes, obtaining the following expression 
\begin{equation}
V_{\mathrm{loc}}=\frac{8 e^K}{\kappa_4^2 l_s^2} t_1 t_2 t_3\left[ (m h_0 + m^i a_i )n_0+(mh_i -  m^j b_{ji} ) n_i) \right].
\end{equation}
The terms inside each parenthesis can be written as linear combinations of some of the $\rho$'s and then this contribution to the scalar potential takes the form
\begin{equation}
\begin{array}{l c l}\vspace*{.2cm}
V_{\mathrm{loc}}&=& \frac{8 e^K}{\kappa_4^2 l_s^2} t_1 t_2 t_3 \Big[n_0  (\rho_m \rho_{h_0} -\tilde{\rho}^i \rho_{a_i})+ n_1 (\rho_m \rho_{h_1} -\tilde{\rho}^i \rho_{b_{i1}}) \\ 
& & \ \ \ +n_2   (\rho_m \rho_{h_2} -\tilde{\rho}^i \rho_{b_{i2}})+n_3 (\rho_m \rho_{h_3} -\tilde{\rho}^i \rho_{b_{i3}}) \Big].\\
\end{array}
\end{equation}

The last term in \eqref{eq:potentialmetric}, the Scherk-Schwarz potential, comes from the dimensional reduction of the purely gravitational part of the action (the curvature scalar) in the presence of metric fluxes. In order to compute it, it is necessary to know the explicit metric of the compact manifold, and this is not the case for a general Calabi-Yau. However, this can be calculated in our toroidal setup since its metric in terms of the geometric moduli is known to be the following\footnote{Here we  denote this metric by  $\tilde{g}_{ij}$ in order to avoid confusions with the previously defined metric on the K{\"a}hler moduli space $g_{ab}$.} \cite{Villadoro:2005cu}:
\begin{equation}
\tilde{g}_{ij}=
\left(
\begin{array}{c c c c c c }
t_1\sqrt{\dfrac{n_0}{n_2 n_3}} & & & & & \\
& t_2\sqrt{\dfrac{n_0}{n_1 n_3}}& & & & \\
& &t_3\sqrt{\dfrac{n_0}{n_1 n_2}} & & & \\
& & & t_1 \sqrt{\dfrac{n_2 n_3}{n_0}} & & \\
& & & & t_2 \sqrt{\dfrac{n_1 n_3}{n_0}} & \\
& & & & & t_3 \sqrt{\dfrac{n_1 n_2}{n_0}}  \\
\end{array}
\right).
\end{equation}
In terms of a general metric $\tilde{g}_{ij}$ and the metric fluxes, the Scherk-Schwarz potential can be written as \cite{Villadoro:2005cu}:
\begin{equation}
\begin{array}{l c l}\vspace*{.2cm}
V_{SS}&=&\dfrac{1}{64\sqrt{-n_0 n_1 n_2 n_3}} \left( \omega^i_{jk} \omega^{i'}_{j' k'} \tilde{g}_{ii'} \tilde{g}^{jj'} \tilde{g}^{kk'}+2\omega^i_{jk} \omega^k_{j'i} \tilde{g}^{jj'}\right)\\ 
&= &\displaystyle\dfrac{1}{64\sqrt{-n_0 n_1 n_2 n_3}} \sum_{i,j,k} \left( \omega^i_{jk} \omega^{i}_{j k} \tilde{g}_{ii} \tilde{g}^{jj} \tilde{g}^{kk}+2\omega^i_{jk} \omega^k_{ji} \tilde{g}^{jj}\right),
\end{array}
\end{equation}
where in the last line the sum is explicitly indicated and applies whenever the metric of the torus is diagonal, as in our case. The result, written in terms of the $\rho_{a_i}$'s, $\rho_{b_{ij}}$'s and the  moduli can be written as the following bilinear:
\begin{equation}
V_{SS}=4 e^K \rho_{v_p} \  M^{p q}   \ \rho_{v_q},
\end{equation}
where we have defined 
\begin{equation}
\rho_{v_p}=
\left(
\rho_{a_1}, \
\rho_{b_{23}}, \
\rho_{b_{32}}, \
\rho_{a_2}, \
\rho_{b_{13}}, \
\rho_{b_{31}}, \
\rho_{a_3}, \
\rho_{b_{12}}, \
\rho_{b_{21}}, \
\rho_{b_{11}}, \
\rho_{b_{22}}, \
\rho_{b_{33}}
\right)
\end{equation}
 and the $12\times12$ matrix $M^{pq}$ is given by
 \begin{equation}
 \begin{split}
 &M^{pq}=  \mathrm{blockdiag}\\ \vspace*{.2cm}
& \left\{
 \left(
 \begin{array}{l c r}
n_0^2 t_1^2 & -n_0 n_3 t_1 t_2 & -n_0 n_2 t_1 t_3 \\
  - n_0 n_3 t_1 t_2  &n_3^2 t_2^2 &-n_2  n_3 t_2 t_3\\
 -n_0 n_2  t_1 t_3 &- n_2 n_3 t_2 t_3 & n_2^2 t_3^2  \\
 \end{array}
 \right),
 \left(
 \begin{array}{c c c}
 n_0^2 t_2^2 & -n_0 n_3 t_1 t_2&-n_0 n_1 t_2 t_3 \\
   -n_0 n_3 t_1 t_2 & n_3^2 t_1^2&-n_1 n_3 t_1 t_3\\
 -n_0 n_1 t_2 t_3 & -n_1 n_3 t_1 t_3 &  n_1^2 t_3^2 \\
 \end{array}
 \right), \right. \\ 
 &\left. \left(
 \begin{array}{c c c}
n_0^2 t_3^2 &- n_0 n_2 t_1 t_3 &-n_0 n_1 t_2 t_3 \\
   -n_0 n_2 t_1 t_3 &n_2^2 t_1^2 &-n_1 n_2 t_1 t_2 \\
 -n_0 n_1 t_2 t_3  &- n_1 n_2 t_1 t_2 & n_1^2 t_2^2  \\
 \end{array}
 \right),
  \left(
 \begin{array}{c c c}
 n_1^2  t_1^2 &- n_1 n_2 t_1 t_2 &-n_1 n_3 t_1 t_3 \\
  - n_1 n_2 t_1 t_2 & t_2^2 n_2^2&-n_2 n_3 t_2 t_3 \\
 -n_1 n_3 t_1 t_3 & -n_2 n_3 t_2 t_3 &n_3^2 t_3^2  \\
 \end{array}
 \right) 
 \right\}. 
 \end{split}
 \label{M4blocks}
 \end{equation}
 One important comment regarding this last result is in order. When computing the scalar potential from the standard $\CN=1$ supergravity formula one obtains a more complicated matrix $M^{pq}$ which further entries than those in \eqref{M4blocks}, and which is not invertible. It is only after applying the Bianchi identities \eqref{BImetric} to constrain the fluxes that $M^{pq}$ becomes block diagonal and invertible. The mixing terms that appear in \eqref{Gmom} will not change this invertibility property, so therefore we find that the bilinear form multiplying the $\rho$'s is only invertible whenever the Bianchi identities have been properly taken into account. Notice that the invertibility of this bilinear form is necessary to have a 4d four-form description of the scalar potential, so it seems that one can only match a Lagrangian of the form \eqref{S4formRR} to the standard F-term potential formula if the Bianchi identities are imposed. It would be interesting to explore the generality of this result and its consequences for further classes of string compactifications. 
  
 Adding all these pieces together, it can be seen that the full potential may be written as a  bilinear in the $\rho$'s, which depend only on the fluxes and the axions, and with bilinear metric depending only on the geometric moduli. The whole scalar potential can be actually written as the following bilinear
\begin{equation}
\label{eq:potmetric}
V=4 e^K \left( (\rho^e)^T \ \ (\rho^{m,h})^T \ \ (\rho^{m,\omega})^T \right) 
\left(
\begin{array}{c|c|c }
G^{e} & 0 &0 \\
\hline
0 & G^{m,h} & 0 \\
\hline
0 & 0 & G^{m, \omega }
\end{array}
\right)
\left( 
\begin{array}{l}
\rho^e \\ \rho^{m,h} \\ \rho^{m,\omega}
\end{array}
\right),
\end{equation}
where we have defined the vectors of $\rho$'s as
\begin{equation}
\rho^e=
\left(
\begin{array}{l}
\rho_0\\
\rho_1\\
\rho_2\\
\rho_3\\
\end{array}
\right), \qquad
\rho^{m,h}=
\left(
\begin{array}{l}
\rho_m\\
\rho_{h_0}\\
\rho_{h_1}\\
\rho_{h_2}\\
\rho_{h_3}\\
\end{array}
\right), \qquad
\rho^{m,\omega}=
\left(
\begin{array}{l}
\tilde{\rho}_1\\
\tilde{\rho}_2\\
\tilde{\rho}_3\\
\rho_{v_p}\\
\end{array}
\right), 
\end{equation}
and the matrices are
\begin{equation}
G^e= 
\left(
\begin{array}{c c c c}
1 & & & \\
 & t_1^2 & & \\
 & & t_2^2 & \\
 & & & t_3^2 \\
\end{array}
\right),
\end{equation}

\begin{equation}
G^{m,h}=
\left(
\begin{array}{c c c c c}
(t_1 t_2 t_3)^2 & n_0 t_1 t_2 t_3 &  t_1 t_2 t_3 n_1  & t_1 t_2 t_3 n_2 & t_1 t_2 t_3 n_3 \\ \\
n_0 t_1 t_2 t_3 & n_0^2 & 0 & 0 & 0 \\ \\
 t_1 t_2 t_3 n_1 & 0 & n_1^2 & 0 & 0 \\ \\
 t_1 t_2 t_3 n_2 & 0 & 0 & n_2^2 & 0 \\ \\
 t_1 t_2 t_3 n_3 & 0 & 0 & 0 & n_3^2 \\ \\
\end{array}
\right),
\end{equation}
\begin{equation}
G^{m, \omega}=
\left(
\begin{array}{ c c c | c c c}
(t_2 t_3)^2 & 0 & 0 & & & \\
0 & (t_1 t_3)^2 & 0 & & (N^T)^{m'n} & \\
0 & 0 & (t_1 t_2)^2 & & & \\
\hline
& & & & & \\
& N^{m n'} & & & M^{mn} & \\
& & & & & \\
\end{array}
\right).
\label{Gmom}
\end{equation}
The indices denoted with primes go from $1$ to $3$ and those without the primes go from $1$ to $12$. The off block-diagonal terms are given by the matrix $N^{m n'}$  and its transpose. It can be read from the potential generated by the localised sources and has the form
\begin{equation}
N^{mn'}= -t_1 t_2 t_3
\left(
\begin{array}{ c c c}
n_0  & 0 & 0 \\
0 & n_3 & 0 \\
0 & 0 & n_2 \\
0 & n_0   & 0 \\
 n_3 & 0 & 0 \\
0 & 0 &  n_1 \\
0 & 0 & n_0 \\
 n_2 & 0 & 0 \\
0 &  n_1 & 0 \\
n_1 & 0 & 0 \\
0 &  n_2 & 0 \\
0 & 0 & n_3 \\
\end{array}
\right).
\end{equation}

Notice that all the off-diagonal terms come from both the localised and the Scherk-Schwarz pieces of the scalar potential. Moreover, all the matrices $G^e$, $G^{m,h}$ and $G^{m,\omega}$ have non-vanishing determinant, so that the matrix that enters eq. \eqref{eq:potmetric} is invertible. This means that the whole scalar potential can be obtained from a 4d effective action like the one in \eqref{S4formRR}. Finally, let us remark that, apart from the bilinear structure of the 4d scalar potential, it is easy to see from the definition of the $\rho$'s in \eqref{ap:rhos} that they can be rotated into a basis in which they are only given by the fluxes, that is, we can find a rotation matrix $\textbf{R}$ that rotates our 4-forms into a basis in which they couple directly to the fluxes, as in all the previous cases. To sum up,  even in the presence of metric fluxes the scalar potential still enjoys the triple factorisation into saxions, axions and fluxes introduced in section \ref{s:4forms}.

For completeness, let us show how we can again express the supergravity auxiliary fields as functions of the $\rho$'s:

\begin{equation}
\begin{array}{r c l}
\bar{F}^{\bar{T}^i}&=&2 e^{K/2} t^i \times \\
		& &\displaystyle  \times \left\{ \left[ \sum_{j\neq i}t^j \rho_j-t^i \rho_i + t^1 t^2 t^3 \rho_m+\sum_{J=0}^3 n^J \rho_{h_J} \right]  \right.\\
		& & \displaystyle  \left. + i 	\left[ \rho_0 - t^j t^k \tilde{\rho}^i + t^i t^j \tilde{\rho}^k+ t^it^k \tilde{\rho}^j-\sum_{j\neq i} n^0 t^j \rho_{a_j} + n^0 t^i \rho_{a_i} 	-\sum_{l=1}^3\sum_{j \neq i} n^l t^j \rho_{b_{jl}} +\sum_{l=1}^3n^l t^i \rho_{b_{il}}\right] 	\right\}, \\ \\
	
	\bar{F}^{\bar{N}^i}&=&2 e^{K/2} n^i \times \\
		& &\displaystyle  \times \left\{ \left[  \sum_{j=1}^3 t^j \rho_j - t^1 t^2 t^3 \rho_m+ n^0 \rho_{h_0}+\sum_{j\neq i}^3 n^j \rho_{h_j}- n^i \rho_{h_i}  \right]  \right.\\
		& & \displaystyle  \left. + i 	\left[ \rho_0 - t^j t^k \tilde{\rho}^i - t^i t^j \tilde{\rho}^k- t^it^k \tilde{\rho}^j-\sum_{j=1}^3 n^0 t^j \rho_{a_j} -\sum_{l=1}^3\sum_{j \neq i} n^j t^l \rho_{b_{lj}} +\sum_{l=1}^3n^i t^l \rho_{b_{li}}\right] 	\right\}, \\ \\

\bar{F}^{\bar{N}^0} & = &2 e^{K/2} n^0 \times \\
		& &\displaystyle  \times \left\{ \left[  \ \sum_{j=1}^3 t^j \rho_j - t^1 t^2 t^3 \rho_m- n^0 \rho_{h_0}+\sum_{j=1}^3 n^j \rho_{h_j}  \right]  \right.\\
		& & \displaystyle  \left. + i 	\left[ \rho_0 - t^j t^k \tilde{\rho}^i - t^i t^j \tilde{\rho}^k- t^it^k \tilde{\rho}^j+\sum_{j=1}^3 n^0 t^j \rho_{a_j} -\sum_{l,j=1}^3 n^j t^l \rho_{b_{lj}}\right] 	\right\}, \\ \\
		
\end{array}
\end{equation}
where all the sums are indicated explicitly and $i\neq j \neq k \neq i$.	

\section{Discrete symmetries in  toroidal $\IZ_2\times \IZ_2$ Type IIA orientifolds}\label{ap:sym}

We describe here how the discrete symmetries discussed in the main text appear as modular symmetries in a simple 
toroidal setting. We consider the Type IIA $\IZ_2\times \IZ_2$ toroidal orientifold discussed in 
 \cite{moredual}, which has in the untwisted sector seven moduli: $T^a$ with $a=1,2,3$ and $N^I$ with $I=0,\dots,4$. The 
 RR fluxes transform as a $(2,2,2)$ representation of the 
tori modular groups $SL(2, \IZ)^3$ under which the three K\"ahler moduli $T^a$ transform 
non-linearly in the usual way.  We can  collect the 8 RR fluxes into  a tensor 
$f^{\alpha \beta \gamma}$, with $\alpha,\beta, \gamma = 1,2$ in the following way:
\begin{eqnarray}
f^{1ab}\ =\ \left(
\begin{array}{c c}
-e_0  &  e_3\\
e_2 &  -m^1
\end{array}
\right) \ ,\  f^{2ab}\ =\ \left(
\begin{array}{c c}
e_1 &  -m^2 \\
-m^3  &  m
\end{array}
\right)  \ .
\end{eqnarray}
Now, the shift generators ${\cal T}^i$ are 
given by
\begin{eqnarray}
{\cal T}^a\ =\ \left(
\begin{array}{c c}
1 &  n^a \\
0 &  1
\end{array}
\right).
\end{eqnarray}
Let us consider now a general shift transformation of this flux tensor, 
\beq
f^{\alpha \beta \gamma} \ \longrightarrow \
{\tilde f}^{\alpha \beta \gamma}={\cal T}^1_{\alpha \rho } {\cal T}^2_{\beta \tau} {\cal T}^3_{\gamma \sigma} f^{\rho \tau \sigma} \ .
\eeq
In particular e.g. for the component  $f^{1ab}$ one obtains
{\footnotesize
\beq
{\tilde f}^{1ab} = {\cal T}^1_{1c}{\cal T}^2_{ad}{\cal T}^3_{br}f^{cdr} =
{\cal T}^1_{11}{\cal T}^2_{ad}f^{1dr}({\cal T}^3)^T_{rb}+{\cal T}^1_{12}{\cal T}^2_{ad}f^{2dr}({\cal T}^3)^T_{rb} \ = \nonumber
\eeq
\begin{eqnarray}
 \left(
\begin{array}{c c}
1   &  n^2\\
0  &   1 
\end{array}
\right) 
\left(
\begin{array}{c c}
-e_0  &  e_3\\
e_2 &  -m^1 
\end{array}
\right) 
\left(
\begin{array}{c c}
1   &  0\\
n^3 &   1 
\end{array}
\right)  + 
 n^1 \left(
\begin{array}{c c}
1   &  n^2\\
0  &   1 
\end{array}
\right) 
\left(
\begin{array}{c c}
e_1  &  -m^2\\
-m^3 &  m
\end{array}
\right) 
\left(
\begin{array}{c c}
1   &  0\\
n^3 &   1 
\end{array}
\right)  \ =   \nonumber
\end{eqnarray}
\begin{eqnarray}
=\left(
\begin{array}{c c}
-e_0+n^ae_a- m^1n^2n^3-m^2n^1n^3-m^3n^1n^2+mn^1n^2n^3   &   e_3-m^1n^2-n^1m^2+mn^1n^2 \\
e_2-m^1n^3-n^1m^3+mn^1n^3  &   -m^1+mn^1  
\end{array}
\right) \ .
\end{eqnarray}
}
This indeed matches the transformations for the RR fluxes described in the main text. One can easily check the transformation for the
other flux components.

In this toroidal case one can also see how the other $SL(2, \IZ)$ generators
\begin{eqnarray}
{\cal S}^i\ =\ \left(
\begin{array}{c c}
0 &   1\\
-1 &  0.
\end{array}
\right).
\end{eqnarray}
act on the fluxes. One finds for the simultaneous duality in all the three complex planes of the torus
{\footnotesize
\beq
{\tilde f}^{1ab} = {\cal S}^1_{12}{\cal S}^2_{ad}{\cal S}^3_{br}f^{2dr} = \left(
\begin{array}{c c}
0   &  1\\
-1  &   0
\end{array}
\right) 
\left(
\begin{array}{c c}
e_1  &  -m^2\\
-m^3 &  m
\end{array}
\right) 
\left(
\begin{array}{c c}
0   &  -1\\
1 &   0 
\end{array}
\right)= 
\left(
\begin{array}{c c}
m &  m^3\\
m^2  &  e_1
\end{array}
\right) 
\eeq
}
which indeed corresponds to the way fluxes transform under  a duality $R^i\rightarrow 1/R^i$  in all three complex
planes, as discussed in \cite{Bielleman:2015ina}. It also corresponds to the duality transformation described 
at the end of section \ref{ss:dss} in the main text.
It would be interesting to explore further the case with NS fluxes.  In this case the fluxes transform in the
$(2,2,2,2,2,2,2,2)$ of $SL(2,\IZ)^7$. One can construct the {\it master polynomial} in this more general case
by setting all saxions to zero in the general superpotential in eq.(6.14) in ref.\cite{moredual}. From here one
can obtain all the shift invariant polynomials associated to all the fluxes, geometric and non-geometric, and write down the potential in terms of them.
An important issue here is the consistency with all the Bianchi identities which would need to be imposed.



\end{document}